\documentclass{aa}
\usepackage{graphicx}
\usepackage{txfonts}
\usepackage{color}
\usepackage{amsmath}
\usepackage{float}
\usepackage[colorinlistoftodos]{todonotes}
\usepackage[normalem]{ulem}
\usepackage{algpseudocode,algorithm,algorithmicx}

\usepackage{multirow, booktabs, colortbl, tabularx, amsbsy}
\usepackage{pifont}
\definecolor{LightGray}{gray}{.8}

\newcommand{\argmin}{\operatorname{argmin}\limits}
\newcommand{\angstrom}{\textup{\AA}}

\newcommand{\norm}[1]{\left\|{#1}\right\|}
\newcommand{\st}{\mathrm{\quad s.t. \quad}}
\newcommand{\abs}[1]{\left|#1\right|}
\newcommand{\snr}{$\mathrm{S/N}$ }

\algblock[AlgBlock]{StartB}{EndB}
\algtext*{StartB}
\algtext*{EndB}

\makeatletter
\newcommand{\Indent}[1]{\hspace{10pt}\State\textbf{#1}}
\makeatother

\begin{document}

   \title{Representation learning for automated spectroscopic redshift estimation}
   \author{J. Frontera-Pons\inst{1,2}, F. Sureau\inst{1}, B. Moraes\inst{3,4}, J. Bobin\inst{1}, F. B. Abdalla\inst{3,5}}
	\institute{Laboratoire AIM, CEA, CNRS, Universit\'e Paris-Saclay, Universit\'e Paris Diderot, Sorbonne Paris Cit\'e, F-91191 Gif-sur-Yvette, France \label{inst1}
	\and DR2I, Institut Polytechnique des Sciences Avanc\'ees, 94200 Ivry-sur-Seine, France \label{inst2}
	\and Department of Physics \& Astronomy, University College London, Gower Street, London WC1E 6BT, UK \label{inst3}
	\and Instituto de Fisica, Universidade Federal do Rio de Janeiro, 21941-972, Rio de Janeiro, Brazil
	\label{inst4}
	\and Department of Physics and Electronics, Rhodes University, PO Box 94, Grahamstown, 6140, South Africa\label{inst5}\\ \email{joana.frontera-pons@cea.fr}
}


\abstract 
{Determining the radial positions of galaxies up to a high accuracy depends on the correct identification of salient features in their spectra. Classical techniques for spectroscopic redshift estimation make use of template matching with cross-correlation. These templates are usually constructed from empirical spectra or simulations based on the modeling of local galaxies.} 
{We propose two new spectroscopic redshift estimation schemes based on new learning techniques for galaxy spectra representation, using either a dictionary learning technique for sparse representation or denoising autoencoders. 
We investigate how these representations impact redshift estimation.}
{We first explore dictionary learning to obtain a sparse representation of the rest-frame galaxy spectra modeling both the continuum and line emissions. Alternately, denoising autoencoders are considered to learn non-linear representations from rest-frame emission lines extracted from the data. In both cases, the redshift is then determined by redshifting the learnt representation and selecting the redshift that gives the lowest approximation error among the tested values.} 
{These methods have been tested on realistic simulated galaxy spectra, with photometry modeled after the Large Synoptic Survey Telescope (LSST) and spectroscopy reproducing properties of the Sloan Digital Sky Survey (SDSS). They were compared to Darth Fader, a robust technique extracting line features and estimating redshift through eigentemplates cross-correlations. We show that both dictionary learning and denoising autoencoders provide improved accuracy and reliability across all signal-to-noise regimes and galaxy types. Furthermore, the former is more robust at high noise levels; the latter is more accurate on high signal-to-noise regimes. Combining both estimators improves results at low SNR.} 
{The representation learning framework for spectroscopic redshift analysis introduced in this work offers high performance in feature extraction and redshift estimation, improving on a classical eigentemplates approach. This is a necessity for next-generation galaxy surveys, and we demonstrate a successful application in realistic simulated survey data.}   
\titlerunning{Representation learning for spectroscopic redshift estimation}

\maketitle


\section{Introduction}

Galaxy redshift surveys are among the main observational tools to probe cosmological models. The leading methods measure the distance scale imprinted in the large-scale distribution of galaxies by oscillations in the primordial baryon-photon plasma \citep{Kazin2014, Alam2017, Bautista2018}. This baryonic acoustic oscillation (BAO) sound horizon can be used as a standard ruler to characterize the expansion rate of the Universe at different times, thereby providing constraints on cosmological parameters such as the total matter and dark energy densities \citep{BlakeGlazebrook2003, SeoEisenstein2003}. A precise measurement of the redshifts of galaxies is fundamental to extract this cosmological information from large galaxy surveys, and it is also key to the supplementary goals of constraining models of galaxy formation and evolution. To achieve these aims, most current and upcoming surveys such as eBOSS and DESI choose to observe the spectroscopic energy distribution (SED) of galaxies in the optical or near-infrared wavelength range with multiplexed fiber spectrographs \citep{Dawson2016, Aghamousa2016}.

Spectroscopic redshift estimation methods are typically based on the identification and fitting of spectral features - such as emission and absorption lines from electronic transitions in different elements - or of distinctive continuum features - such as the 4000\angstrom{} break due to the absorption of high-energy photons from metals in stellar atmospheres and the reduced number of hot blue stars in old galaxies \citep[e.g.][]{Baldry04, Hutchinson2016}. Especially for bluer, higher-redshift galaxies without a prominent continuum break, one of the main hurdles for redshift estimation is the identification of relevant spectral features; high noise levels may introduce features that could be interpreted as physical lines if the analysis is too sensitive to noise, or true features might not be identified if attempts are made to mitigate false positive detections \citep{Machado13}.

Another significant difficulty is ensuring that the spectral templates upon which many fitting methods depend are physically consistent and sufficiently representative of the observed galaxy population in a particular survey \citep{Bautista2018}. Redshifts are generally determined by cross-correlation or $\chi^2$ fitting between observed spectra and a reference set of spectroscopic templates \citep{Glazebrook98}. These template-matching methods strongly rely on a catalog of galaxy spectra at zero redshift to which the unknown redshift galaxies will be compared to, namely the template set. A template set can be prohibitively large if we wish to ensure correct retrieval of most of the significant features of an observed spectrum; additionally, there might be many degeneracies in the information spread throughout the full template set.

Former approaches exploit Principal Component Analysis (PCA) to reduce the dimensionality of the problem and summarize the most relevant signal features in a set of principal components \citep[most famously][]{Glazebrook98}. One can then choose to retain a certain number of derived eigentemplates based on an ``energy'' metric that summarizes the amount of information captured. However, the resulting representation would only be efficient if the whole catalog shared common features (continuum, emission lines) that could be encoded efficiently with a few orthogonal templates. In other words, each eigentemplate used to represent a galaxy spectrum will probably contain a combination of features not specific to the galaxy we need to represent.

Moreover, modern large-scale surveys observe increasingly large data sets of galaxy spectra. In this context, although visual identification of the key features in the spectrum constitutes the most common method for validating galaxy redshift estimation \citep{Hinton2016}, the huge amount of data impels the development of robust and fully automated data-processing schemes to analyze the data and extract useful information such as the redshift.

We explore in this article unsupervised feature-extraction from galaxy spectra through modern learning techniques. Notably, we investigate dictionary learning for sparse decomposition and denoising autoencoders for spectra representation. 
Compared to PCA, dictionary learning with sparse representation is much more efficient to capture features that are not shared among the training data (such as combination of emission lines for instance) or different structures in the data (e.g. lines and breaks). It is therefore a good candidate for robust representation of structures specific to the tested spectra, leading to robust redshift estimation. Denoising autoencoders were selected for their ability to capture complex non-linear features present in the data, as already illustrated in \citep{frontera2017unsupervised}
Ultimately, we exploit these new representations for spectroscopic redshift estimation. We assess the relative performance of the two resulting algorithms by comparing them with the redshift estimation code Darth Fader \citep{Machado13}, based on cross-correlating estimated line features with eigentemplates learnt from a training set. We also investigate whether combining the results of both proposed estimators improves the performance in redshift estimation.

This paper is organised as follows. In section \ref{sec:DF}, the Darth Fader algorithm that is used for comparison is briefly recapped. The section \ref{sec:dicolearn} is devoted to presenting dictionary learning for spectra representation and its application for redshift estimation. In section \ref{sec:dae}, we detail the denoising autoencoder architecture and its corresponding redshift estimation scheme. 
Section \ref{sec:data} describes the simulated galaxy spectroscopic data used in the analysis. Section \ref{sec:results} describes the different code configurations and analysis the results of the runs, comparing the performances of the different methods. Section \ref{sec:conclusion} summarizes the results of the paper. 

\section{Redshift estimation with Darth Fader}\label{sec:DF}
{
Darth Fader is a robust redshift estimation code \citep{Machado13}. Line features are firstly extracted from the galaxy SED, and then cross-correlated with eigentemplates as in traditional redshift estimation methods \citep{Glazebrook98}. The robustness of the method comes from the robustness of the line feature extraction step, as well as a criterion to estimate redshift only when a sufficient number of line features are detected. This significantly improves redshift measurement performance and is one of the main advantages of Darth Fader over the alternatives \citep{Machado13}. Lines are estimated from the spectra using wavelet filtering and sparsity to remove continuum emission and represent line features in a galaxy SED \citep{Machado13}. Wavelets are particularly suited for these tasks, given that measured SEDs are composed of a slowly-varying continuum with mostly-uncorrelated high-frequency noise and a few very sharp emission and absorption features. In the following we describe the main steps and features used in this approach to estimate the redshift.
}

\subsection{Spectra modeling}
{
To perform feature extraction and denoising, Darth Fader assumes that it can model spectra as a combination of lines, noise and continuum:
\begin{equation}
\label{LNC}
S = L + N + C \, ,
\end{equation}
where after continuum subtraction lines can further be broken down between emission and absorption lines, $L_e > 0$ and $L_a < 0$. It further assumes that line features are only important on small and intermediate wavelength scales and that the continuum possesses solely large scale information. These assumptions are not rigorously true; in particular, strong lines can spread to larger scales and contribute to the continuum. Conversely, weak lines can be confused with noise in a low signal-to-noise regime. In the next paragraphs, we describe how Darth Fader deals with these issues.
}
\subsubsection{Continuum subtraction}\label{app:continuum}
{
To subtract the continuum, Darth Fader first identifies strong emission and absorption lines, and extracts them from the original spectrum using a pyramidal median transform \citep{Starck1996}. The reason for this choice of transform is that strong lines will be flux outliers compared to the continuum, hence the advantage of a median transform. Furthermore, it is a multiscale transform, which means that it filters features of varying widths. Applying the transform, outliers are identified and removed. The remaining spectrum is a good representation of continuum and noise. Darth Fader then applies a starlet transform \citep{Starck2015} to this representation to identify and remove the continuum. The starlet transform is a particular form of a wavelet transform - an undecimated isotropic wavelet transform - which decomposes the signal as follows:
\begin{equation}\label{starlet}
S_{\lambda} = c_{J}(\lambda) + \sum_{j=1}^{J} w_j(\lambda) \;,
\end{equation}
where the $w_j$ are the details at scale $2^{-j}$. The largest scale coefficient, $c_J$ will be the best representation of the continuum, which can then be subtracted from the original SED.
}
\subsubsection{Line feature denoising}
{
Once the continuum is subtracted, the key step is to separate lines from noise. Darth Fader employs sparsity constraints in a wavelet representation to achieve this goal in low signal-to-noise regimes. The shape of the spectral line features suggests that a particular choice of basis functions - in this case, a family of wavelets - will ensure that the decomposed signal contains at most a few significantly non-zero coefficients. Imposing this sparsity condition by minimizing a $\ell_1$ norm on the wavelet-transformed line signal, and enforcing additional constraints on line and input wavelet coefficients, we can reconstruct the solution $\hat{L}$ that best matches the input data. More rigorously, we consider the following minimization problem to consecutively estimate emission and absorption lines:
\begin{equation}
\min_{L} \norm{\mathbf{\hat{\mathcal{W}}} L}_1 \quad \st \quad S \in \mathcal{C},\\
\end{equation}
where $\mathbf{\hat{\mathcal{W}}}$ is the wavelet transform operator, $\norm{.}_1$ is the $\ell_1$ norm promoting sparsity - so that overall we look for a sparse solution in the wavelet domain - and $\mathcal{C}$ is a convex set of constraints. $\mathcal{C}$ is the intersection of a set of positive (resp. negative) constraints on emission lines $L_e$ (resp. absorption lines $L_a$) and a data fidelity constraint built as follows:
\begin{equation}
\abs{ { w_j^{[S]}(\lambda) - w_j^{[L]}(\lambda)}} \le \varepsilon_{j} , \; \forall \; (j,\lambda) \in \mathcal{M} \; ,
\end{equation}
where $\cal{M}$ is the multi-resolution support. This multi-resolution support is first built by detecting significant wavelet coefficients at scale j and wavelength $\lambda$ in the continuum-free spectrum using a prescribed threshold based on false discovery rate (FDR) for each wavelet scale \citep{Starck2006}. This method ensures that, on average, false positives generated by noise will be kept at a chosen level, which has been shown to be a more efficient method for detection of features in low signal-to-noise data regimes than alternatives such as K$\sigma$ clipping.
}
\subsection{Eigentemplates construction and redshift estimation}
{
To estimate redshift using these extracted line features, eigentemplates are first constructed from a training set of rest-frame line features, by removing the continuum of a set of noise-free galaxy spectra as previously described. The tested spectra processed for line feature extraction as described above are then cross-correlated with these eigentemplates to derive a redshift estimate as in traditional redshift estimation methods  \citep{Glazebrook98}. For the purposes of providing benchmark results for the new methods developed in this paper, we will not preselect the galaxies by counting the number of line features in the spectra as was  done in \citep{Machado13}, to illustrate the raw performance of the methods.
}

\section{Redshift estimation with dictionary learning}\label{sec:dicolearn}

	The first learning technique we propose for redshift estimation relies on learning a representation for the full galaxy spectrum (i.e. continuum and emission lines) with a dictionary learning approach, assuming that spectra can be sparsely decomposed in such dictionary (i.e. only a few atoms are used  or similarly a few coefficients are non-zero for each decomposition). Estimating the redshift is then performed by finding the redshift where the  sparse decomposition in this dictionary leads to the lowest approximation error. In the following, we first present our motivation for using dictionary learning to represent galaxy spectra, then describe how we perform dictionary learning, and explain how such adaptive dictionary can be used for redshift estimation.
    
\subsection{Motivation}

    Dictionary Learning techniques have been proposed in the early 2000s \citep{Olshausen97, Engan99, Aharon06} and have since been applied to many restoration problems \citep[e.g.][]{Elad06, Mairal08a,Mairal09,Zhang10}. Contrary to methods relying on principal component analysis (PCA) (e.g. \citet{Glazebrook98,Machado13}) where template information is compressed in several orthogonal eigentemplates learnt from data or simulations, these techniques rather learn correlated templates assuming the observed spectra can be sparsely represented in a dictionary obtained from the data. Such techniques are therefore adapted to learn features (such as combination of emission lines for instance) or different structures in the data (e.g. lines and breaks) that are not necessarily common to all data but representative of a subset of it and are potentially correlated, whereas PCA would rather extract orthogonal features common to all data. 
    This makes dictionary learning a good candidate for galaxy spectrum representation and redshift estimation, considering that it is unlikely that we can obtain a close sparse approximation of a tested spectrum if we redshift this learnt representation using an incorrect redshift value.

 \subsection{Dictionary learning for galaxy spectrum representation}    
 
      To fix our notations, a spectrum $\mathbf{x}\in \mathbb R^{w_s}$ is approximated by a sparse decomposition $\mathbf D \boldsymbol{\alpha}$ in a dictionary $\mathbf{D}\in\mathbb R^{w_s \times n_a}$ with $n_a$ atoms and with only a few coefficients of $\boldsymbol{\alpha}$ different from zero. $\mathbf{D}$ is derived from a training set of $n_t$ examples $\mathbf X \in  \mathbb R ^{w_s \times n_t}$ by solving the following bilinear minimization problem:
\begin{equation}
\label{eq:dictionarylearning}
\underset{\mathbf D \in \mathcal D, \mathbf A}{\argmin} \, \,||\mathbf X - \mathbf D \mathbf A ||^2_F \,\, \text{s.t.}\,\, \forall i, \,\, ||\boldsymbol{\alpha}_i ||_0 \leq \tau
\end{equation}
where ${\mathbf A} \in \mathbb R^{n_a \times n_t}$ is the matrix containing the coefficients $\{ \boldsymbol{\alpha}_i\}_{i=1..n_t}$ as columns for each training example. $|| \cdot ||_F$ denotes the Frobenius norm, $|| \cdot ||_0$ counts the number of non-zero entries of a vector, $\tau$ enforces a targeted sparsity degree, and $\mathcal{D}$ designates the set of dictionaries with atoms in the unit $\ell_2$ ball. 

The training set used for learning can either be derived from real or simulated data. In practice, the critical point lies on the choice of a representative training set in terms of spectral variety and an observed wavelength range large enough to encompass the band of wavelength for testing the entire probed redshift range. The training set is then constructed by blueshifting these spectra to obtain rest-frame data that is used to learn a dictionary. 

In the training phase, the joint nonconvex problem described in
Eq.~\ref{eq:dictionarylearning} is typically handled by using an alternate minimization strategy, alternating sparse coding steps with dictionary updating steps as illustrated in Algo.~\ref{Algo:DL}. The former is performed by minimizing Eq.~\ref{eq:dictionarylearning} with respect to $\mathbf A$ for $\mathbf D$ fixed to its previous estimate. The latter corresponds to minimizing Eq.~\ref{eq:dictionarylearning} with respect to $\mathbf D$ for $\mathbf A$ fixed to the previously estimated codes. Both steps can be achieved using standard algorithms. In this work, we will use the classical dictionary learning Method of Optimal Directions (MOD) as detailed in \citet{Engan99} with Orthogonal Matching Pursuit (OMP) \citep{Mallat93,Pati93} as a sparse coder.
 
Because the problem described in Eq.~\eqref{eq:dictionarylearning} is non-convex, initialization of the dictionary is important so as not to obtain a non-meaningful local minimum. Furthermore, we would like to learn both continuum and line features on spectra to preserve as much information as possible (to be robust to noise and to reduce line confusion). In order to capture such a variety of features, we propose to use the  procedure described in Algo.~\ref{Algo:DLSpectra} to initialize the learning algorithm. We first separate lines from continuum in our rest-frame training set by masking known line emission bands and extrapolating the resulting data in the region of the mask (step 1). A dictionary is then learnt for the line features and a second one for the extracted continuum  (step 2). Finally, we concatenate the two dictionaries to initialize the learning procedure for the dictionary to represent both continuum and lines (step 3). The global dictionary is learnt, with a targeted sparsity degree $\tau$ given by the sum of the targeted sparsity degrees selected to derive the two sub-dictionaries.

\begin{algorithm}
  \caption{Dictionary Learning with MOD \citep{Engan99} \label{Algo:DL}}
  \begin{algorithmic}[1]
  \State Initialization: Choose the number of atoms $n_a$, the targeted sparsity degree $\tau$, initialize the dictionary. Choose the number of iterations $N_{it}$.
  \For{$n=0$ to $N_{it}$} \Comment{\textbf{Main Learning Loop}}
  \For{ $i=1..n_t$}\Comment{\textbf{Sparse Coding}} 
        \State Compute the sparse code $\boldsymbol{\alpha}_i$ using OMP with stopping criterion $\|\boldsymbol{\alpha}_i\|_0<\tau$
  \EndFor
  \State Update $\mathbf{D}$ using MOD
  \Comment{\textbf{Dictionary Update}}
  \EndFor\\
  \Return $\mathbf{D}$
  \end{algorithmic}
\end{algorithm}

\begin{algorithm}
  \caption{Dictionary Learning for galaxy spectra representation\label{Algo:DLSpectra}}
  \begin{algorithmic}[1]
  \StartB \textbf{Initialization step:}
   \Indent{Line/Continuum separation:} From the original training set $\mathbf X$, obtain two training sets:  $\mathbf{X_{L}}$ for line features and $\mathbf{X_{C}}$ for continuum extraction. 
   \Indent{Sub-dictionary learning:} Choose the number of atoms $M_{L}$ (resp. $M_{C}$), a targeted sparsity degree $\tau_L$ (resp. $\tau_C$), and a number of iterations  $N_L$ (resp $N_C$). Use Algo.~\ref{Algo:DL} to learn a dictionary for lines $\mathbf{D_{L}}$ based on $\mathbf{X_{L}}$ and a dictionary for continuum $\mathbf{D_C}$ based on $\mathbf{X_C}$, with a dictionary initialized by randomly picking training examples.
    \Indent{Concatenation:} concatenate $\mathbf{D_L}$ and $\mathbf{D_C}$ to obtain a dictionary $\mathbf{D_T}$ with $M_L+M_C$ atoms.
 \EndB
 \StartB \textbf{Dictionary Learning:}
 \Indent{} Use Algo.~\ref{Algo:DL} to learn a dictionary $\mathbf{D}$ from the original training set $\mathbf X$, setting the number of atoms $M_L+M_C$ and the targeted sparsity degree $\tau_L+\tau_C$, with $N_T$ iterations, with the initial dictionary $\mathbf{D_T}$.
 \EndB
 \Return $\mathbf{D}$
 \end{algorithmic}\end{algorithm}

\subsection{Redshift Estimation}

Once the dictionary has been built, the redshift for the tested spectra can be estimated using a cross-matching procedure. For a tested redshift value, the atoms of the dictionary are redshifted and the best sparse decomposition is computed using the same targeted sparsity degree as in the training phase. Finally, for each spectrum, the redshift is chosen as the one providing the best sparse approximation among all the evaluated redshift values.

More precisely, for an observed spectrum $\mathbf{x}_z$, at a certain redshift ${z}$, our estimate is \\ 
\begin{equation}\label{eq:zest}
\hat{z} =  \,  \underset{t\in\mathcal{T},\boldsymbol{\alpha}^{(t)}}{\argmin}\, \, \cfrac{ \,|| \mathbf x_z - \mathbf{D}^{(t)} \boldsymbol{\alpha}^{(t)}||_2^2}{|| \mathbf x_z ||_2^2} \,\,\,\mathrm{s.t.} \,\, ||\boldsymbol{\alpha}^{(t)} ||_0 \leq \tau
\end{equation}
where $\mathbf{D^{(t)}}$ is the dictionary computed previously whose atoms have been redshifted by $t$ accounting for the observed wavelength range,  $\boldsymbol{\alpha}^{(t)}$ are the corresponding sparse coefficients estimated using OMP with the same targeted sparsity degree $\tau$ as in the training phase, and $\mathcal{T}$ is a grid of tested redshift values that should be sufficiently finely sampled to typically avoid line confusion. 

This approach therefore assumes that whenever the tested redshift is incorrect, a sparse decomposition in the redshifted dictionary cannot adequately approximate the signal, because all features captured in the dictionary do not match the observed spectrum. Note that the extreme case of a sparsity degree of one in the testing phase (which we will not use) would lead to select the best matching atom in our dictionary.

\section{Redshift estimation with denoising autoencoders}\label{sec:dae}

In this section, we investigate another type of representations for spectroscopic data built with a deep learning architecture, namely the denoising autoencoders. Autoencoder architectures define a direct encoding function that transforms the input into a more suitable representation and a decoding function that reconstructs the corresponding input signal \citep{bourlardkamp88}. More suitable representations preserve a significant amount of information which allow reconstruction of the original signal. We detail in this section the application of denoising autoencoders for unsupervised line feature extraction in spectroscopic data. Ultimately, the learnt features will be used for redshift estimation.

\subsection{Motivation}

\noindent Recent advances in machine learning and deep learning techniques have shown their  capability in solving supervised tasks. They provide state of the art results in classification for computer vision (e.g., \citep{krizhevsky2012imagenet}), speech recognition (e.g. \citep{hinton2012deep}; \citep{dahl2012context}), natural language processing (e.g., \citep{collobert2011natural}), galaxy surface analysis (e.g., \citep{tuccillo2017deep}); among other applications.
Moreover, representation learning methods have been praised as a powerful tool to derive unsupervised data-driven representations \citep{bengio2013representation}. 
These methods allow to design features that efficiently unfold complex underlying structures contained in the data. Unsupervised feature-extraction techniques such as denoising autoencoders have been successfully exploited and compared to PCA for galaxy spectral energy distribution (SED) representation in \citep{frontera2017unsupervised}. In this work, the denoising autoencoders ability to capture useful information, such as the redshift, has been highlighted which motivates the study of this model for galaxy spectrum representation and redshift estimation. 

\subsection{Denoising autoencoders for template representation}

In the classical autoencoder framework, the \emph{encoder} $f_{\boldsymbol{\theta}}$, provides the representation vector from the input galaxy spectrum $ \mathbf h_i = f_{\boldsymbol{\theta}} (\mathbf x_i) $, where $\mathbf x = [x_1, \dots, x_m]^T \in \mathbb R^m $ corresponds to the spectrum with only extracted line features for each galaxy in the considered population $\{\mathbf x_1, \dots \mathbf x_N \}$, $\mathbf h_i \in \mathbb R^{nhid}$ is the feature vector or code and $nhid$ is the number of hidden units or the dimension of the representation vector. Analogously, the \emph{decoder} $g_{\boldsymbol{\theta}}$, projects from the code space back into the input space, yielding a reconstruction of the original spectrum, $\hat{\mathbf x}_i = g_{\boldsymbol{\theta}} (\mathbf h_i)$. 
More specifically, these functions are usually written as affine transformations, typically followed by a non-linearity, $  f_{\boldsymbol{\theta}} (\mathbf x) = s_f(\mathbf b_f + \mathbf W_f \,\mathbf x)$ and $ g_{\boldsymbol{\theta}} (\mathbf h) = s_g (\mathbf b_g + \mathbf W_g \,\mathbf h)$ where $s_f$ and $s_g$ are the encoder and decoder activation functions and ${\boldsymbol{\theta}}$ is the set of parameters that characterise the encoder and the decoder.
Common options for the activation functions include the element-wise sigmoid, the hyperbolic tangent non-linearity or the identity function, if staying linear, among others. 
The parameters $\mathbf b_f$ and $\mathbf b_g$ are the bias vectors of the encoder and decoder respectively and $\mathbf W_f$ and $\mathbf W_g$ are the encoder and decoder weight matrices. Different weight matrices in the encoder and decoder are permitted in the architecture. However, weight-tying, in which one defines $\mathbf W_g = \mathbf W_f^T$, is most often adopted and so it will be assumed hereafter. Moreover, the bias vectors $\mathbf b_f$ and $\mathbf b_g$ have not been considered in this work to construct the representation.

The optimization of the parameters is performed to minimize the reconstruction error for the galaxy spectra, $L(\mathbf x,\hat{\mathbf x})$ over all the samples in the training population. Therefore, the cost function can be written according to: 
\begin{equation}\label{J_AE}
 J_{AE} (\boldsymbol{\theta}) = \displaystyle \sum_{i = 1}^N L ( \mathbf x_i, g_{\boldsymbol{\theta}}(f_{\boldsymbol{\theta}}(\mathbf x_i)))  
\end{equation}
This minimization is generally carried out by stochastic gradient descent. The choice of the reconstruction error measure $L(\cdot)$ depends on the input data domain range and nature. In other words, it is selected so that $L(\cdot)$ returns a negative log-likelihood for the observed value of $\mathbf x$. The mean squared error loss has been used for feature extraction, 
$L(\mathbf x,\hat{\mathbf x}) =|| \mathbf x - \hat{\mathbf x}||^2$.\\
It is worth mentioning that with this configuration the basic autoencoders could learn the identity function to perfectly reconstruct the input. In order to avoid this trivial solution, some regularization constraints should be included during the training stage. Some studies, like \cite{rifai2011contractive} and \cite{alain2014regularized}, underline the improvement brought by regularized autoencoders compared to the basic autoencoders framework. The purpose of this regularization is to render the representation invariant to local variations in the input.

In this article we focus on denoising autoencoders. In this case, the regularization by denoising makes the whole transformation robust and insensitive to small random perturbations in the input. Other variations could be explored such as under-complete representations that allow for a compression of the input data, or over-complete representations imposing sparsity on the code \citep{coates2011analysis}.

\begin{figure*}
\centering
\includegraphics[width=0.48\textwidth]{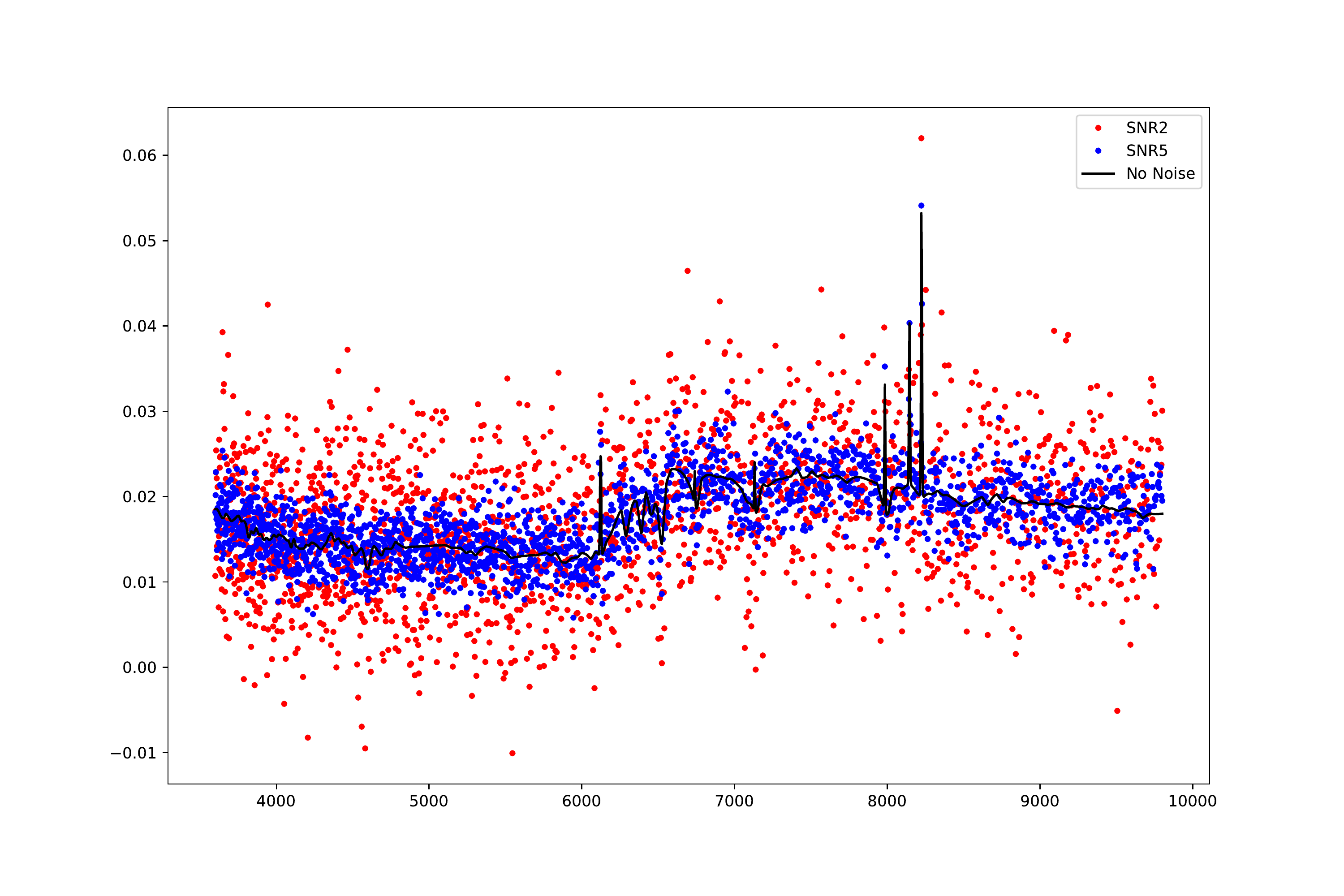}
\includegraphics[width=0.48\textwidth]{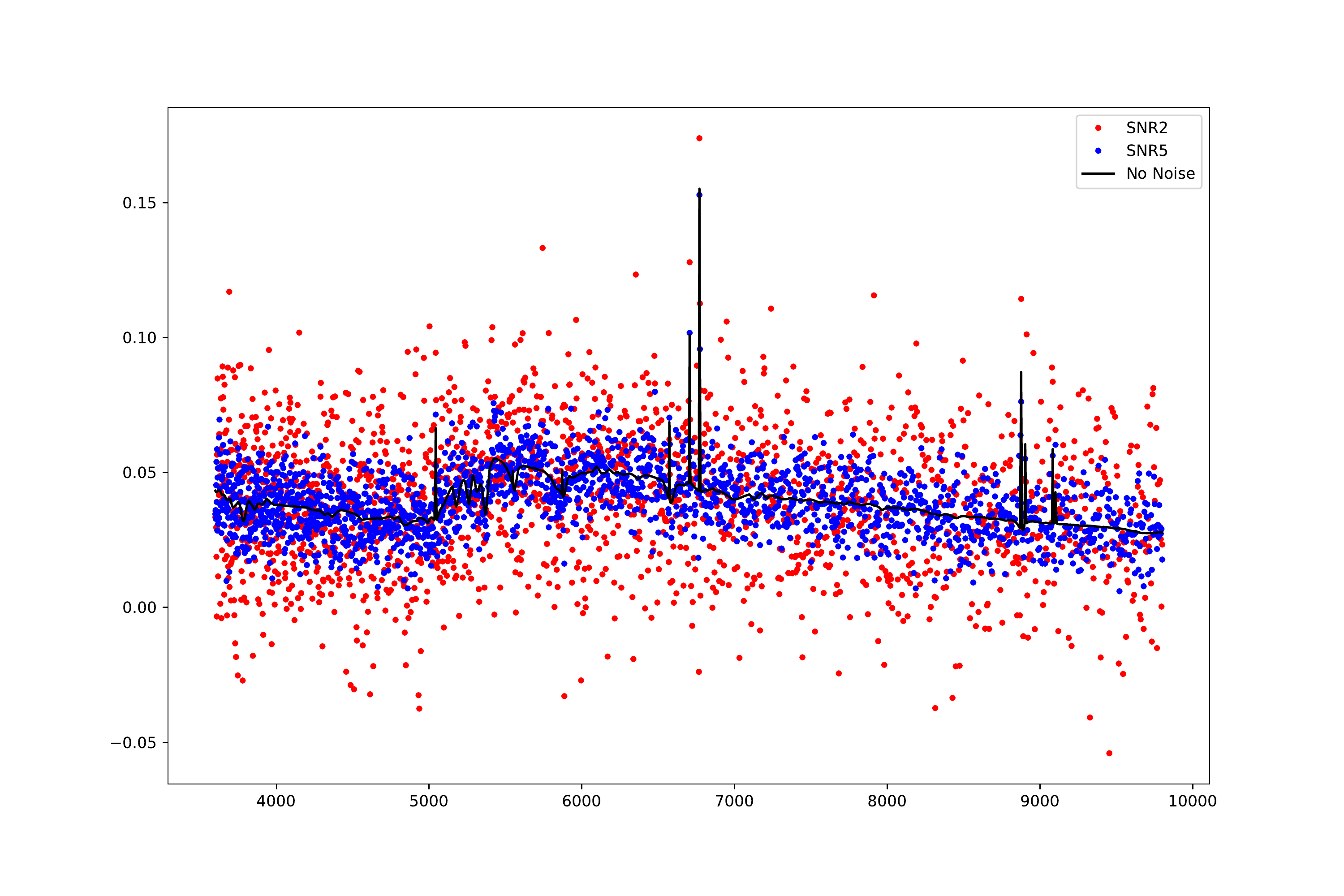}
\includegraphics[width=0.48\textwidth]{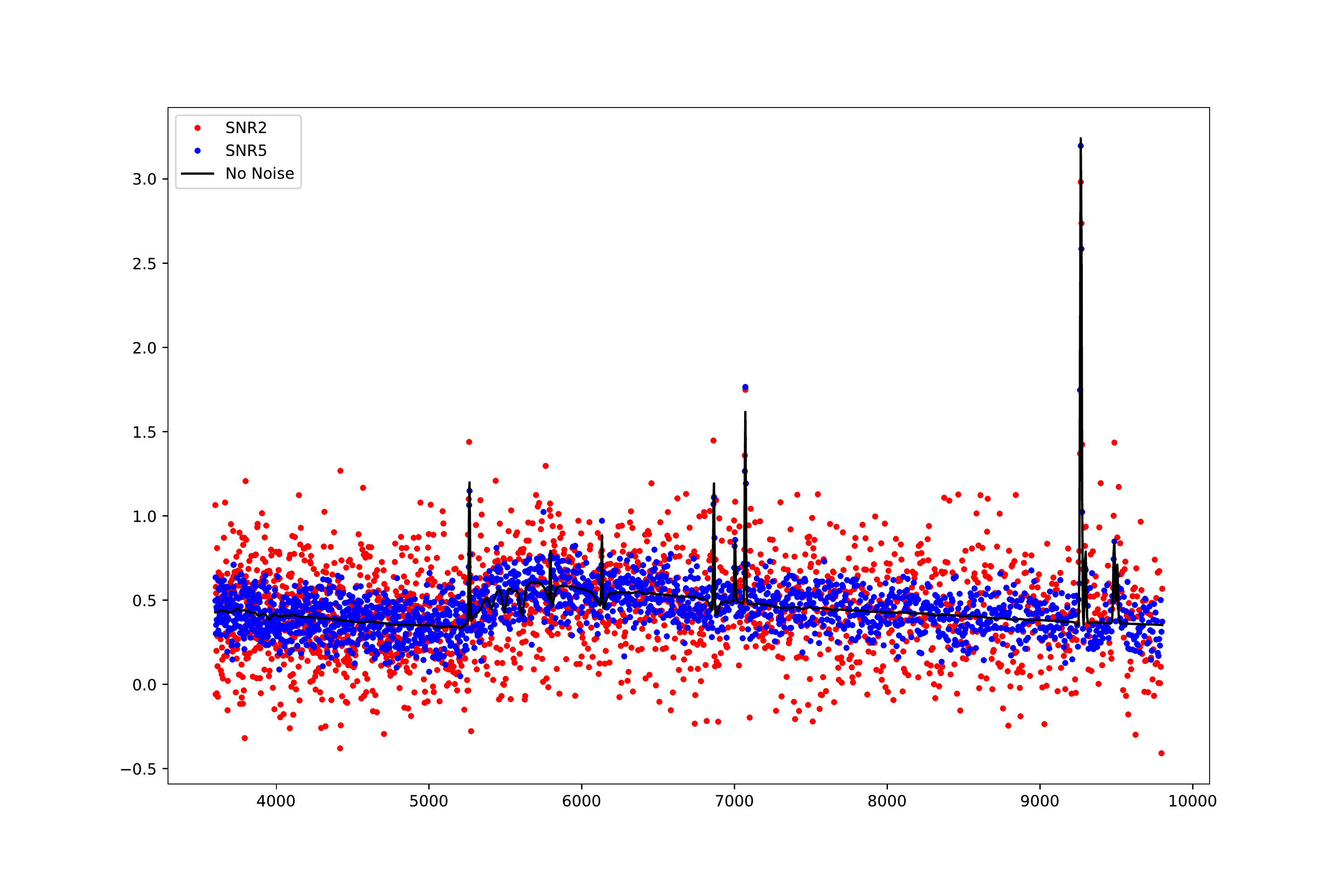}
\includegraphics[width=0.48\textwidth]{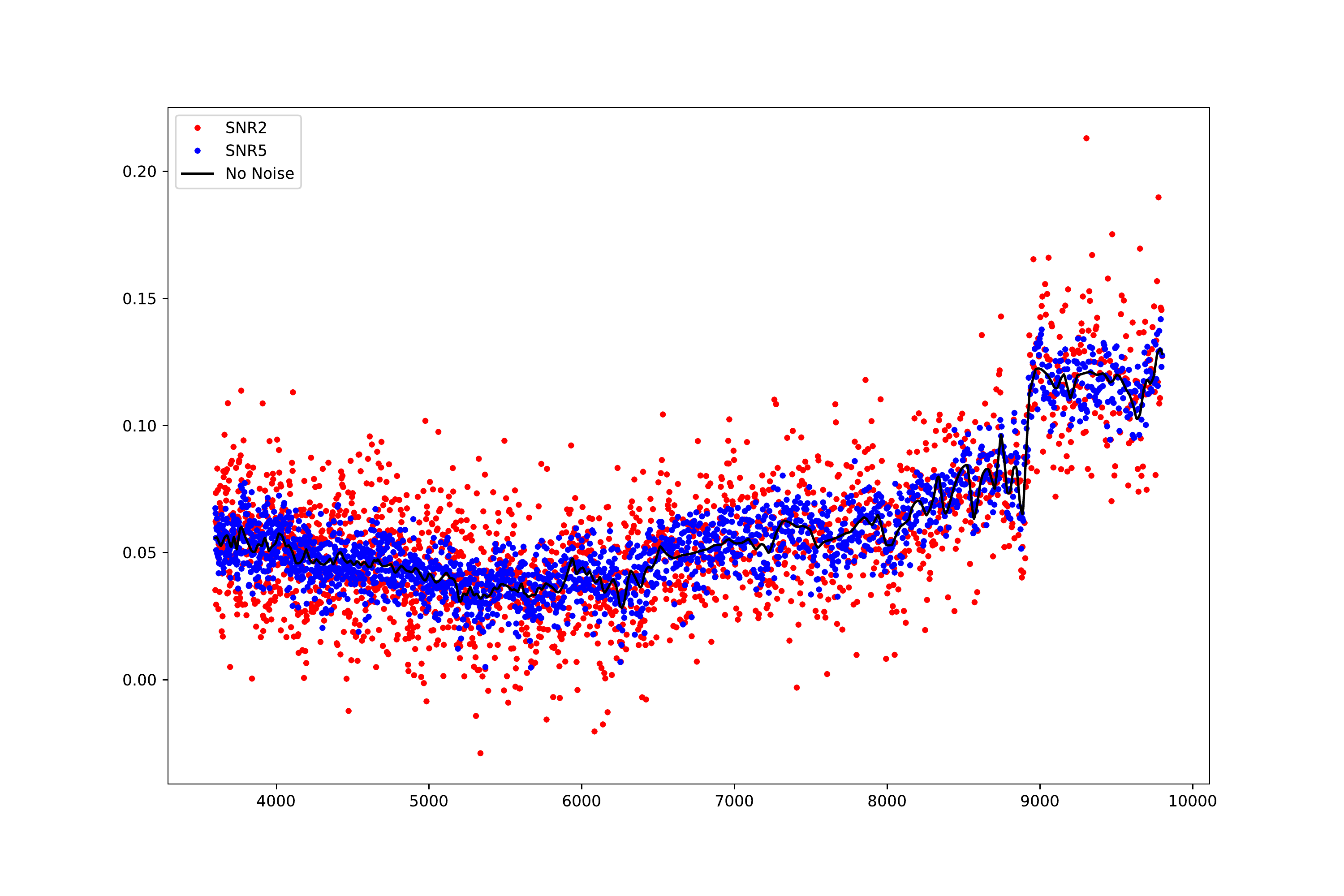}
\caption{Example of test set spectra with various SNR (measured as in Darth Fader paper) in the white Gaussian noise scenario.}
 \label{fig:filters}
\end{figure*}

Denoising autoencoders were originally introduced by \cite{vincent2008extracting}. In this approach, the training objective in Eq. \eqref{J_AE} is modified to recover a clean input spectrum from an artificially corrupted version of it. Specifically, the cost function to be minimized becomes,
\begin{equation}
 J_{DAE} (\boldsymbol{\theta}) = \displaystyle \sum_{i = 1}^N  \mathbb E_{q(\tilde{\mathbf x}| {\mathbf x_i})} [ L ( \mathbf x_i, g_{\boldsymbol{\theta}}(f_{\boldsymbol{\theta}}(\tilde{\mathbf x}))) ] 
\end{equation}
where $\mathbb E_{q(\tilde{\mathbf x}| {\mathbf x_i})}[\cdot]$ denotes the expectation over all the corrupted samples in the training population and  $J_{DAE}$ is optimized by stochastic gradient descent. Therefore, the recovered signal does not seek a perfect reconstruction of the original galaxy spectrum $\mathbf x$, but to retrieve the mean of the distribution that might generate $\mathbf x$. The different corruption processes discussed in \cite{vincent2008extracting, vincent2010stacked} involve additive Gaussian noise, salt and pepper noise, or masking noise. If some prior knowledge about the kind of perturbation the data might encounter is available, it can be incorporated in the corruption stage to make the model robust against this perturbation. Otherwise, the above mentioned corruptors are useful in most scenarios. Moreover, the underlying structure and the information contained in the galaxy population have to be retained by the scheme in order to undo the effect of the corruption process, i.e. perform denoising. Good generalization of the model translates to a low reconstruction error for galaxies with similar characteristics to those in the training population, while yielding high reconstruction error for most other configurations.\\


\subsection{Redshift estimation}

The denoising autoencoders presented above are used to learn representations from rest-frame spectroscopic data. Then, the redshift is estimated for each spectra as the value providing the smallest reconstruction error from the model that has been redshifted in order to match the observed test spectra. Specifically, the model is built from a catalogue of galaxy spectra at zero redshift denoted the training set as for the dictionary learning framework. The denoising autoencoder architecture is optimized in order to minimize the reconstruction error for the samples in the training set. These samples have to be representative of the expected galaxy population in the analysis.

After training, for every redshift value evaluated, the model parameters are redshifted. Thereupon, the tested spectra are projected to the representation space through the encoder function defined by the denoising autoencoder and projected back to the input space with the decoder, leading to an approximation of the input signal. The redshift is estimated as the value minimizing this approximation error. In other words, we hope that, when the discriminating features of the test spectra will be aligned with their rest-frame counterparts used for the training stage, the model will be able to reconstruct the input signal with small error while yielding a large reconstruction error in any other cases. The error is computed using an Euclidean metric, and for each test sample the redshift is obtained according to:

\begin{equation}\label{eq:zest_dae}
\hat{z} =  \,  \underset{t\in\mathcal{T}}{\argmin}\, \,  \cfrac{ \,|| \mathbf x_z - g_{\boldsymbol{\theta}^{(t)}}(f_{\boldsymbol{\theta}^{(t)}}({\mathbf x}_z)) ||_2^2}{|| \mathbf x_z ||_2^2}
\end{equation}
where ${\boldsymbol{\theta}^{(t)}}$ denotes the denoising autoencoder model redshifted at $t$ and $\mathcal{T}$ is the grid of tested redshift values. In other words, the columns of the weighting matrix $W$ are redshifted and treated similarly as the atoms in a dictionary described in Section \ref{sec:dicolearn}.

\section{Data}\label{sec:data}

In order to assess how these two new methods perform for spectroscopic redshift estimation in a realistic setting, we wish to use a simulated data set consisting of a combination of photometric and spectroscopic data mimicking modern galaxy surveys. We require that this data obey the following constraints:

\begin{enumerate}[(i)]
  \item There is a realistic distribution of galaxy types, photometric properties and redshifts, corresponding to an idealised selection function for a state-of-the-art photometric galaxy survey.
  \item Each galaxy is consistently matched to a corresponding SED from a template library containing realistic continuum, emission and absorption features. The matching must ensure that the integrated flux through broadband filters corresponding to the original photometric sample is consistent with the original observations.
  \item The SEDs are resampled, integrated and noise is added to simulate realistic spectra from existing or planned galaxy spectroscopic surveys.
\end{enumerate}

\begin{figure*}
\centering
\includegraphics[width=0.48\textwidth]{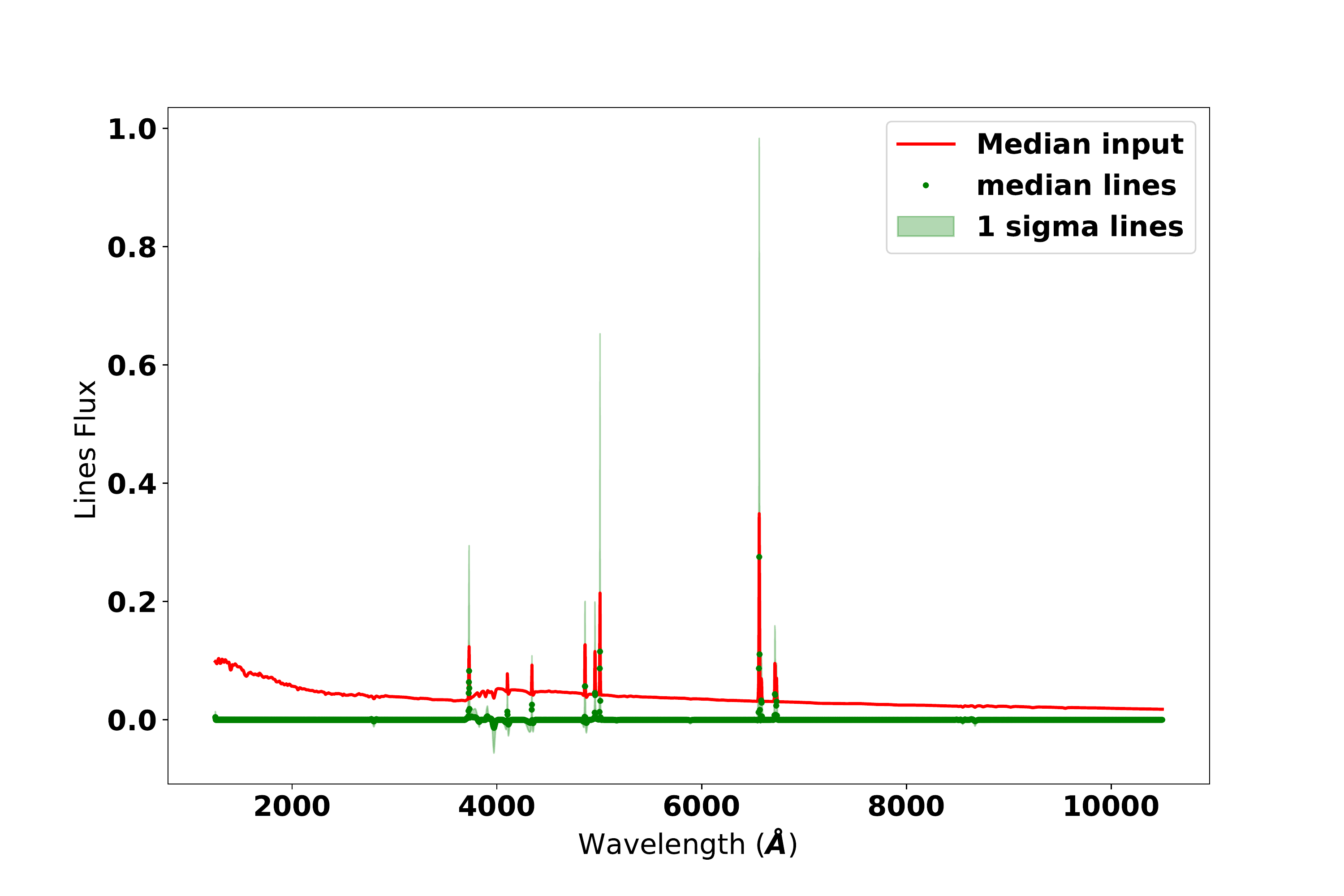}
\includegraphics[width=0.48\textwidth]{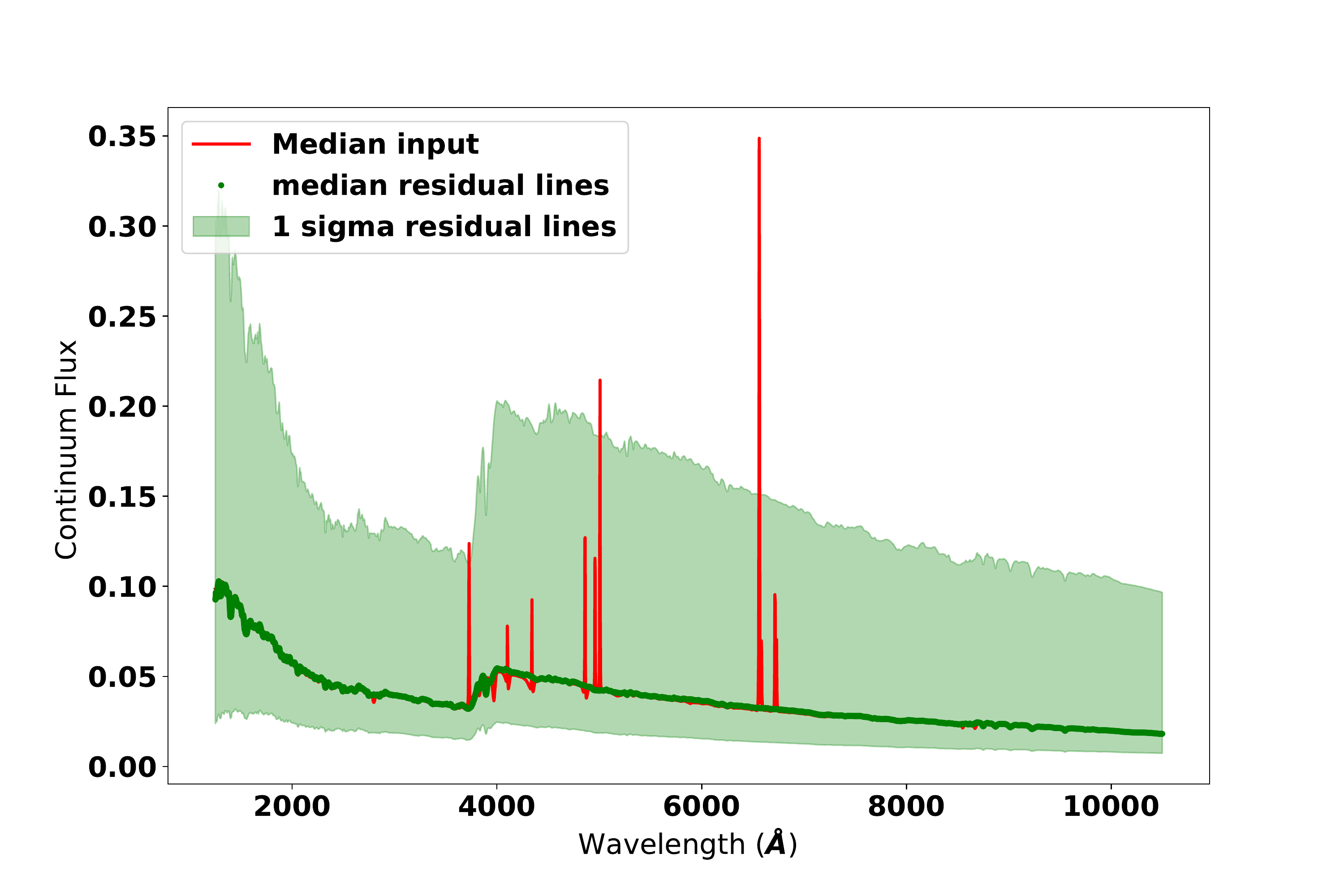}
\caption{Statistics on extracted line (left) and continuum (right) features. The input median signal in the original training set is displayed, along with the median and standard deviation computed over the line (left) and continuum (right) training data sets. In the first case, the emission lines are well extracted, with signal variability only in regions of emission or absorption lines and without offset due to continuum leakage into line estimation. In the second case, the continuum does not display neither discontinuities nor acute signal variability in regions of strong emission or absorption lines.}
 \label{fig:seplines_cont}
\end{figure*}

To generate a realistic distribution of galaxies with simulated redshift values, SEDs and broadband photometry, we employ the COSMOSSNAP simulation code \citep{jouvel2009designing}. COSMOSSNAP uses real data from the 30-band COSMOS photometric redshift catalog as a basis \citep{ilbert2008cosmos}, thereby ensuring that realistic relationships between galaxy type, color, size and redshift are taken into account. The catalog was originally generated from a combination of observations from astronomical surveys covering the spectral range from the UV (GALEX), through the optical (Subaru) and to near- and far-infrared bands (CFHT, UKIRT, Spitzer). This data set is matched to Hubble ACS imaging data, thus including realistic size-magnitude distributions from high-quality shape measurements originally made for weak lensing applications \citep{leauthaud2007}.

From this seed catalog, and assuming that the measured photometric redshifts are “true” redshifts, COSMOSSNAP can create simulated catalogs for any broadband photometric survey. The procedure is as follows: based on each galaxy’s properties, COSMOSSNAP chooses a spectral template from a predefined library, such that the integrated fluxes through the 30 broadband filters provide the best-fit to the observations. It uses the Coleman Extended library, which includes four spectral types - Elliptical, Sbc, Scd and Irregular \citep{Coleman1980}. It extends the spectral range into the UV and IR using synthetic spectra from the GISSEL library \citep{Bruzual1993} and adds an extra fifth type to represent starburst galaxies. To add realistic spectral features atop the original templates, galaxy emission line fluxes are calculated based on continuum properties of each galaxy. From the UV rest-frame luminosity of a given galaxy, a star formation rate is inferred using a calibration from \cite{Kennicutt1998}. This is then translated into an [OII] line flux, a relation which is valid for different galaxy types. Additional emission line fluxes are calculated relative to the [OII] flux, based on observations \citep{Moustakas2006}. The final SED of each galaxy is then corrected by host extinction (i.e. dimming due to dust within the galaxy itself, estimated from the photometric properties) and redshifted following the best-fit photometric redshift value.

At the end of this process, each galaxy has a "true" redshift and its associated SED model. Given an arbitrary choice of broadband filter and a model of the full filter throughput - including atmospheric transmission, telescope optical effects and more - the SED can be integrated to calculate simulated noiseless magnitudes. A realistic two-component model of magnitude errors with tunable observational properties is applied for each galaxy in the catalog. The resulting magnitude and error distributions can reproduce closely those of current and future large-scale galaxy surveys. For this analysis, we decide to create a photometric catalog modeled after the expected throughput of the 6 Large Synoptic Survey Telescope\footnote{http://www.lsst.org/} (LSST) broadband filters commonly referred to as ‘$ugrizY$’, and  with the expected depth properties of the science-ready ‘Gold’ sample \citep{abell2009lsst}. Hence we exclude from the catalog galaxies fainter than $25.3$ AB magnitude and with signal-to-noise $(S/N) < 10$ in the $i$-band. We obtain 218966 galaxies in an effective area of $1.24\;\text{deg}^2$ with realistic photometric properties, together with best-fit spectral templates with realistic continuum and emission line properties.

\begin{figure*}
\centering
\includegraphics[width=0.48\textwidth]{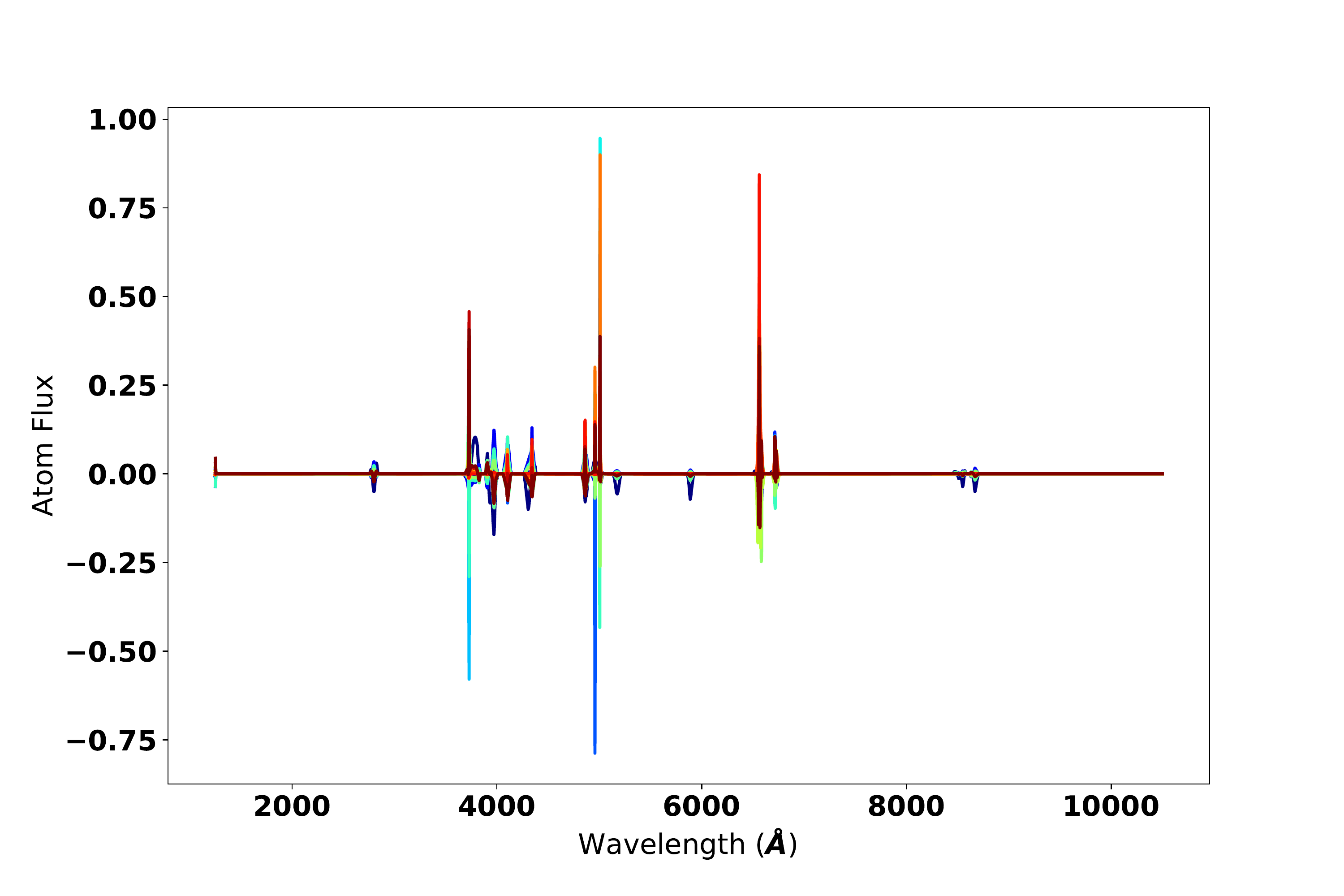}
\includegraphics[width=0.48\textwidth]{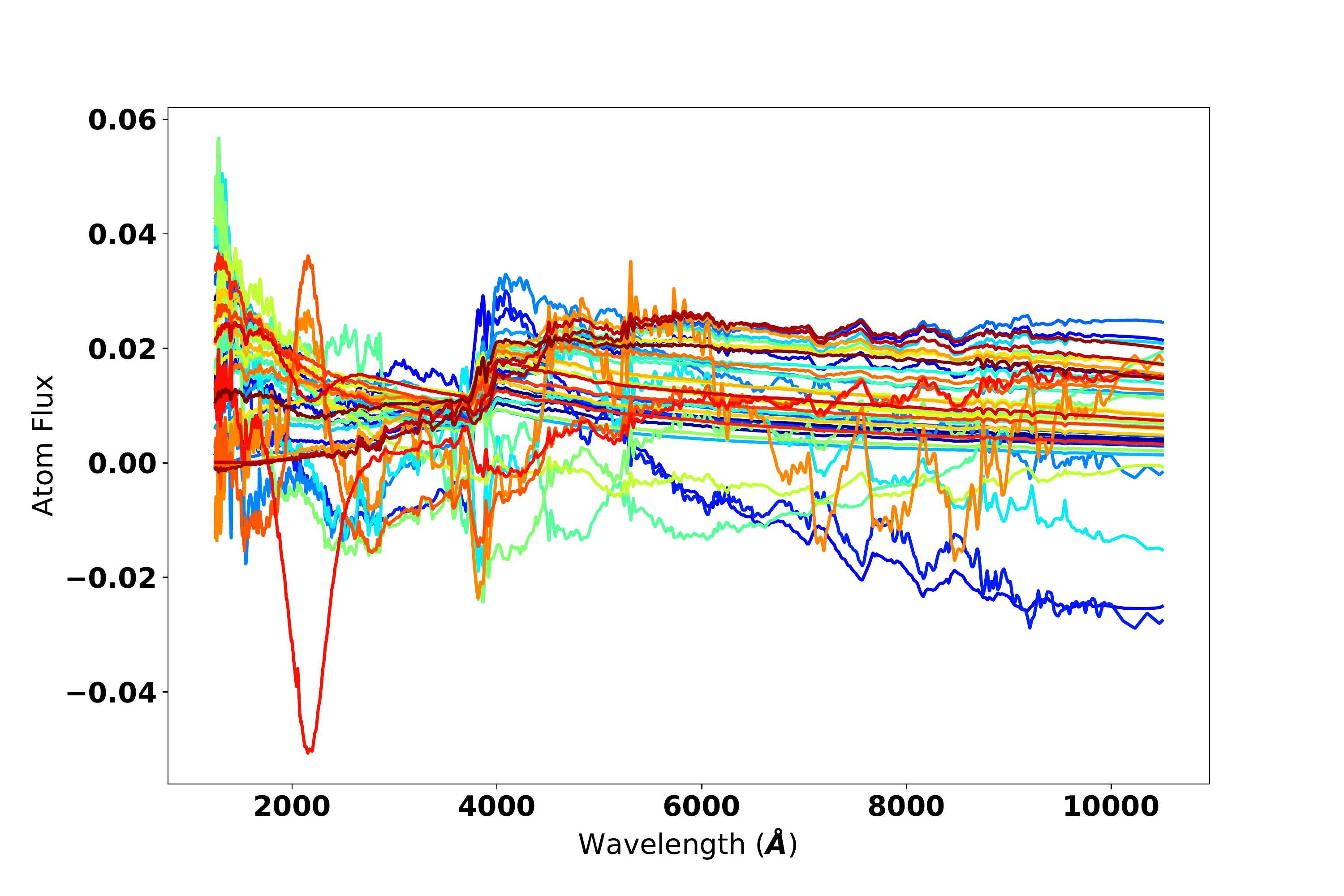}\\
\includegraphics[width=0.48\textwidth]{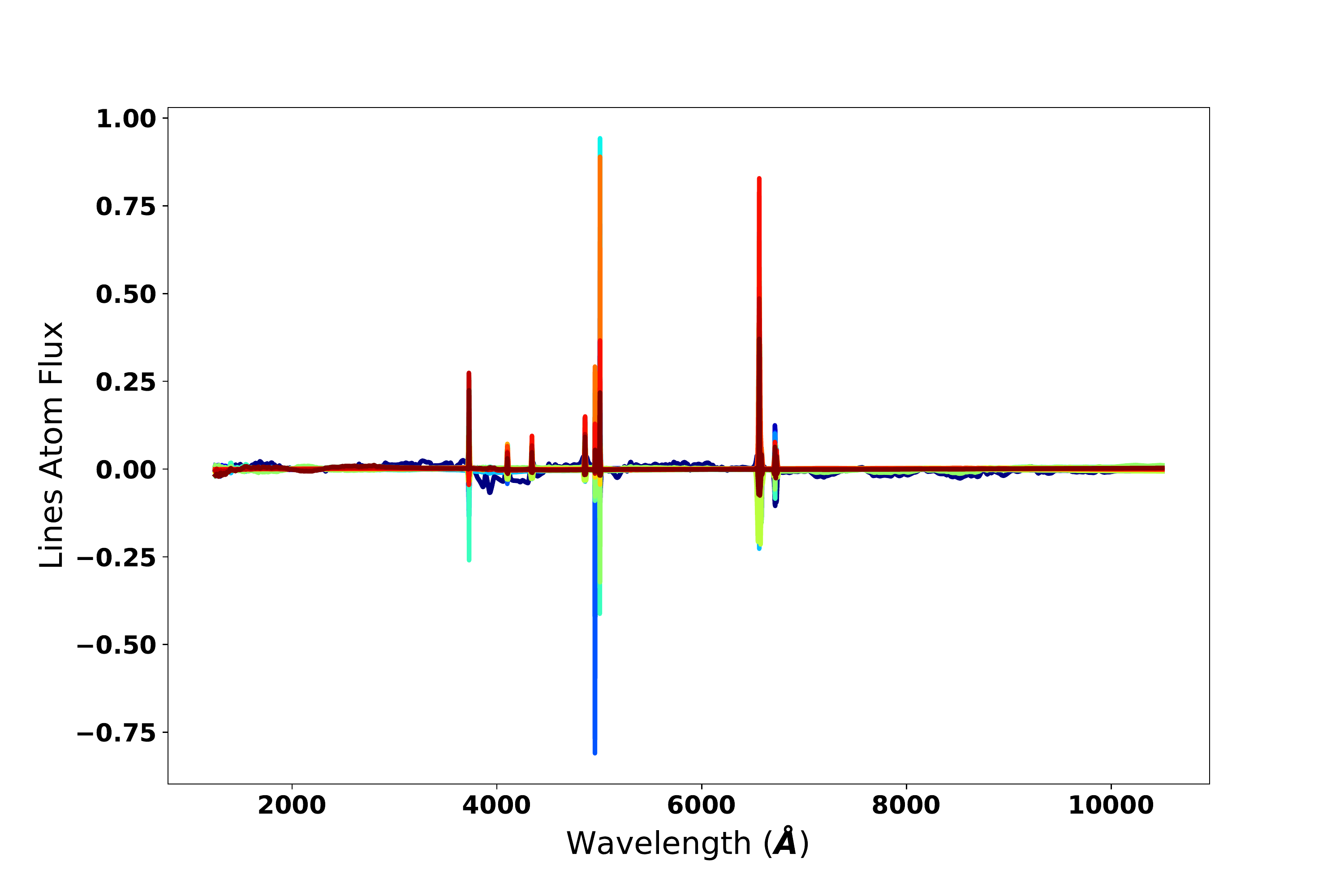}
\includegraphics[width=0.48\textwidth]{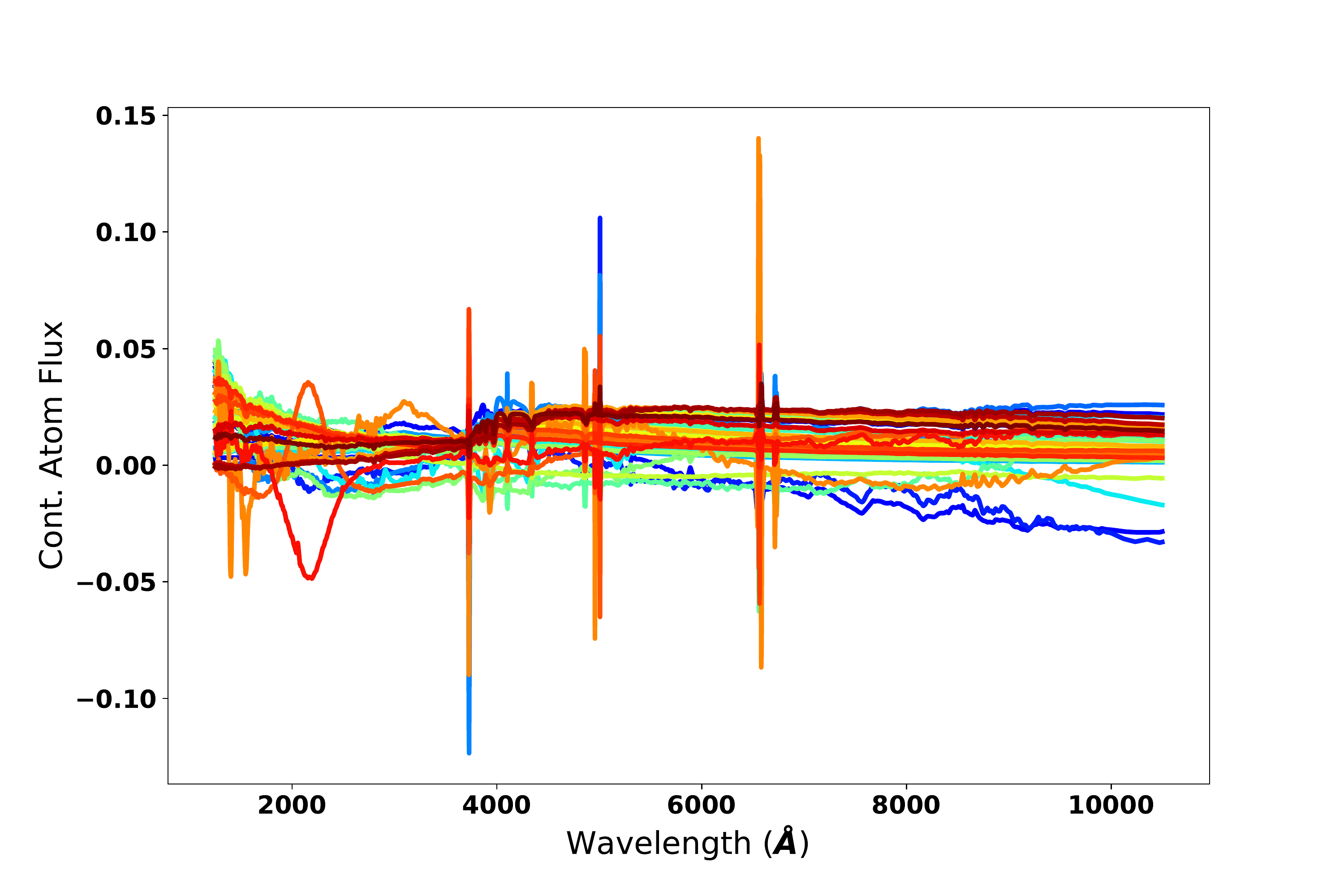}
\caption{(top left) Dictionary learnt from training data set with line features. Atoms typically contain a mixture of line emissions, with different sign and resolution. (top right) Dictionary learnt from training data set with continuum features. Atoms are correlated, with breaks such as the $4000\angstrom$ break. (bottom left) Component of the final dictionary corresponding to the original line dictionary. Some fluctuations from the continuum are added. (bottom right) Component of the final dictionary corresponding to the original continuum dictionary. Extra line emissions are added to the continuum.}
\label{fig:linesdico_contdico_linesdico2_contdico2}
\end{figure*}


COSMOSSNAP produces SEDs with a chosen wavelength resolution for the continuum and absorption lines. The emission lines are added at higher resolution, to ensure that their shapes and amplitudes are fully characterized. To work with realistic spectra, we need to resample, integrate and add noise to the best-fit SEDs.  On a real fiber-fed spectrograph such as the ones designed for the BOSS \citep{Smee2013} and DESI \citep{Aghamousa2016} surveys, the resolution is a variable property that depends on the characteristics of the instrument, in particular on the interplay between the 1D point spread function FWHM of the spectrograph and the pixel size of the CCD. Noise on the 1D spectra is mainly due to Poisson sampling of photons from the source and CCD readout noise, among other effects. For our purposes of evaluating the performance of redshift algorithms, we will assume a constant resolution $R\equiv\lambda/\Delta\lambda$ - which implies a wavelength binning constant in logarithmic scale - and uncorrelated gaussian noise with constant variance $\sigma$ on all wavelength bins. The SEDs are log-binned with a constant bin size of $2.17\,10^{-4} \log10 \angstrom$ - corresponding to a resolution of $R\sim 850$ - and integrated within those bins. We add noise of different $\mathrm{S/N}$ levels, where the level is defined according to the spectral energy flux integrated in the $r$-band\footnote{The LSST $r$-band filter is not identical to the SDSS one. Therefore, our definition of \snr is not identical to that in \cite{Machado13}. This does not have any impact in our analysis and conclusions, as they are self-consistent.}, following \cite{Bolton2012} and \cite{Machado13}:
\begin{equation}
\mathrm{S/N}\,_{r} = \textrm{median}\,\left[\,\frac{\textrm{flux}}{\sigma}\,\right]\,^{6760\,\textup{\AA}}_{5600\,\textup{\AA}} \: .
\end{equation}
We work with $\mathrm{S/N}\in\{2, 5, 20\}$, where $\mathrm{S/N}=20$ is our "clean" case.

For the remainder of this paper, we work with a training data set which includes 2000 clean resampled spectra randomly sampled from the original COSMOSSNAP data from a redshift range [0, 1.7]. These training SEDs are blueshifted to the rest-frame and cropped to the wavelength range $[1250 \angstrom, 10500\angstrom]$. For testing, we randomly sample galaxies of different types. COSMOSSNAP classifies its spectra in 36 classes, organised in 4 groups: \texttt{EllS0}, \texttt{SaSc}, \texttt{SdSm} and \texttt{SB}. We build a test set by randomly sampling 5000 galaxies from each of the groups to investigate potential systematic effects in our methods depending on the galaxy type. This galaxy type information was not included in the training phase. All test galaxy spectra are cropped to the wavelength range $[3000 \angstrom, 10500\angstrom]$.


\section{Experimental Results}

\label{sec:results}
Let us now present the results for redshift estimation obtained with the proposed dictionary learning and denoising autoencoder representation learning frameworks. These two approaches have been compared to the Darth Fader algorithm \citet{Machado13}, based on extracting robustly line features and cross-correlating them with eigentemplates to infer a redshift. We start by describing how we set this algorithm up for this comparison and then describe the parameters for the two proposed methods.

\subsection{Darth Fader}

We run Darth Fader on the $4\times 5000$ test galaxy spectra for all SNR levels. Our configuration choices mostly follow the standard setup. We briefly describe them now, and refer the reader to the section \ref{sec:DF} and to \cite{Machado13} for more details. We derive eigentemplates from the clean training data set described in section \ref{sec:data}. We set the threshold for the principal components so that the eigentemplates retained contain 99.93\% of total eigenvalue weight, as in \cite{Machado13}. We keep 26 eigentemplates as a result. {If the preserved percentage of the variance and the number of retained eigentemplates are not sufficient for reconstruction, the performances of the redshift estimation scheme can degrade. On the other hand, if all the energy was intended to be retained by the representation, a larger number of eigentemplates may yield a drop in performance for noisy scenario. According to our experiments, these factors worsen the redshift estimation but not significantly.}
For denoising the test spectra, we restrict the multiscale transform to six scales and keep the regularisation at 0.01. For redshift estimation, we cross-correlate the eigentemplates with the denoised version of the test spectra to avoid the misclassification of noise features as spectral lines, even though denoising may result in removing physical features in low SNR regimes. Contrary to what is recommended for optimal use of Darth Fader, we do not clean the resulting catalogue with FDR thresholding. Although this results in sub-optimal performance metrics, we wish to compare the performance of the algorithms on the full galaxy set, and not only on those relatively few galaxies where Darth Fader successfully retrieves many features.

\subsection{Dictionary Learning}

As illustrated in Algo~\ref{Algo:DLSpectra}, the first step in our analysis involves constructing a meaningful initial dictionary for the subsequent learning phase. This implies constructing two sub-dictionaries for line features and continuum and therefore the separation between line features and continuum emissions in the rest-frame training set. The line features have been removed with a mask centered on known wavelength value for each emission line, and extended to a width of $80\angstrom$ in order to ensure that no emission line energy is leaking into the continuum. Then, a multiscale iterative inpainting technique is used to extrapolate the continuum inside this region: we iteratively keep only low frequency scale coefficients in the mask while enforcing values outside the masked region not to be affected by the procedure. Some statistics on the separation are summarized in Fig. \ref{fig:seplines_cont}, which illustrates its good performance. 

\begin{figure}[htb]
\centering
\includegraphics[width=0.43\textwidth]{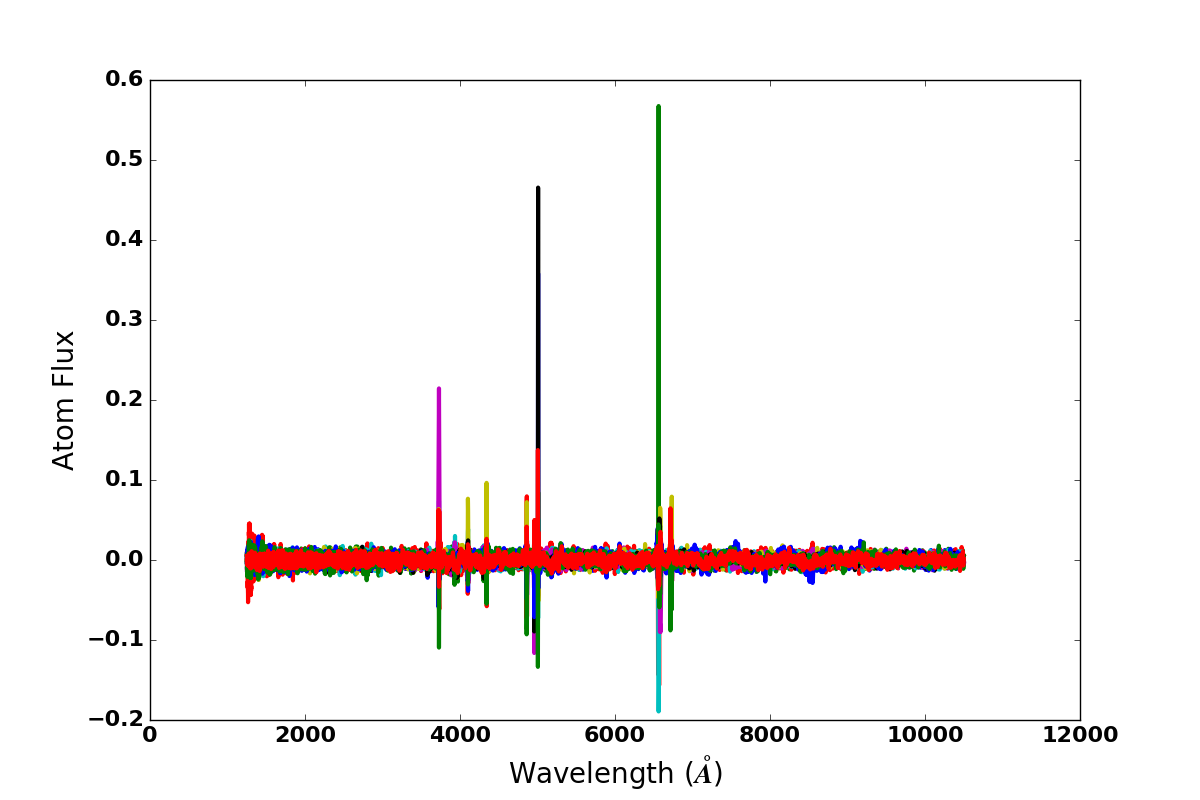}
\caption{Weight filters learned by the DAE for nhid = 100 and SNR = 5.}
\label{fig:weightdae}
\end{figure}

\begin{figure*}
\centering
\begin{tabular}{ccc}
\includegraphics[width=0.33\textwidth]{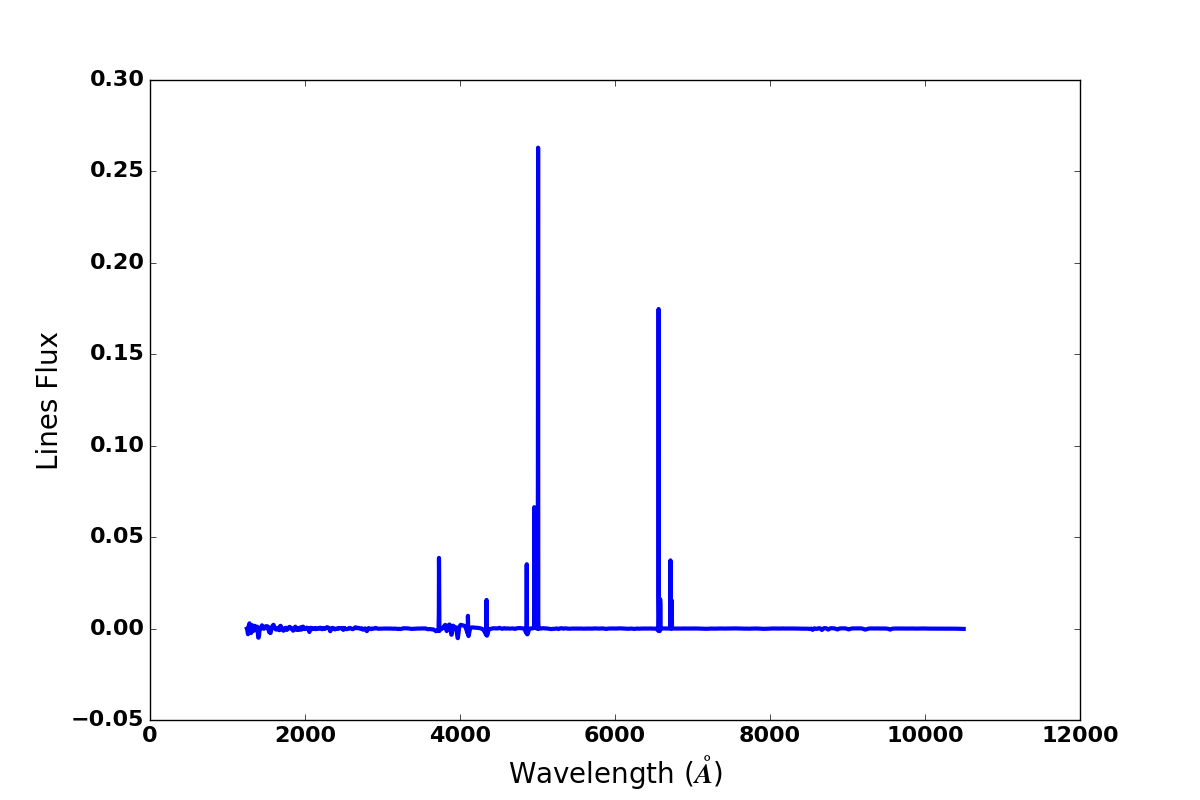}&
\includegraphics[width=0.33\textwidth]{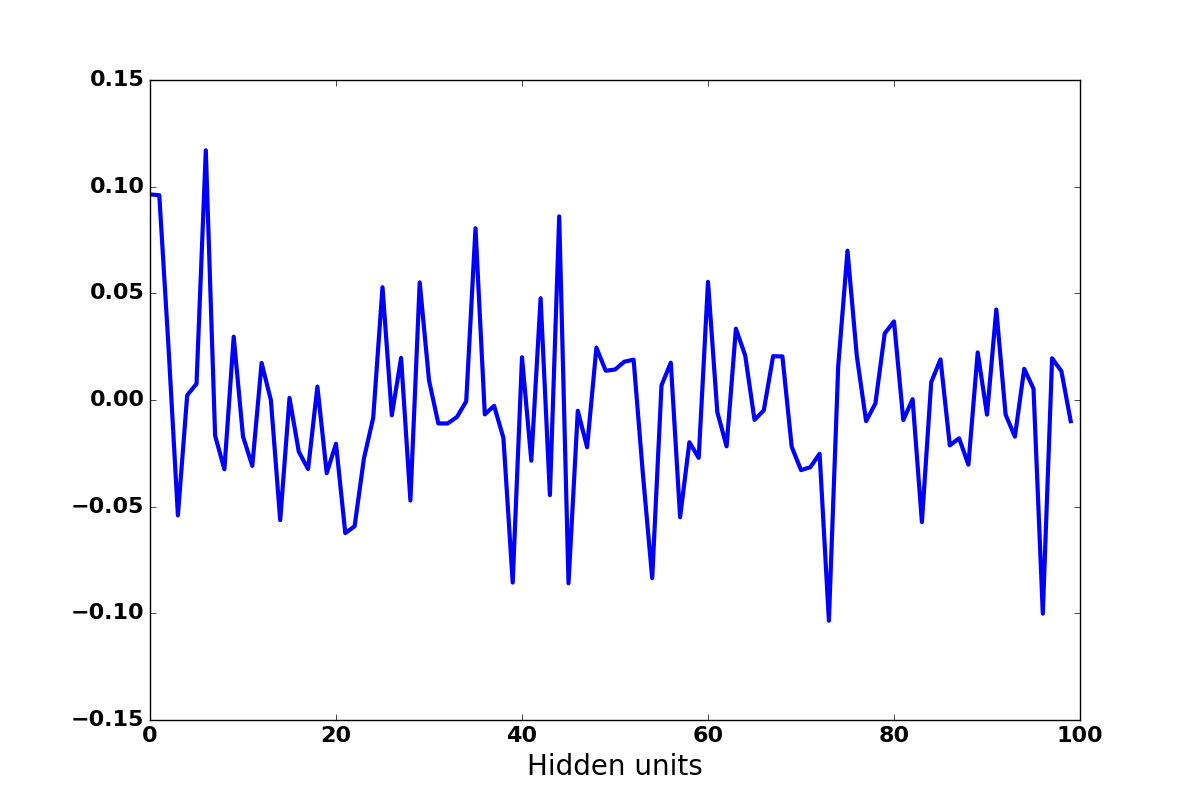}&
\includegraphics[width=0.33\textwidth]{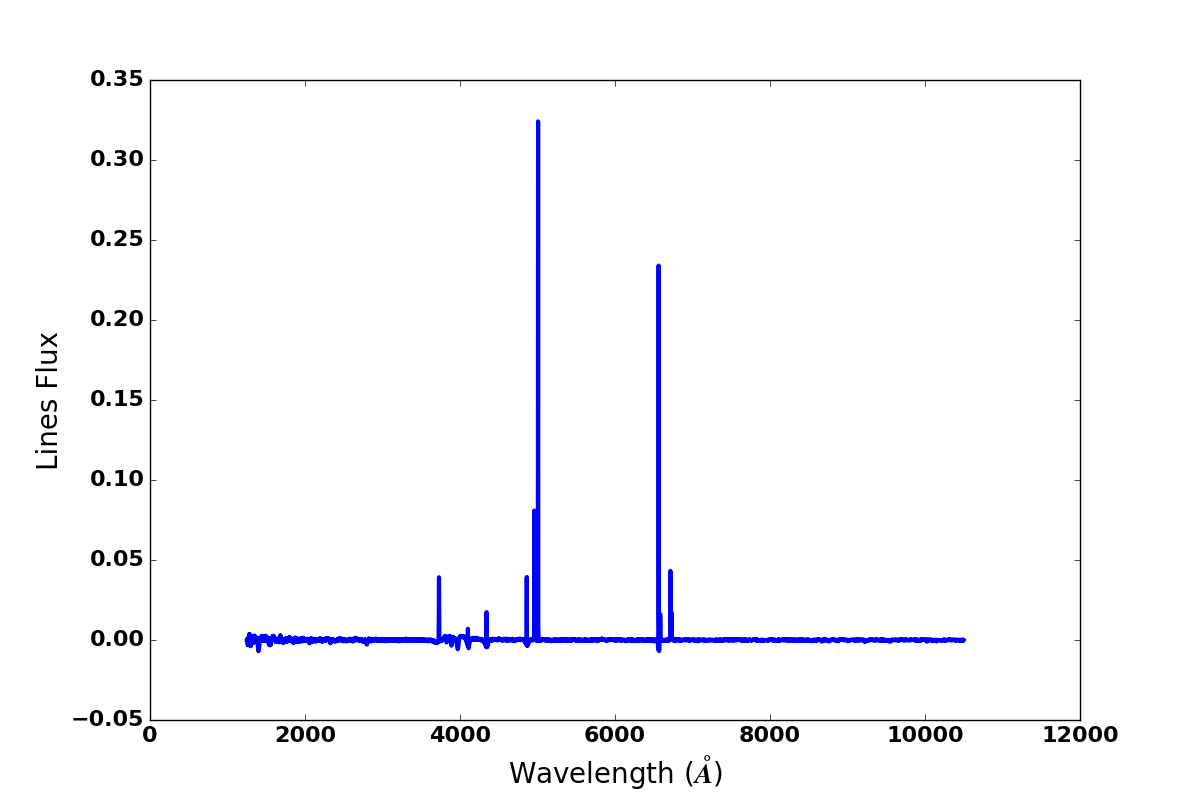}\\
{(a)}&{(b)}&{(c)}\\
\end{tabular}
\caption{(a) Training sample, (b) Representation, (c) Reconstructed spectra.}
\label{fig:daerecons}
\end{figure*}

A "continuum" template dictionary of $40$ atoms with a targeted sparsity degree of $3$ and $100$ iterations was learnt using the training data set with continuum features. Similarly for the set with line features, a "line" template dictionary of $20$ atoms with a targeted sparsity degree of $3$ and $100$ iterations was learnt. {The number of atoms and sparsity degree were heuristically fixed from the overall complexity observed in the training data (using more atoms leads to high correlation preventing the learning) and as a trade-off between estimation error, efficient learning and robust subsequent redshift estimation. Indeed, even though increasing the sparsity degree would decrease the overall approximation error it would also increase the risk of line confusion by potentially selecting as atoms more isolated features in the spectra.} 

The atoms learnt for the two sub-dictionaries are represented in the upper panels of Figure \ref{fig:linesdico_contdico_linesdico2_contdico2}. Several remarks can be made from this figure: first, in both cases, the atoms are correlated contrary to what would have been obtained with PCA ; second, in particular because of the small targeted sparsity degree, the atoms learnt for lines all contained a combination of line features, which would help avoiding line confusion ; third: the 4000 \AA\, break feature is captured by most but not all atoms in the continuum subdictionary.
Learning was finally performed on the original training data set by initializing the dictionary with the $60$ atom dictionary obtained by concatenating the two previous ones, using $10$ iterations and a sparsity degree of $6$. The two components of the final dictionary are displayed in the lower panels of Figure \ref{fig:linesdico_contdico_linesdico2_contdico2}, showing that some leakage between the two components has been introduced during the final learning phase, which was actually beneficial to reduce approximation errors. This also leads the continuum part to contain more high frequency features, which could result in better redshift estimation by combining several correlated line and continuum features in the atoms. 

For spectroscopic redshift estimation, the redshift grid $\mathcal{T}$ tested was built by uniformly sampling the range from $0$ to $1.7$ with a density of $0.001$ and sparse coding was performed for all galaxy spectra at the given sampled redshifts.

\subsection{Denoising autoencoders}
\label{sec:daeresults}

Firstly, the model was constructed using TensorFlow \citep{abadi2016tensorflow}. TensorFlow is an open source software library for numerical computation using data flow graphs. It was developed by Google and tailored for machine learning. Extensive documentation describing all the functionalities can be found in \footnote{https://www.tensorflow.org/}.

The denoising autoencoder was trained with the training set described above composed of N = 2 000 samples spanning from 1250 \AA\, to 10500 \AA \, resulting in a vector of dimension m = 4 258. The training samples have followed the same preprocessing as in the Darth Fader scheme for continuum line subtraction, the continuum of the spectra has been removed through wavelet filtering as explained in section \ref{app:continuum}. The number of visible units is fixed to m =  4 258 in agreement with the size of the input data and the number of hidden units is a free parameter. 

{The representation size have to be large enough in order to retain sufficient information for reconstruction.
As the representation size increases, the approximation error will decrease. This improvement will be significant only for the galaxy true redshift and will not impact on its estimation. As the computation time for the redshift estimation algorithm is proportional to $nhid$, we chose a value providing a sufficiently reliable reconstruction of the original signal for the redshift estimation, but greater values could be used instead. We have investigated $nhid = 20, 100, 200, 1000$ and chosen an architecture with $nhid = 100$. Smaller values did not provide a useful reconstruction of the spectra and greater values did not improve the redshift estimation step.
Thus, this choice is a trade-off between approximation error and computation time.}

The weights are randomly initialized from a uniform distribution and are tied across all the experiments. 
Moreover, the activation function for both the encoder and the decoder is set to be a hyperbolic tangent. The input is artificially contaminated with a Gaussian noise such that the SNR will be constant and equal to 5 through all the training set. Furthermore, the optimization of the parameters is performed through stochastic gradient descent; and the reconstruction cost criteria to be minimized is the squared reconstruction error. The learning rate has been set to $10^{-4}$ and the batch size to 100. The training stops after 500 epochs. The choice of the learning rate, the batch size and the number of epochs is done in order to ensure a convergence on the training stage.
Fig. \ref{fig:weightdae} displays the encoder weights learned by the denoising autoencoder. This illustrates what kind of features the model is sensitive to and gives an idea about how the input data are coded. The earmarks highlighted by the filters are in agreement with the position of the absorption and emission lines dominating the training samples.
Moreover, Fig. \ref{fig:daerecons} displays one sample belonging to the training set (a), its representation (b), and its corresponding reconstruction (c). 
From the way the information is coded in Fig. \ref{fig:daerecons} it is hard to give a straightforward interpretation. Due to the mixing performed by the encoder, the information is distributed over all the code and nothing can be said about hidden units individually.

\begin{figure}[htb]
\centering
\includegraphics[width=0.43\textwidth]{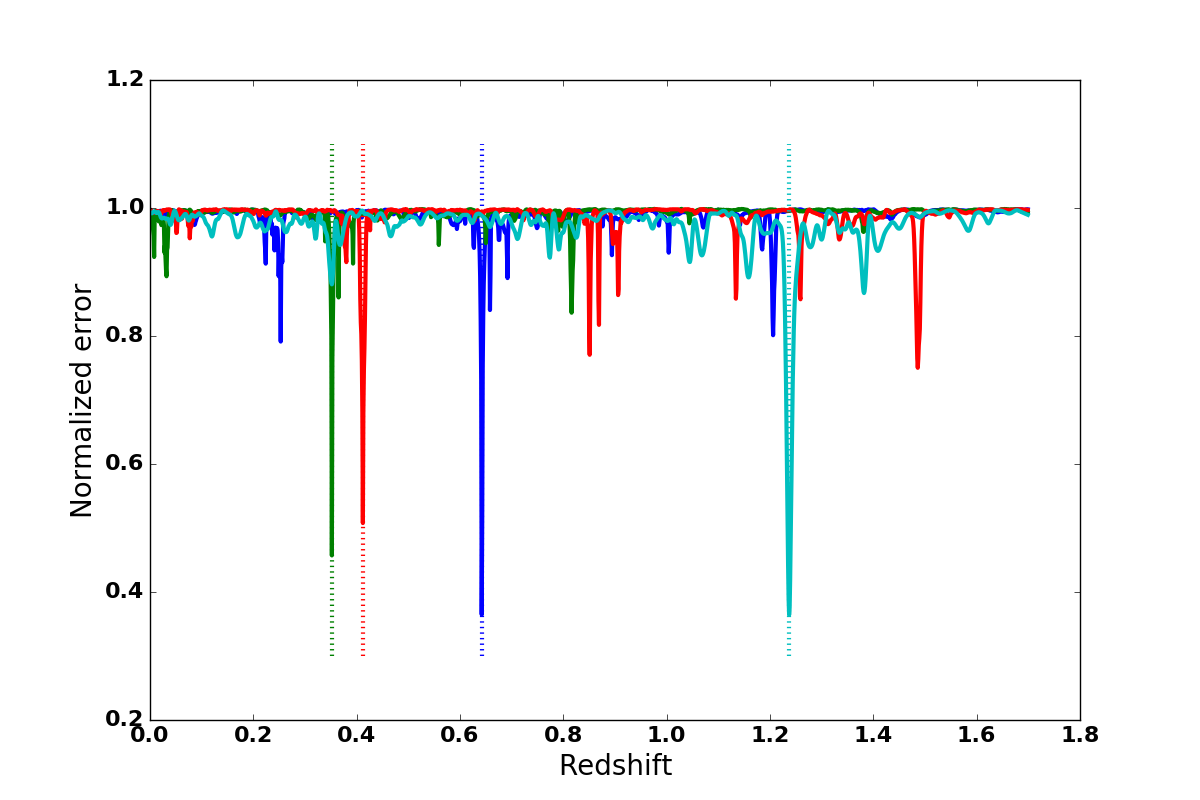}
\caption{Approximation error profiles as a function of the redshift for four galaxies of the test set. The true redshift is displayed as a dotted vertical line for each galaxy.}
\label{fig:errorprofiles}
\end{figure}

Let us now illustrate the results for redshift estimation with the denoising autoencoders. Fig. \ref{fig:errorprofiles} illustrates some approximation error profiles over all the investigated redshift values. The redshift has been uniformly sampled from 0 to 1.7 with 0.001 steps. From this figure, it is clear that the proposed method correctly finds the redshift in the studied scenario. Moreover, the technique described in this work is highly sensitive to the true redshift value and its precision depends on the redshift sampling grid. 
However, some strong oscillations due to the matching of the different features present in the spectra show that some line confusion could occur in the subsequent redshift estimation. 

\begin{figure*}
 \includegraphics[width=0.33\textwidth]{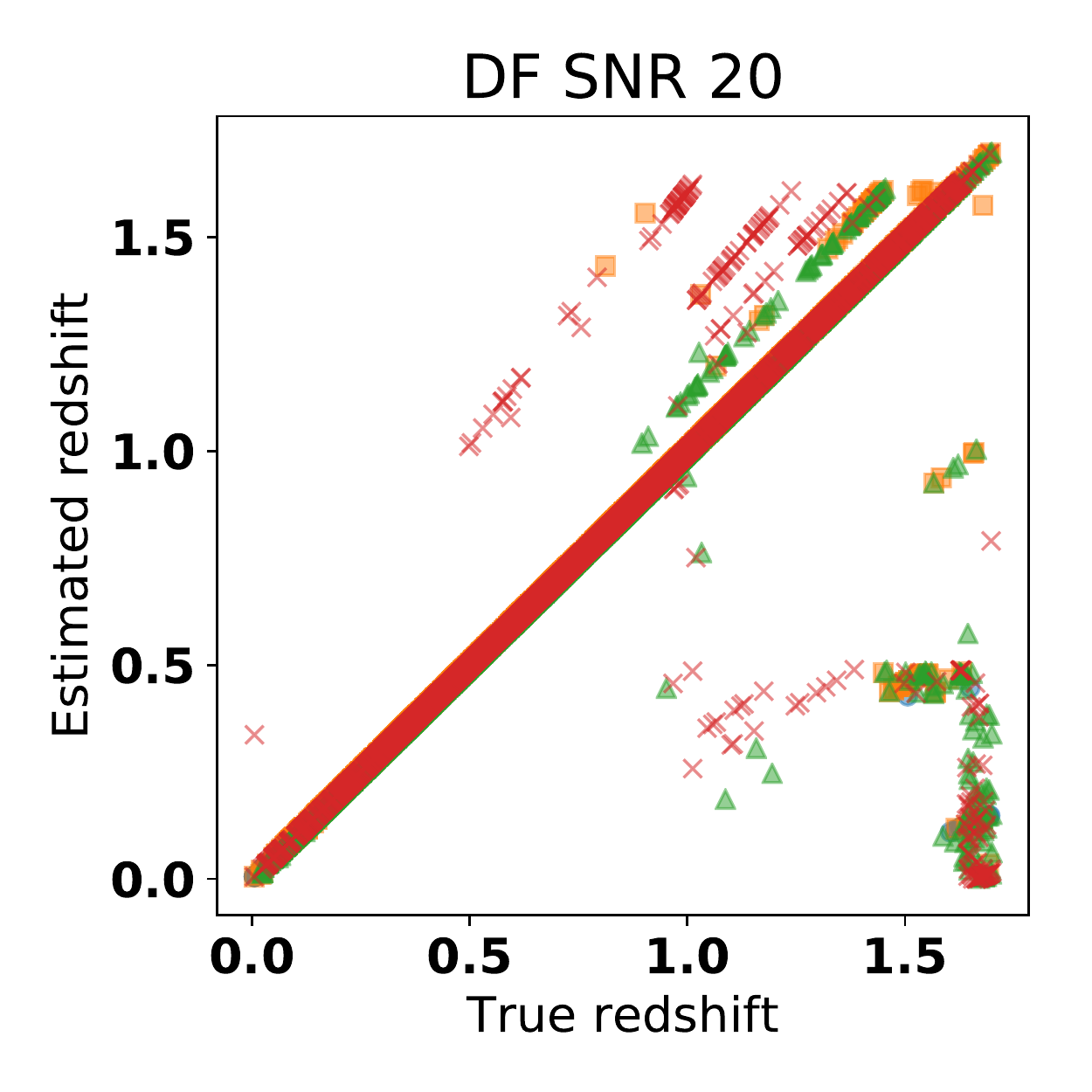}
 \includegraphics[width=0.33\textwidth]{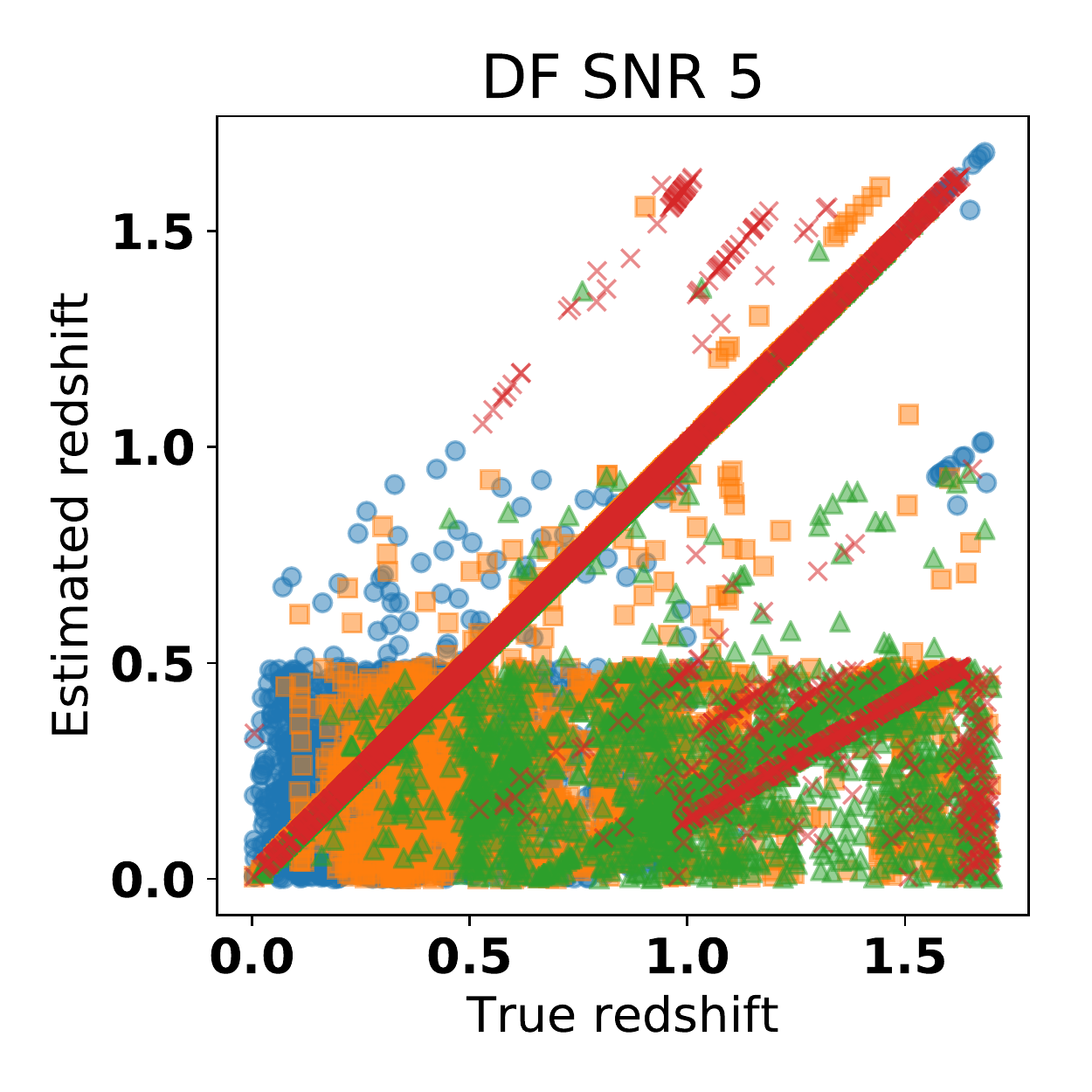}
 \includegraphics[width=0.33\textwidth]{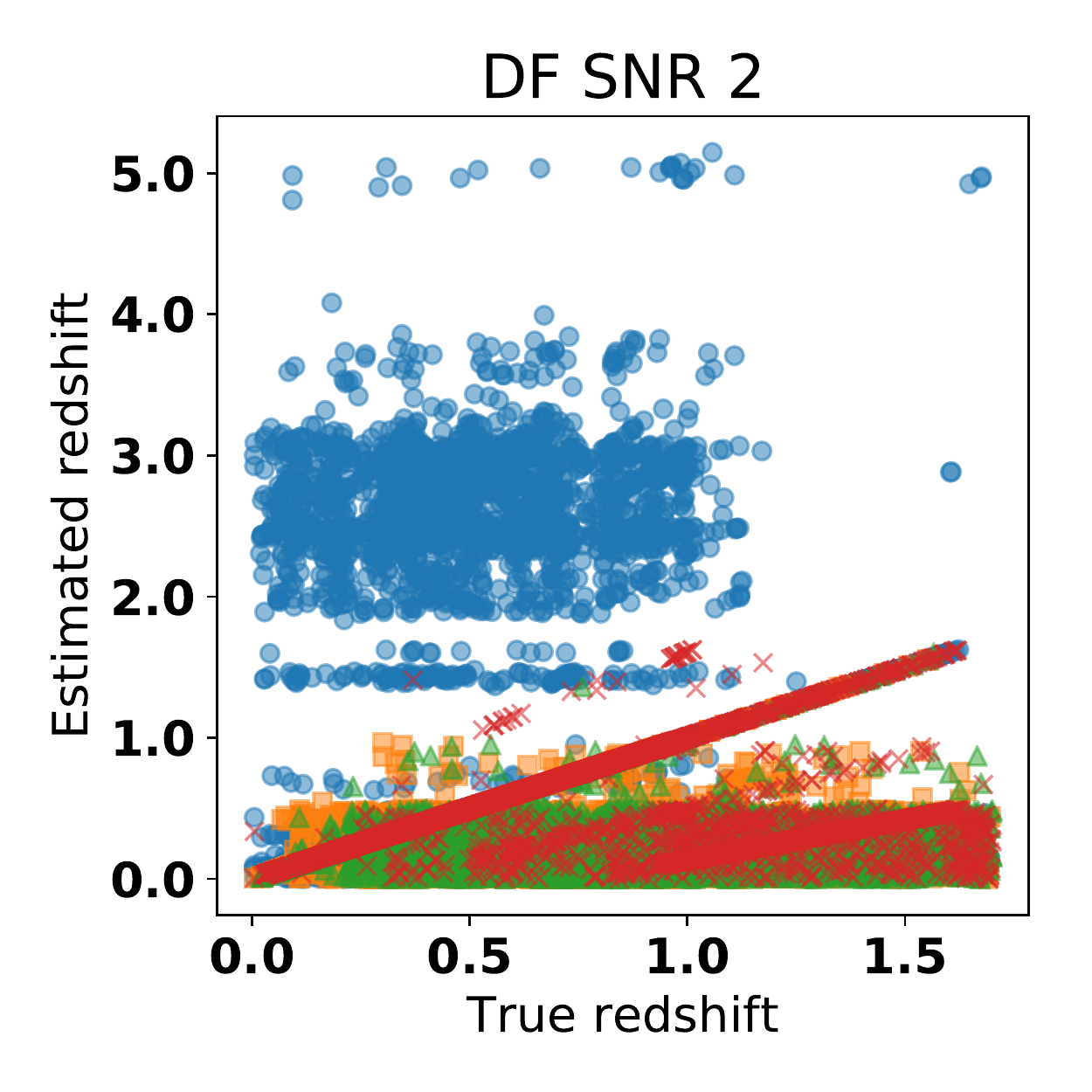}\\
 \includegraphics[width=0.33\textwidth]{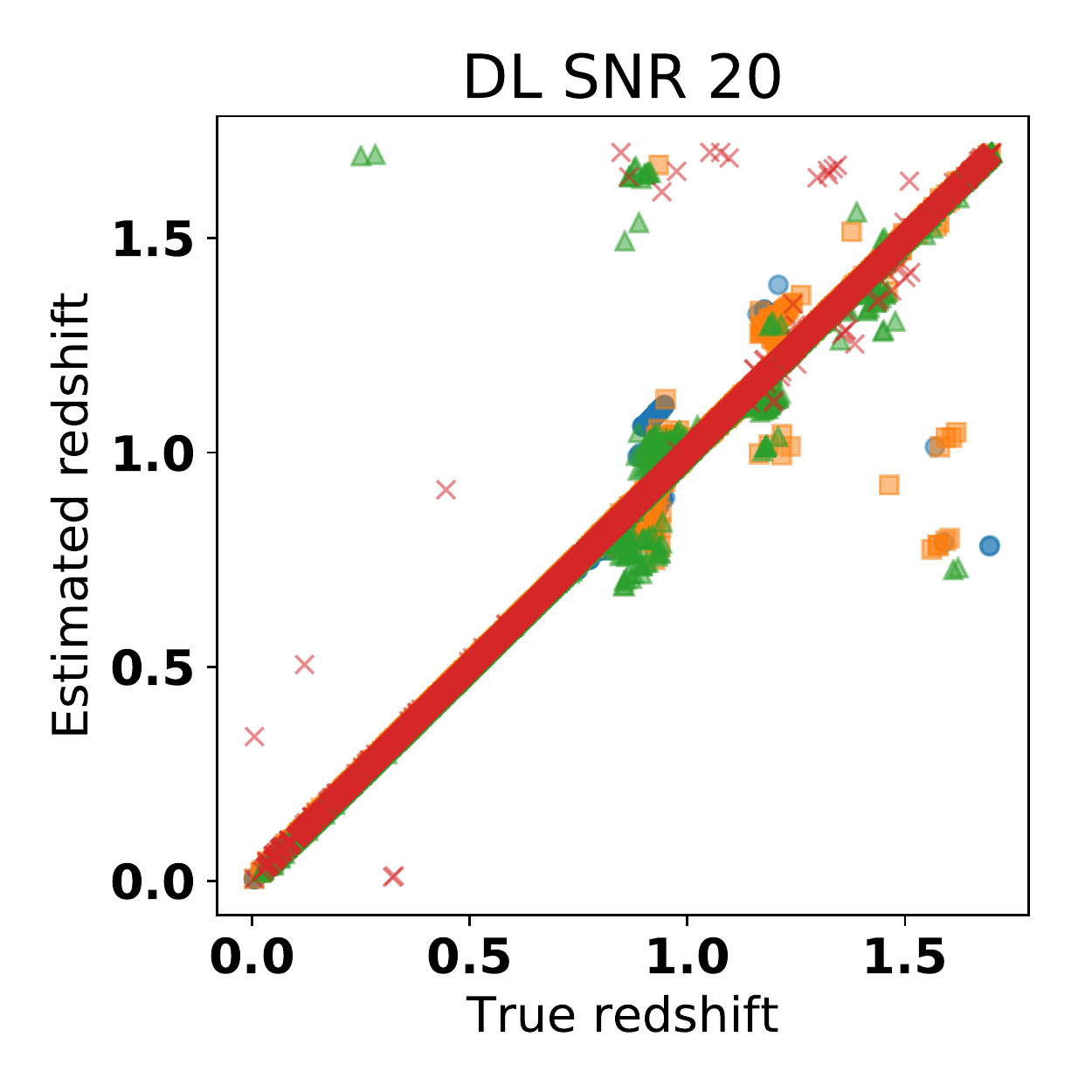}
 \includegraphics[width=0.33\textwidth]{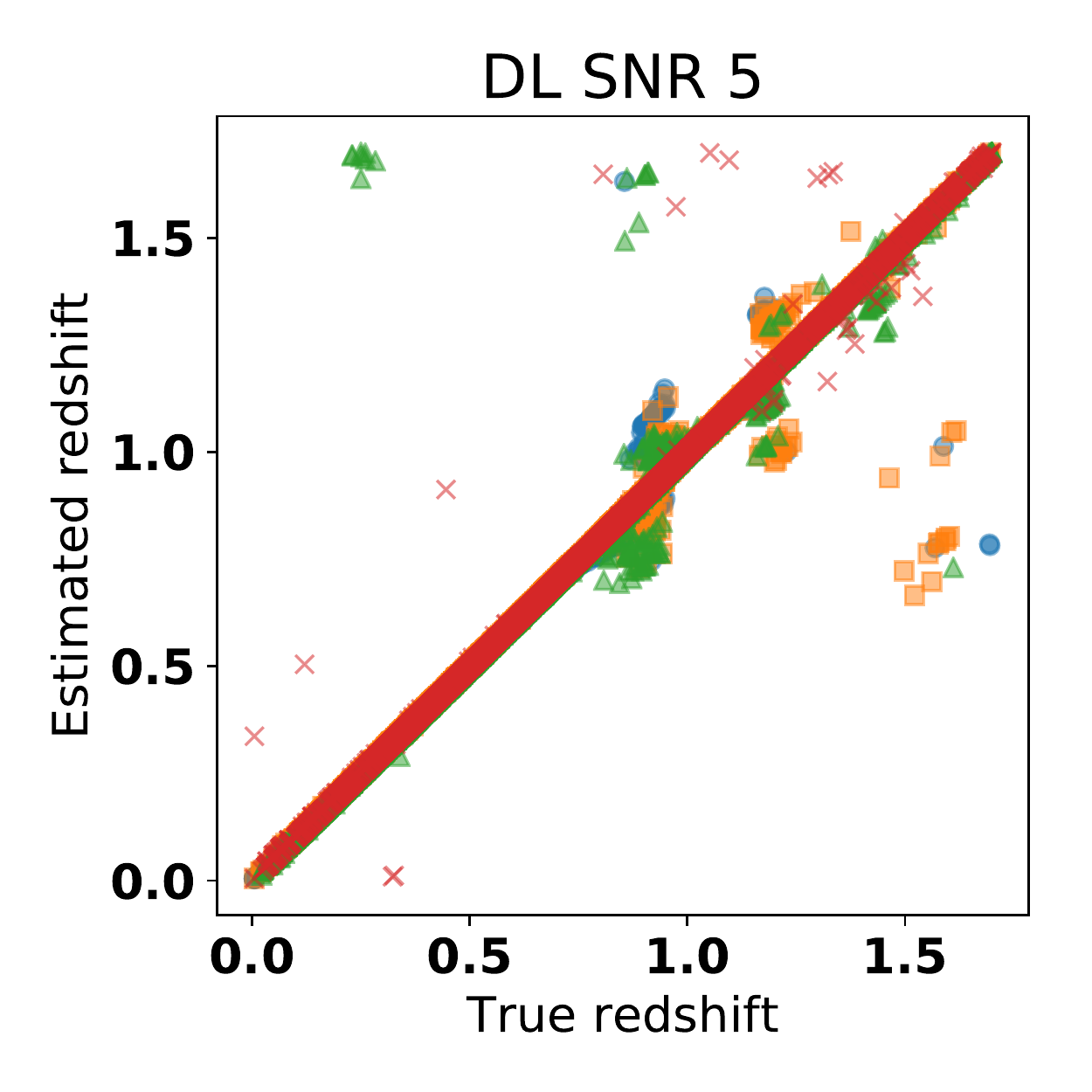}
 \includegraphics[width=0.33\textwidth]{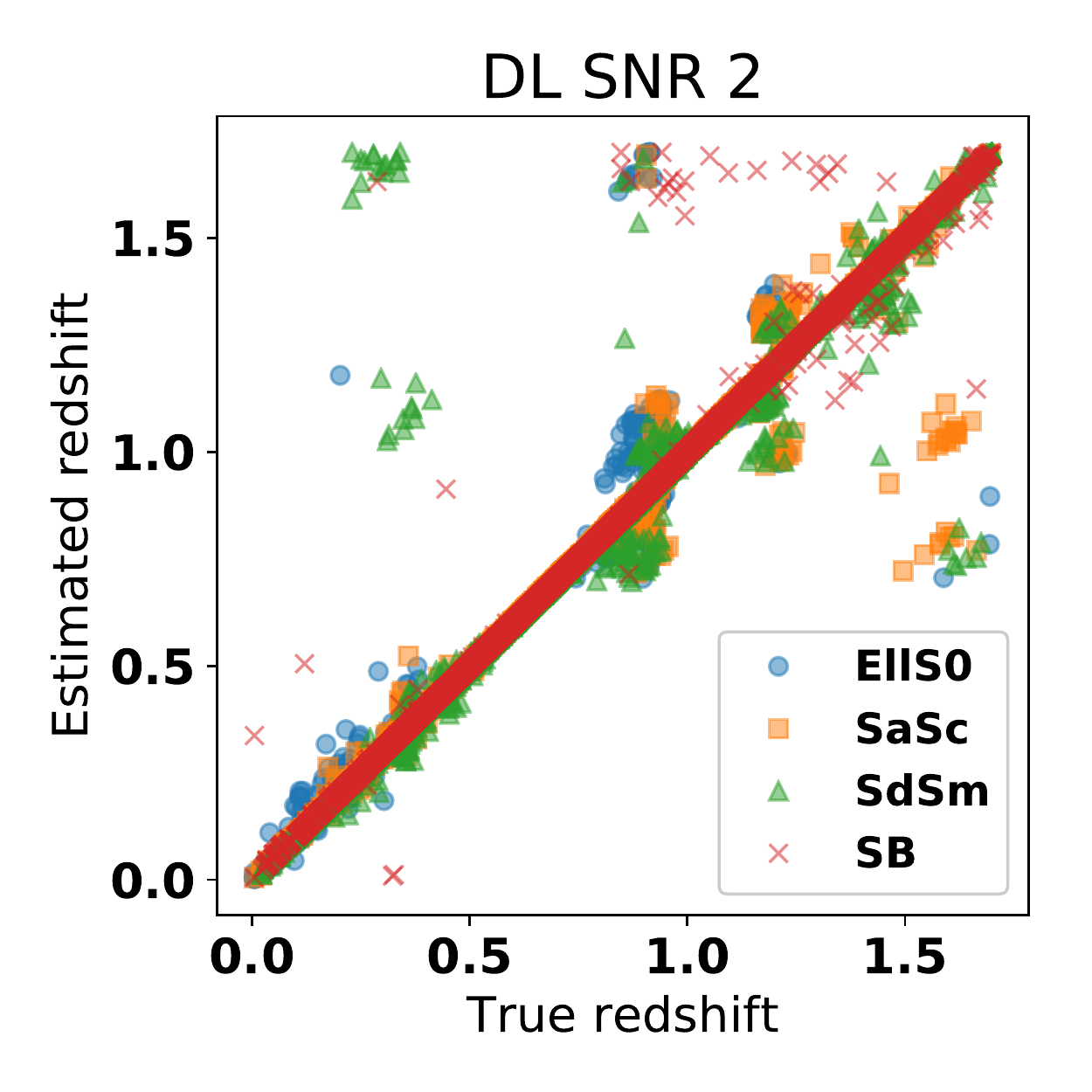}\\
 \includegraphics[width=0.33\textwidth]{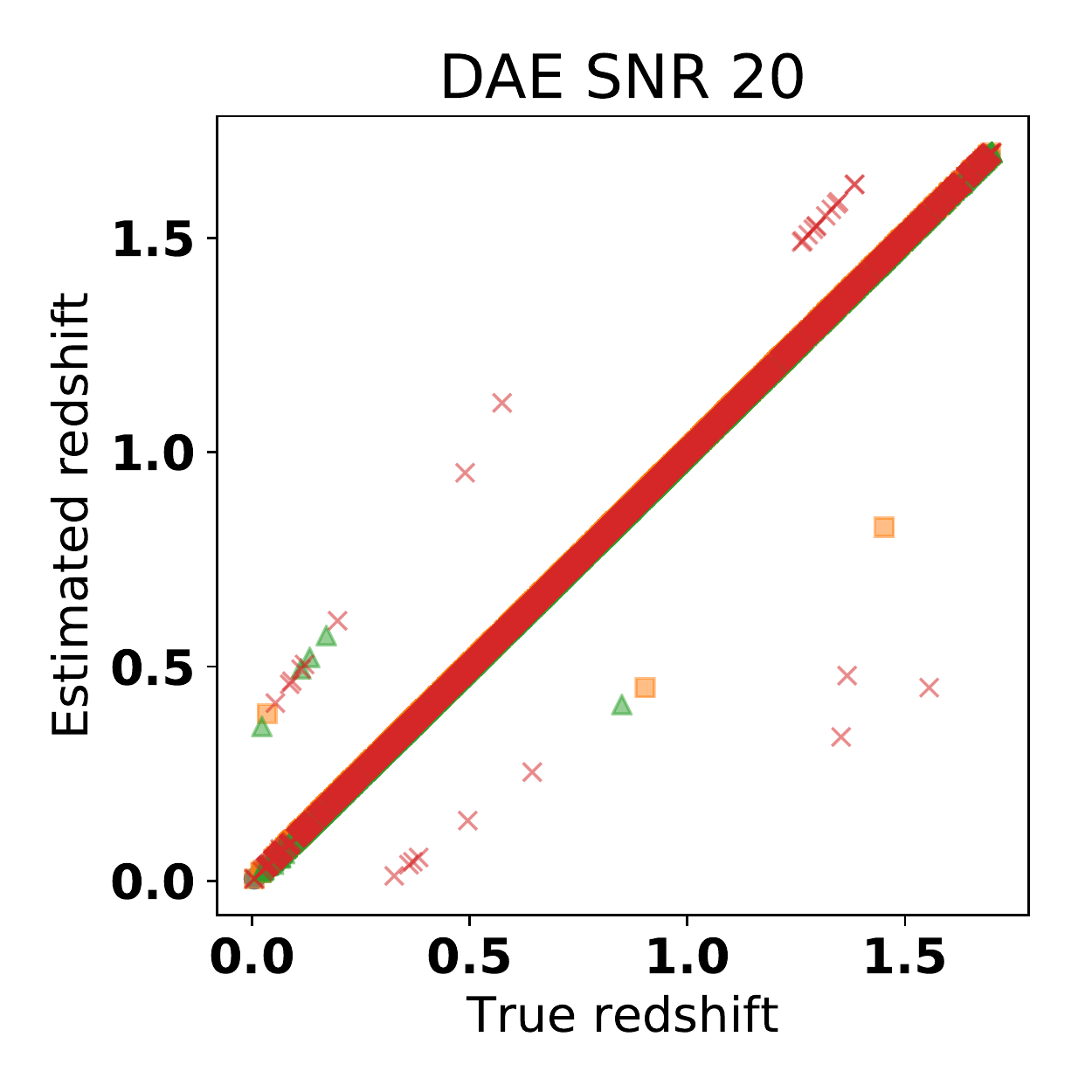}
 \includegraphics[width=0.33\textwidth]{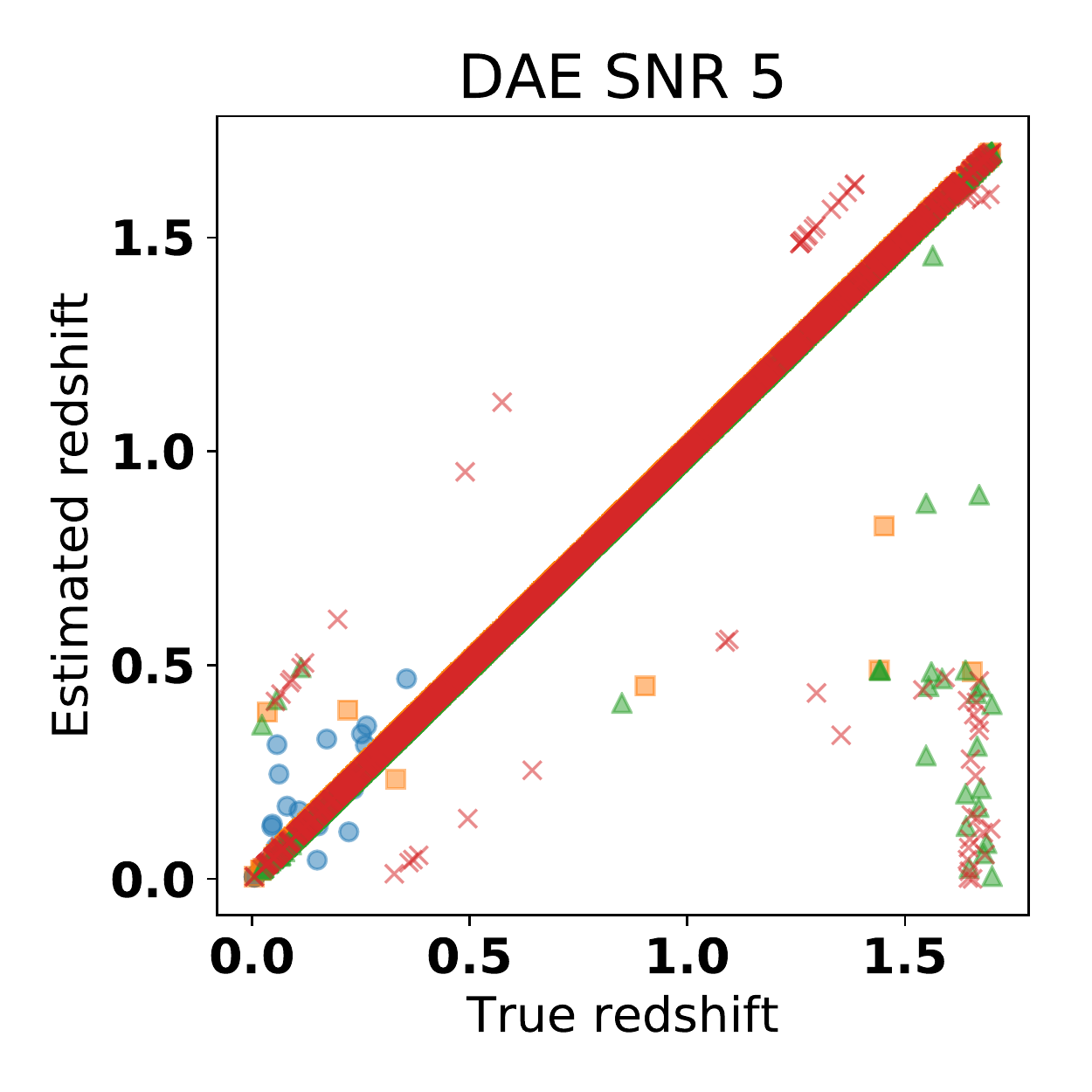}
 \includegraphics[width=0.33\textwidth]{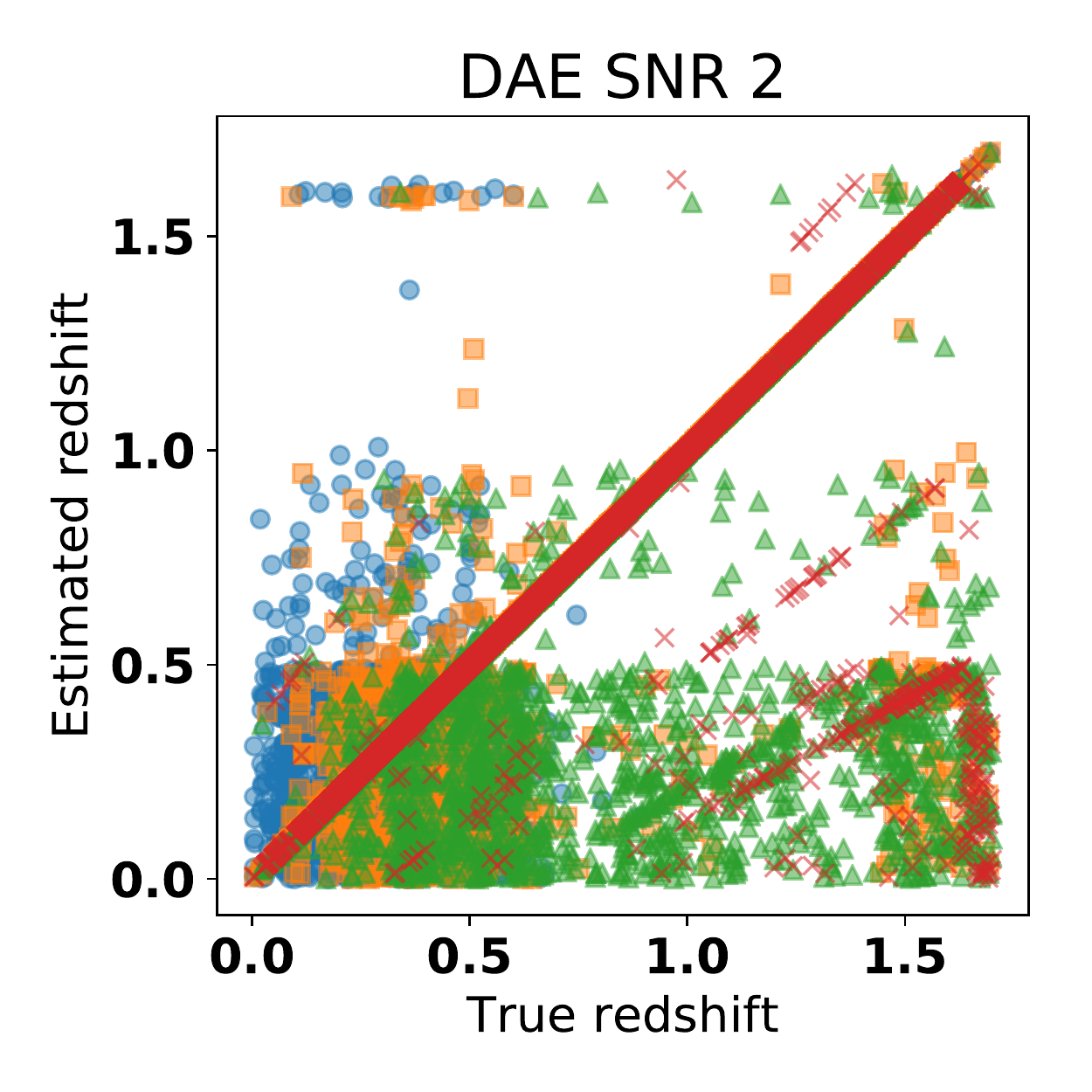}
 \caption{True vs estimated redshift values for Darth Fader (top row), dictionary learning (middle row) and denoising autoencoders (bottom row) in the SNR = 20, 5 and 2 cases (left, middle and right columns, respectively). Galaxies are divided by type: ellipticals and lenticulars (EllS0, blue circles), regular spirals (SaSc, orange squares), broken/irregular spirals (SdSm, green triangles) and starbursts (SB, red crosses).}
 \label{fig:df_scatter_all}
\end{figure*}

\begin{figure*}
\centering
\includegraphics[width=0.33\textwidth]{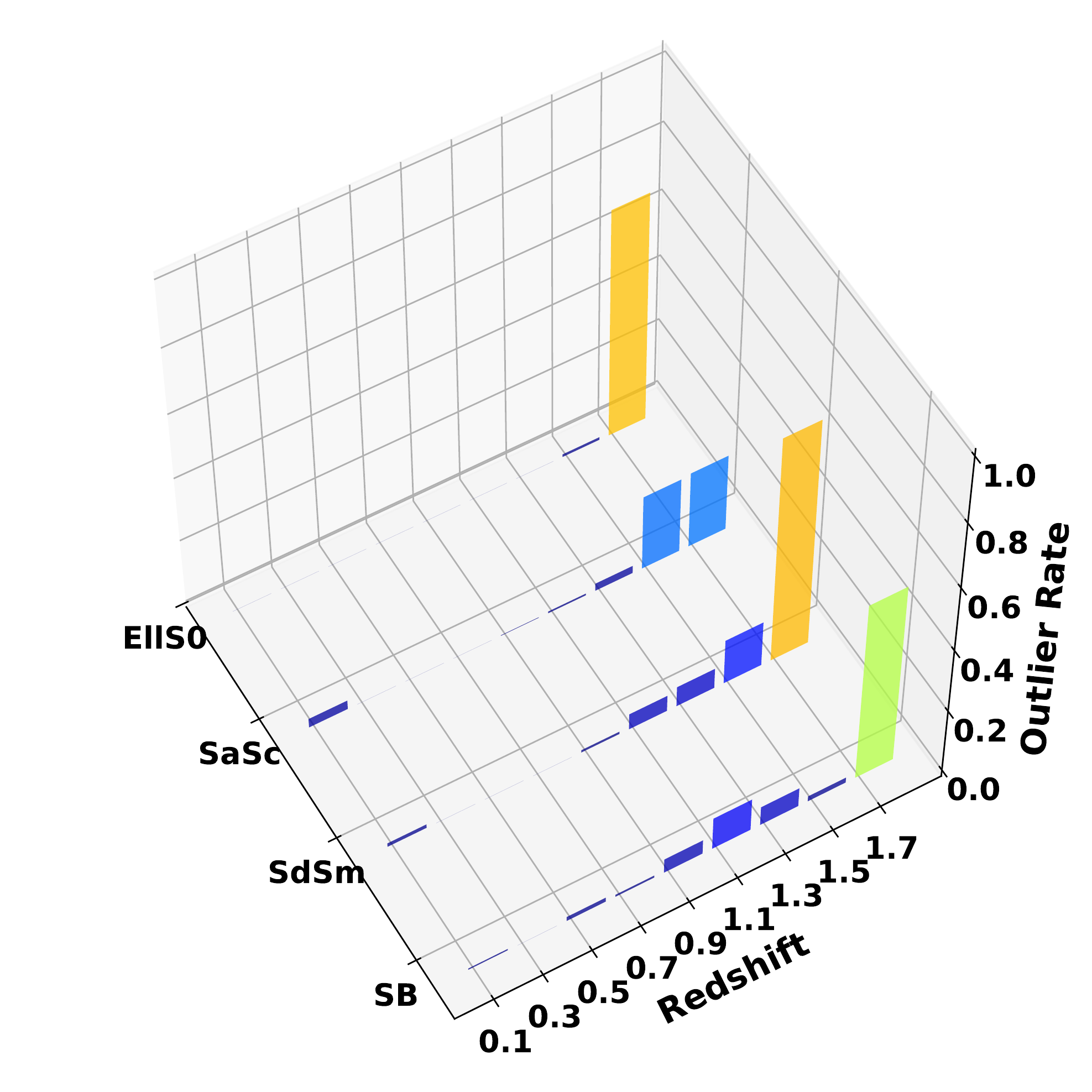}
\includegraphics[width=0.33\textwidth]{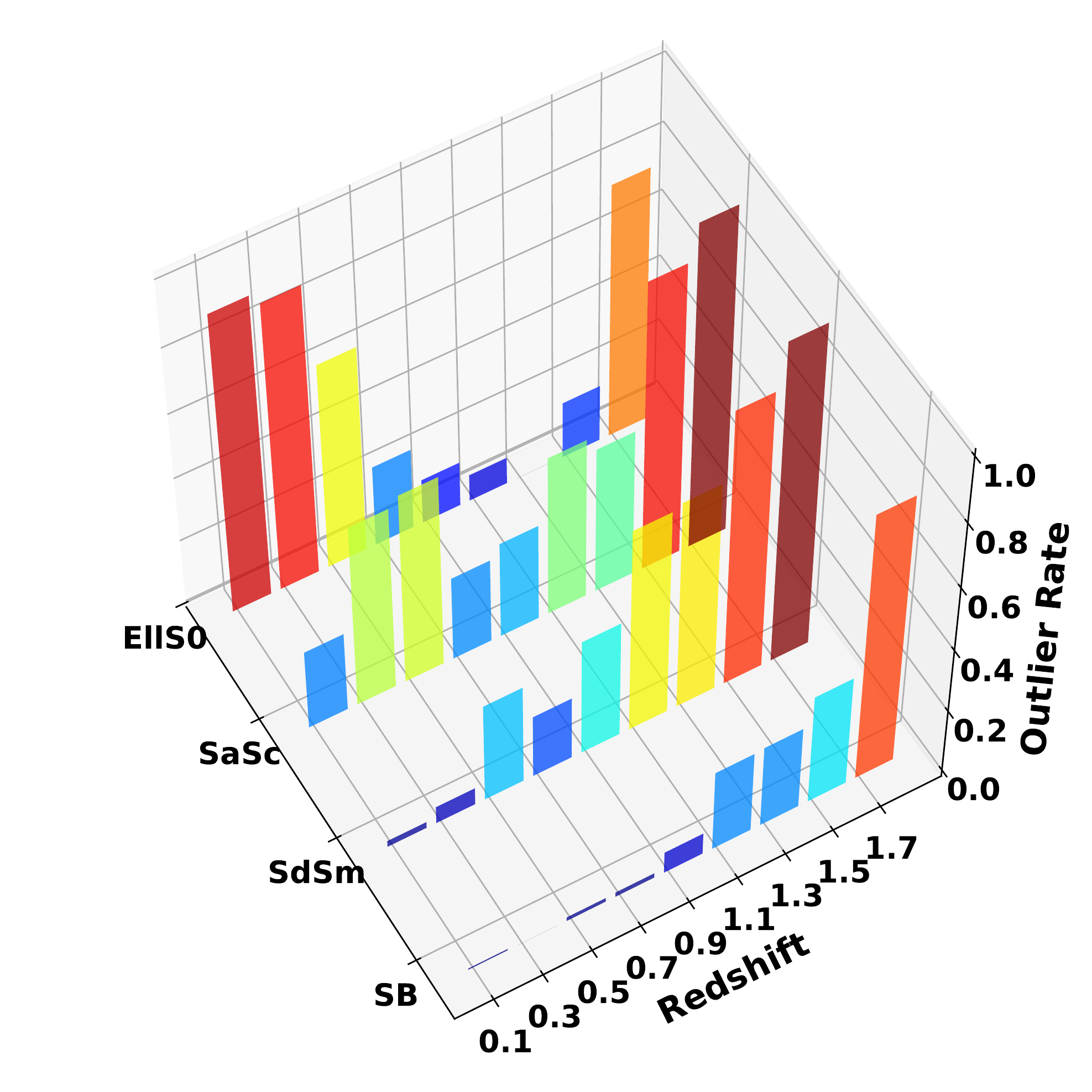}
\includegraphics[width=0.33\textwidth]{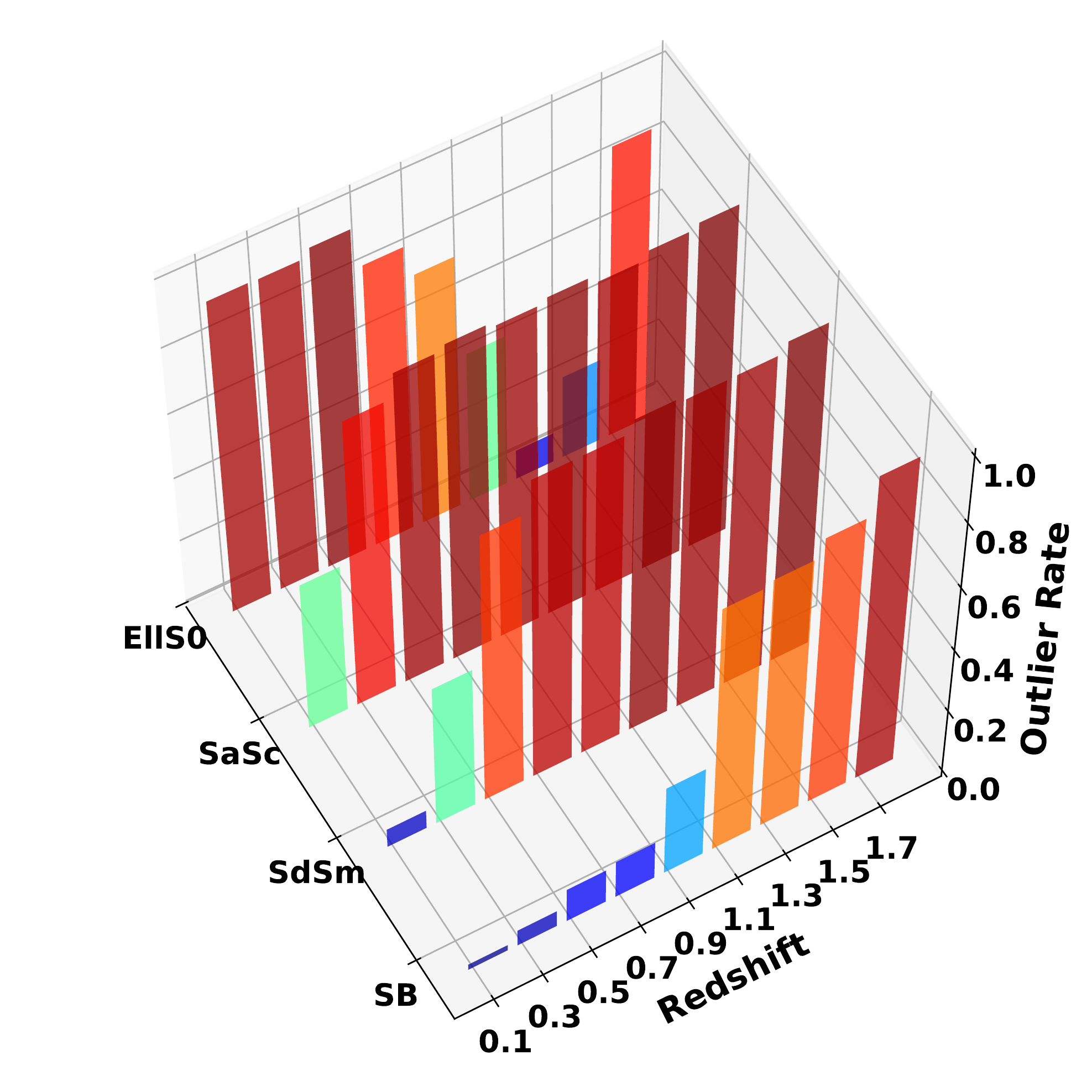}\\
\includegraphics[width=0.33\textwidth]{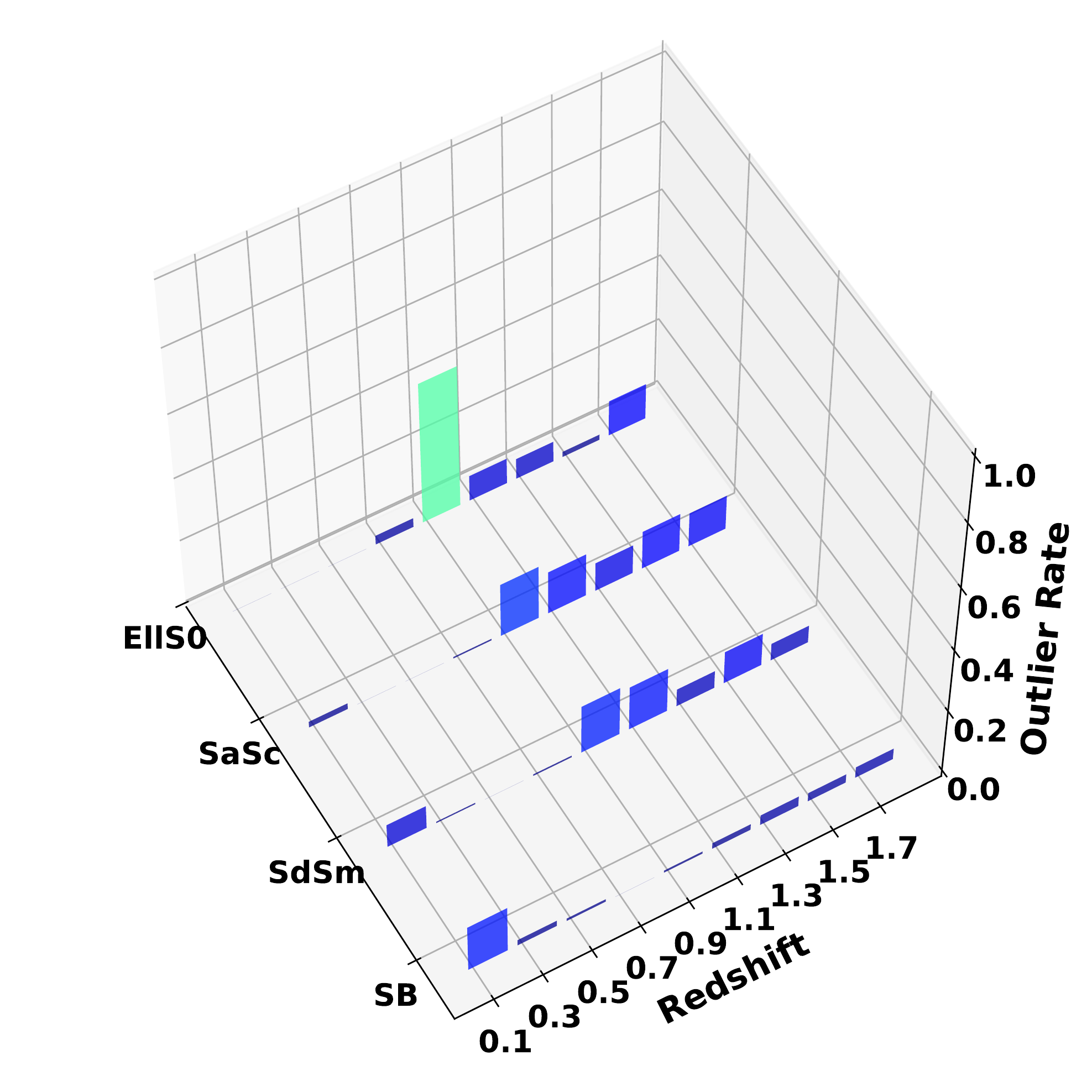}
\includegraphics[width=0.33\textwidth]{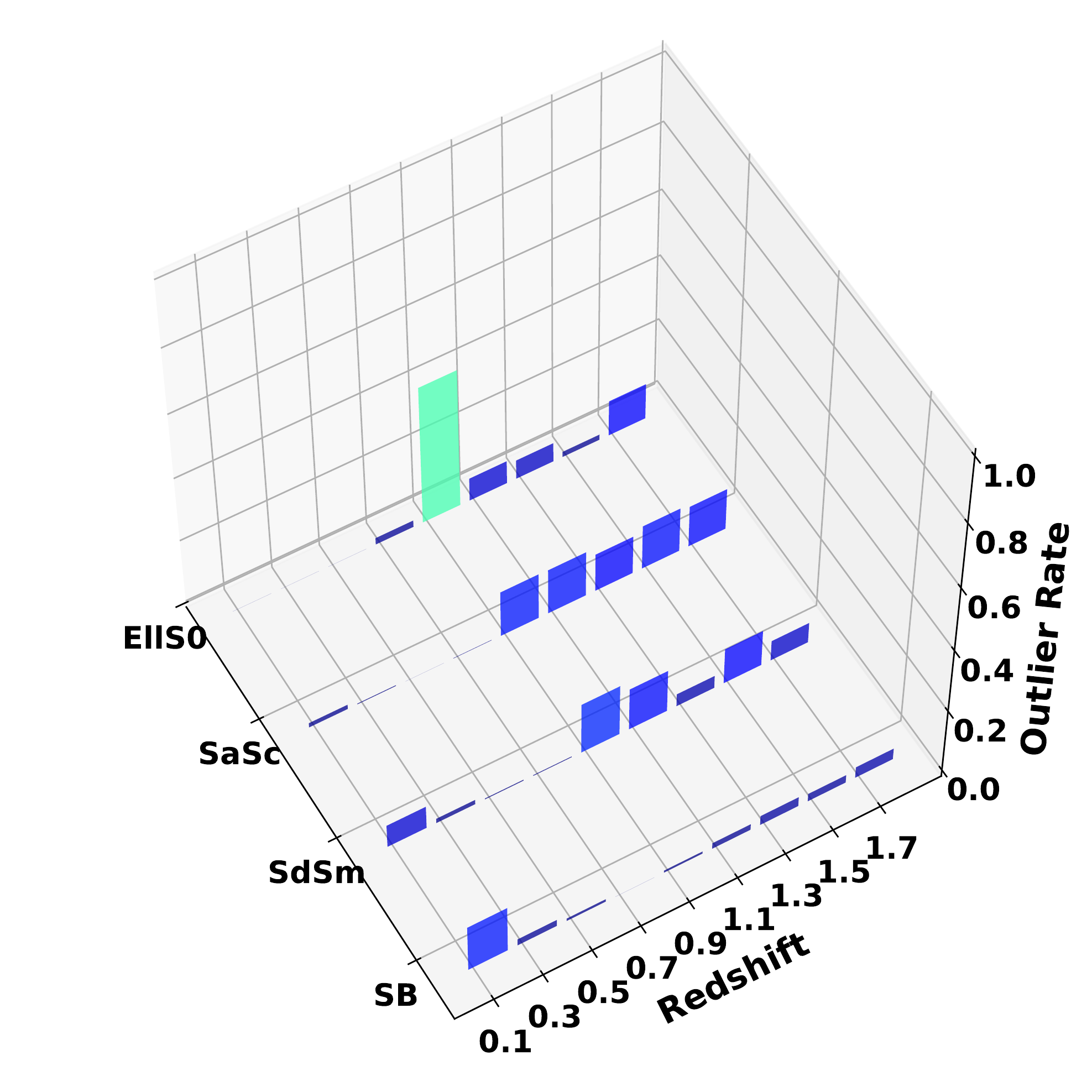}
\includegraphics[width=0.33\textwidth]{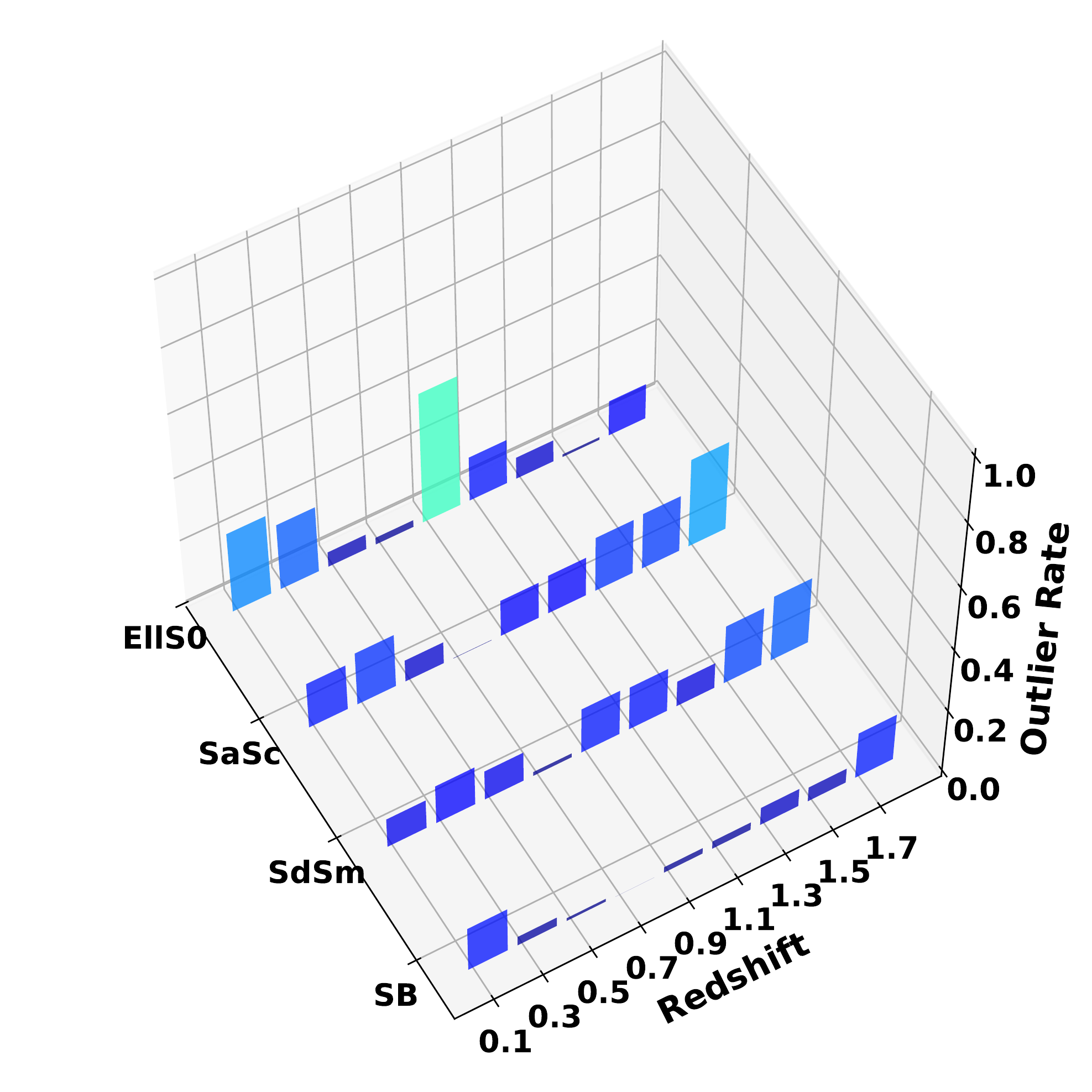}\\
\includegraphics[width=0.33\textwidth]{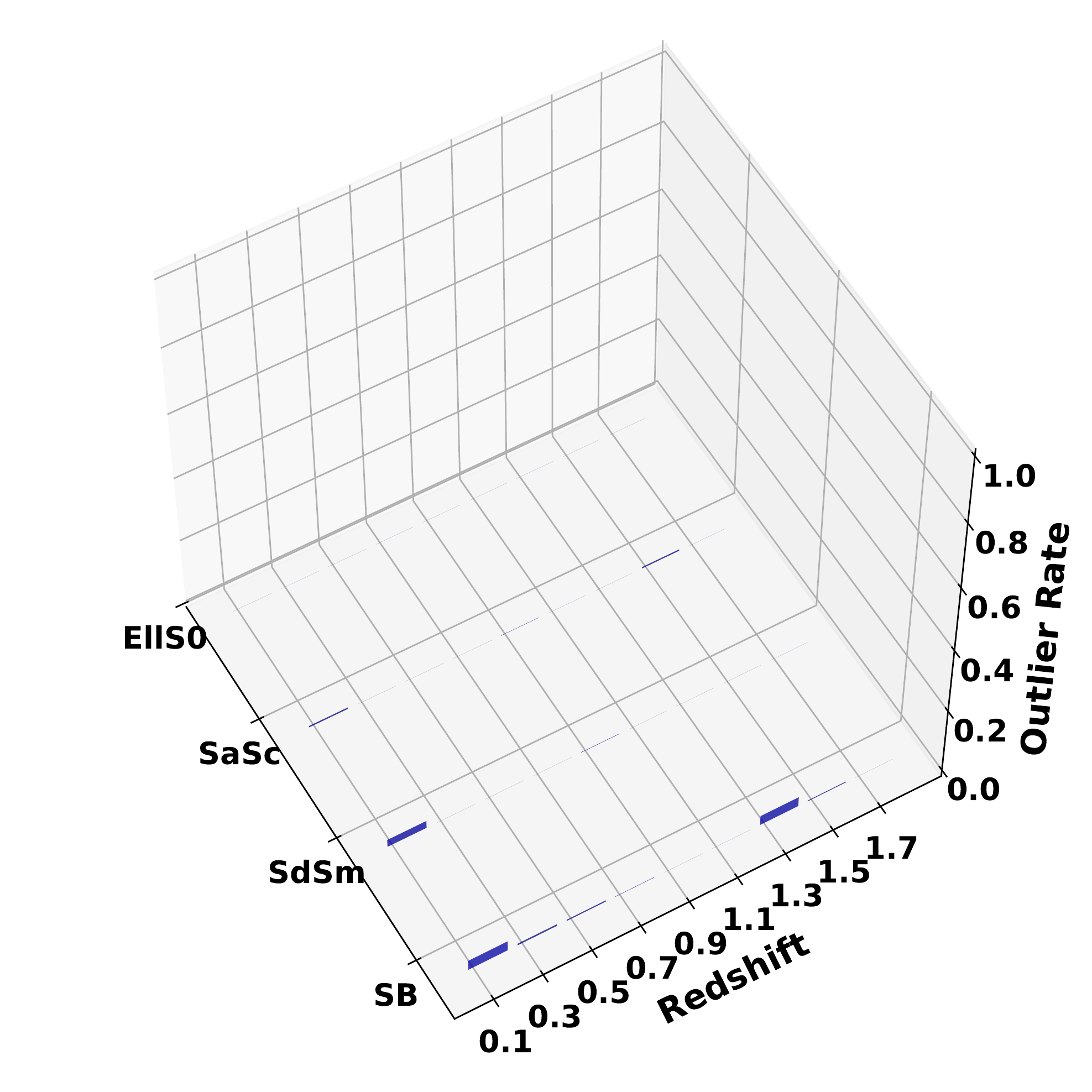}
\includegraphics[width=0.33\textwidth]{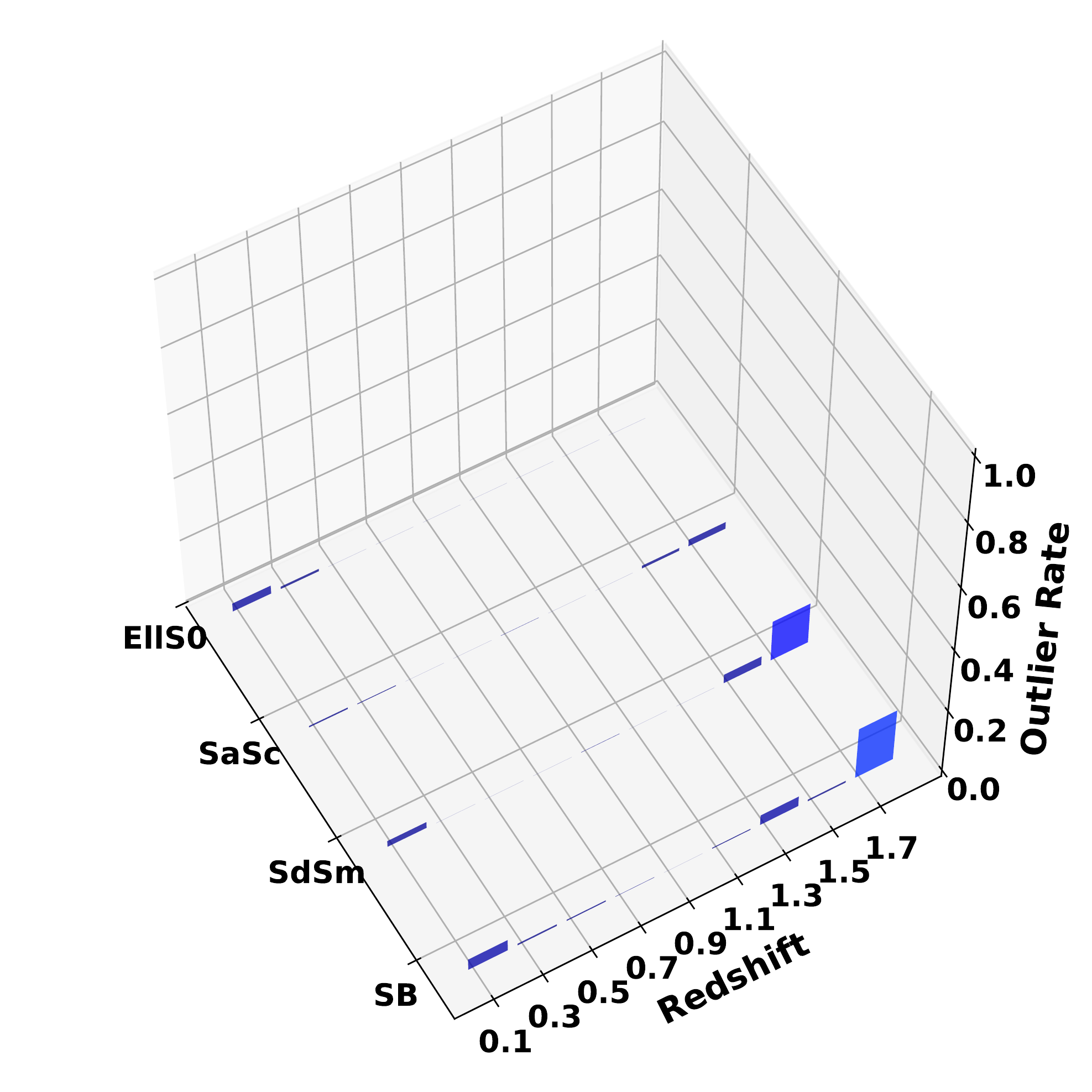}
\includegraphics[width=0.33\textwidth]{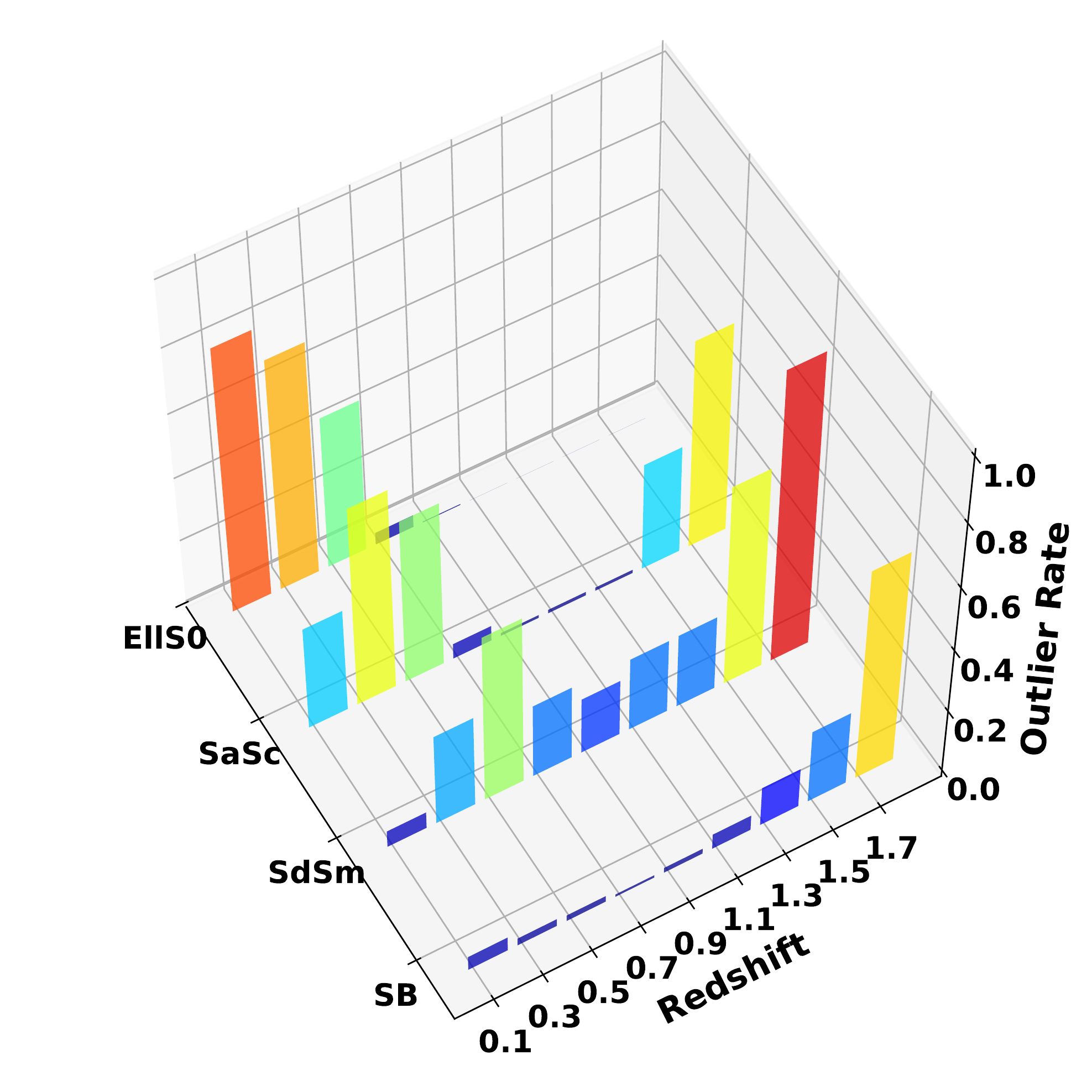}
\caption{Catastrophic failure rate (measured as $|\Delta z / (1+z)|>0.003$) by galaxy type and redshift bins, Darth Fader (top row), dictionary learning (middle row) and denoising autoencoders (bottom row) in the SNR = 20, 5 and 2 cases (left, middle and right columns, respectively). Performance degrades as noise levels increase for most galaxy types and redshifts, especially in Darth Fader's case. Denoising autoencoders perform particularly well in the high SNR regimes, while dictionary learning is more stable across SNR regimes.}
 \label{fig:df_galtypes}
\end{figure*}

\subsection{Comparison of results}
\label{subsec:comparison}
In this section, we will compare the results of all three methods, paying special attention to dictionary learning and denoising autoencoders. These methods improve performance in all SNR regimes and for all galaxy types when compared to Darth Fader, but each has its own advantages and drawbacks.

Figure \ref{fig:df_scatter_all} compares the estimated to true redshift values for all three methods in three different SNR regimes. For SNR = 20, all methods perform well, with small dispersion around the truth and very few catastrophic outliers. Nonetheless, there are qualitative differences in the distribution of these outliers. Dictionary learning encounters small difficulties with most types of galaxies, in particular at intermediate redshifts. Darth Fader and DAE, on the other hand, show a common pattern of secondary linear features, which is due to feature confusion: when an insufficient number of features is present, the redshift estimation process confuses between features in a predictable matter. For example, an H$\alpha$ hydrogen transition line can be misidentified as a OII oxygen line. The phenomenon is stronger with Darth Fader, which in addition has another cluster of outliers at low-estimated/high-true redshift.

At SNR 5, Darth Fader performance is significantly degraded. DAE and DL outlier rate increases slightly, but the patterns of errors remain the same for them. However, DAE starts showing a similar cluster of outliers that were already present in Darth Fader's SNR 20 results. This trend is even stronger when DAE reaches SNR 2. There are striking similarities between its results in this regime and Darth Fader's at SNR 5, seeming to indicate that both algorithms are reacting to the same underlying patterns in the data. The situation is markedly different for Dictionary Learning. In the SNR 2 regime, the number of outliers increases but they still follow a pattern of clustered groups of outliers relatively close to the true redshift values.

\begin{table*}[t]
\centering
   \begin{tabular}{cccccccccccc}
\toprule
\multicolumn{4}{c }{SNR = 20} & \multicolumn{4}{c }{SNR = 5} & \multicolumn{4}{c }{SNR = 2}\\
\midrule
\multicolumn{4}{c }{Total DF success: 19479 (97.4 \%)} & \multicolumn{4}{c }{Total DF success: 13125 (65.6 \%)} & \multicolumn{4}{c }{Total DF success: 5586 (27.9 \%)}\\[0.075cm]
\multicolumn{4}{c }{Total DL success: 18668 (93.3 \%)} & \multicolumn{4}{c }{Total DL success: 18694 (93.5 \%)} & \multicolumn{4}{c }{Total DL success: 17822 (89.1 \%)}\\[0.075cm]
\multicolumn{4}{c }{Total DAE success: 19954 (99.8 \%)} & \multicolumn{4}{c }{Total DAE success: 19880 (99.4 \%)} & \multicolumn{4}{c }{Total DAE success: 15545 (77.7 \%)}\\[0.075cm]
\cmidrule(lr){1-4} \cmidrule(lr){5-8} \cmidrule(lr){9-12}
\multicolumn{4}{c }{Total 'best' success: 19945 (99.7 \%)} & \multicolumn{4}{c }{Total 'best' success: 19895 (99.5 \%)} & \multicolumn{4}{c }{Total 'best' success: 18630 (93.1 \%)}\\[0.075cm]
\cmidrule(lr){1-4} \cmidrule(lr){5-8} \cmidrule(lr){9-12}
DF&DL&DAE&&DF&DL&DAE&&DF&DL&DAE&\\
\midrule
\ding{52} & \ding{52} & \ding{52} & 18185 &
\ding{52} & \ding{52} & \ding{52} & 12238 &
\ding{52} & \ding{52} & \ding{52} & 5203\\
\ding{52} & \ding{52} & \ding{54} & 14 &
\ding{52} & \ding{52} & \ding{54} & 18 &
\ding{52} & \ding{52} & \ding{54} & 87\\
\ding{52} & \ding{54} & \ding{52} & 1264 &
\ding{52} & \ding{54} & \ding{52} & 856 &
\ding{52} & \ding{54} & \ding{52} & 256\\
\ding{54} & \ding{52} & \ding{52} & 455 &
\ding{54} & \ding{52} & \ding{52} & 6357 &
\ding{54} & \ding{52} & \ding{52} & 9082\\
\ding{52} & \ding{54} & \ding{54} & 16 &
\ding{52} & \ding{54} & \ding{54} & 13 &
\ding{52} & \ding{54} & \ding{54} & 40\\
\ding{54} & \ding{52} & \ding{54} & 14 &
\ding{54} & \ding{52} & \ding{54} & 81 &
\ding{54} & \ding{52} & \ding{54} & 3450\\
\ding{54} & \ding{54} & \ding{52} & 50 &
\ding{54} & \ding{54} & \ding{52} & 429 &
\ding{54} & \ding{54} & \ding{52} & 1004\\
\ding{54} & \ding{54} & \ding{54} & 2 &
\ding{54} & \ding{54} & \ding{54} & 8 &
\ding{54} & \ding{54} & \ding{54} & 878
\end{tabular}
   \vspace*{0.1cm}
   \caption{Success rate of the redshift estimation algorithms in the different SNR cases, where failure is defined as $z_{\mathrm{est}} - z_{\mathrm{true}} > 0.003\,(1 + z_{\mathrm{true}})$. The 'best' algorithm combines DL/DAE results as described in section \ref{sec:zbest} and shown in Fig \ref{fig:zbest_scatter_all}. The lower half of the table shows the partial statistics for different combinations of algorithm success; for example, the 1st line counts the number of galaxies for which the 3 algorithms succeeded in retrieving the true redshift, the 2nd line counts the number of galaxies where DF and DL succeeded while DAE failed, and so forth.}
   \label{tab:catout_rates}
\end{table*}

Figure \ref{fig:df_galtypes} investigates the distribution of catastrophic outliers per redshift range and galaxy type for the same three SNR cases. Failure rates are clearly dependent on these variables. In particular, galaxy type strongly influences the performance of the algorithm. This is to be expected: the SED of starburst galaxies, for example, typically includes very strong emission lines due to intense star-formation activity, whereas the SED of elliptical galaxies includes a strong break on the 4000\angstrom{} rest-frame continuum as its most prominent feature.  At $\mathrm{SNR} = 20$, most DF failures take place at the highest redshifts, for most types. DAE demonstrates almost perfect performance, while DL has a relatively low rate of failure, mostly concentrated on redshifts larger than one. Ellipticals and lenticulars (EllS0) at intermediate redshifts are a curious case: Darth Fader and - in particular - DAE maintain a low rate of catastrophic outliers for this subclass. However, DL shows bad performance for this subclass, even at high SNR = 20. What we observe for this type of galaxy and this redshift bin is confusion of features in the continuum in the representation learnt via Dictionary Learning. Indeed, these galaxies represent only 5\% of the training set, dominated by SB galaxies (about 86\%), so learning is mainly driven by approximation of SB galaxies rather than elliptical galaxies. Errors in modeling the continuum of EllS0 is expected therefore to be larger, which could translate to incorrect redshift estimation if this affects discriminative features for redshift estimation. Furthermore, for this redshift bin, the 4000\angstrom{} break is close to the end of the observable range, and confusion of features arise with other discontinuities in the continuum located before or after this break.

For $\mathrm{SNR} = 5$ and $\mathrm{SNR} = 2$, both DAE and DL markedly outperform Darth Fader in all galaxy classes. Comparison between DAE and DL methods is more ambiguous. DAE achieves excellent results for all galaxy classes in higher SNR regimes, as seen in the top and middle panels of the middle column in Figure \ref{fig:df_galtypes}; catastrophic outlier rates are nearly zero everywhere. However, performance rapidly degrades when SNR decreases, as seen in the lower panel. Moreover, comparison to the middle lower panel of Figure \ref{fig:df_scatter_all} shows that those failures are random within a band of estimated redshift, which signals a pathological behaviour. DL, on the other hand, keeps a consistent pattern of failures, maintaining lower than 20\% catastrophic outlier rates for most galaxy types and redshift bins. In that sense, both methods are complementary. In terms of galaxy types, all methods perform best with starburst galaxies; even in the low signal-to-noise $\mathrm{SNR} = 2$ regime, Darth Fader still succeeds in keeping the catastrophic outlier rate of this class below 20\% for redshifts smaller then $z=1$.

Table \ref{tab:catout_rates} presents a quantitative overview of the catastrophic outlier rates. In addition to the total success rates for each algorithm in the different SNR regimes, we also investigate the partial rates when different combinations of algorithms succeed in measuring precise redshifts. We see, for example, that it is exceedingly rare for Darth Fader to succeed when both Dictionary Learning and Denoising Autoencoders fail. Even if only one of them succeeds, it is still quite rare for Darth Fader to succeed also. In other words, once we have DL and DAE results, DF results are superfluous; they measure accurate redshifts mostly when the other two methods also do. Focusing on the $\mathrm{SNR} = 2$ case, we see the potential advantage of combining the methods. The total success rate when both DL and DAE succeed is $71.4\,\%$ (i.e. $(5203 \,+\, 9082)/20000$), which is already significant for a low SNR regime. However, if we identify a way of selecting the best method when one \textit{or} the other succeeds, we can reach $18739/20000 = 93.7\,\%$ success rate. In the next subsection, we will develop an algorithm to select the best redshift possible in each case and study its accuracy.

\begin{figure*}
\centering
\includegraphics[width=0.24\textwidth]{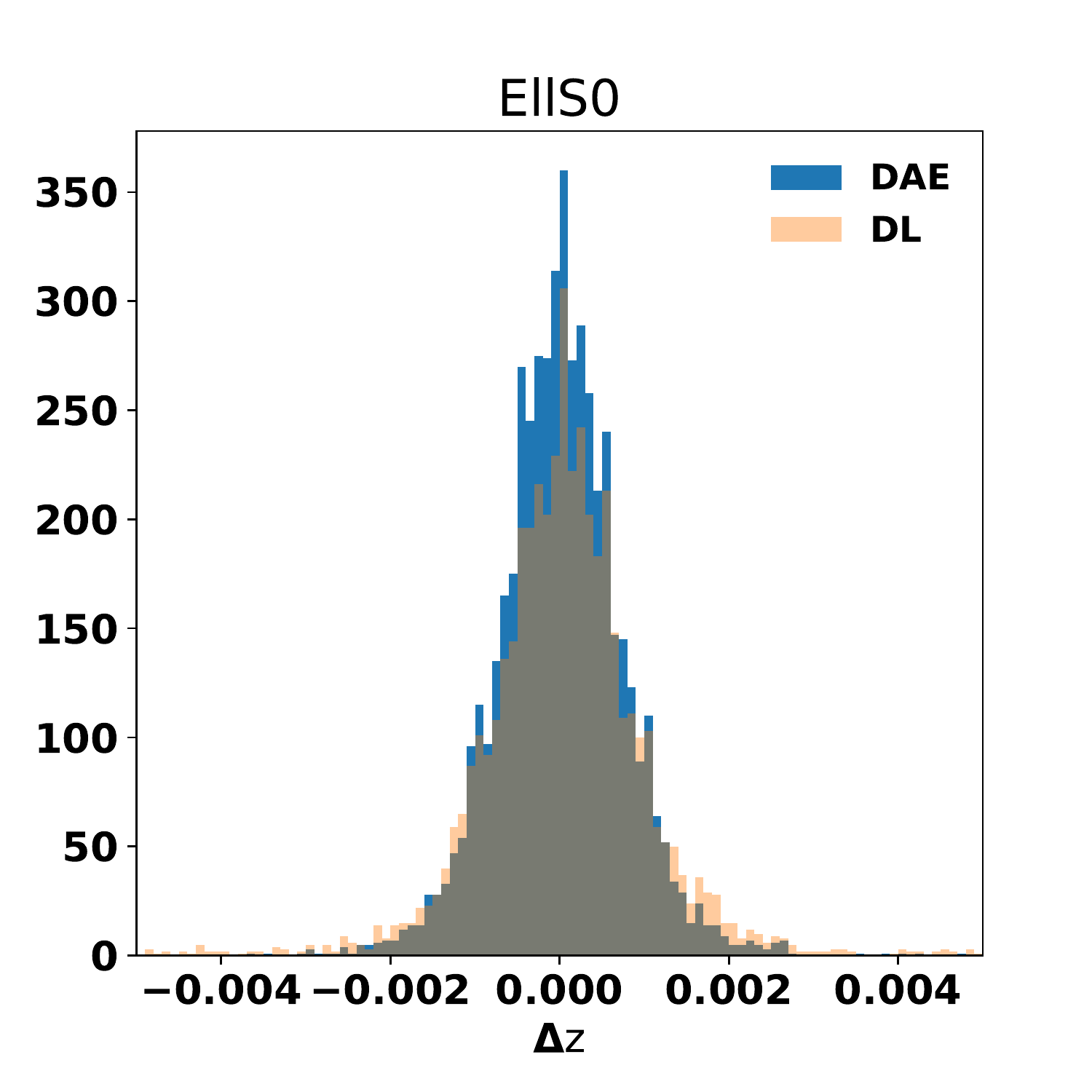}
\includegraphics[width=0.24\textwidth]{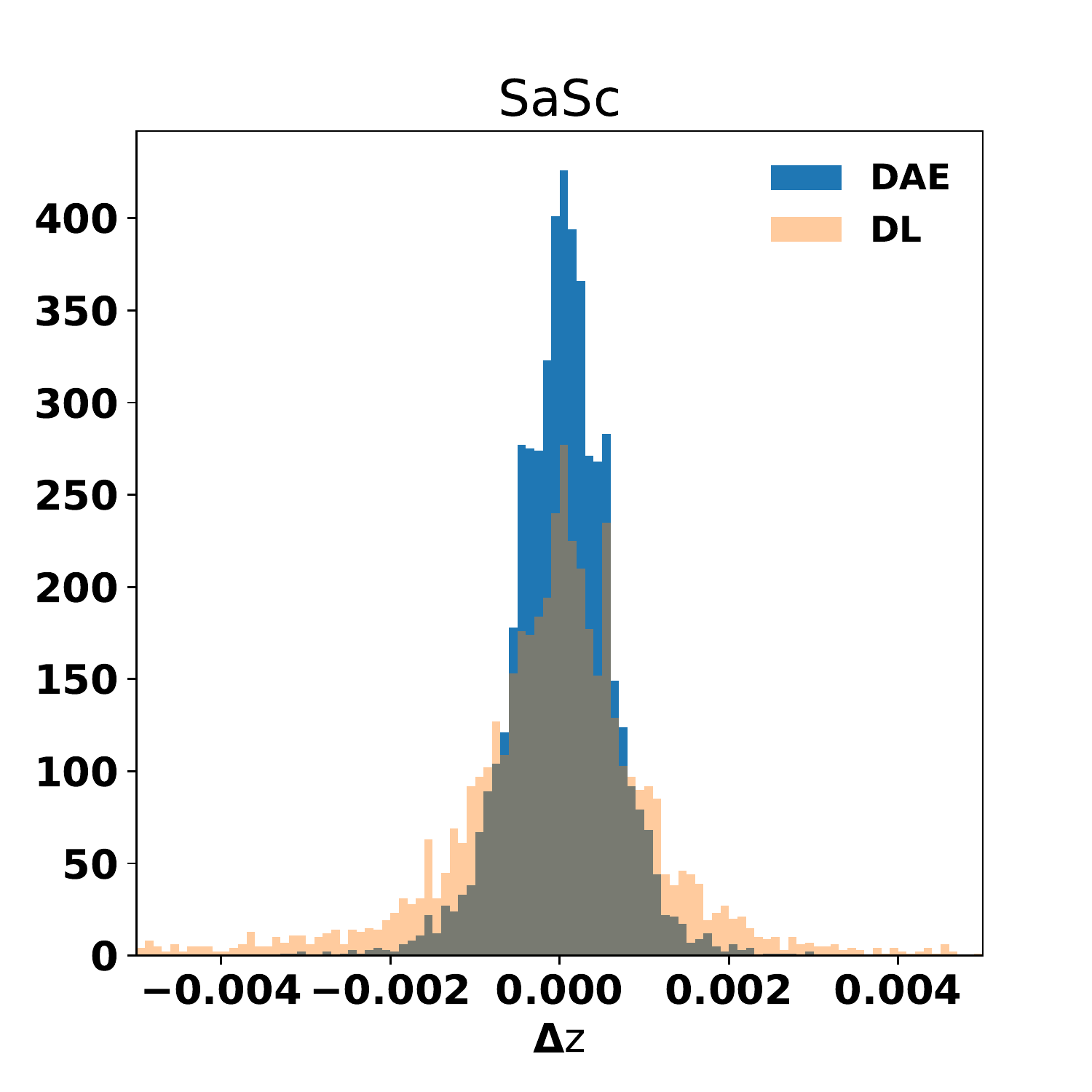}
\includegraphics[width=0.24\textwidth]{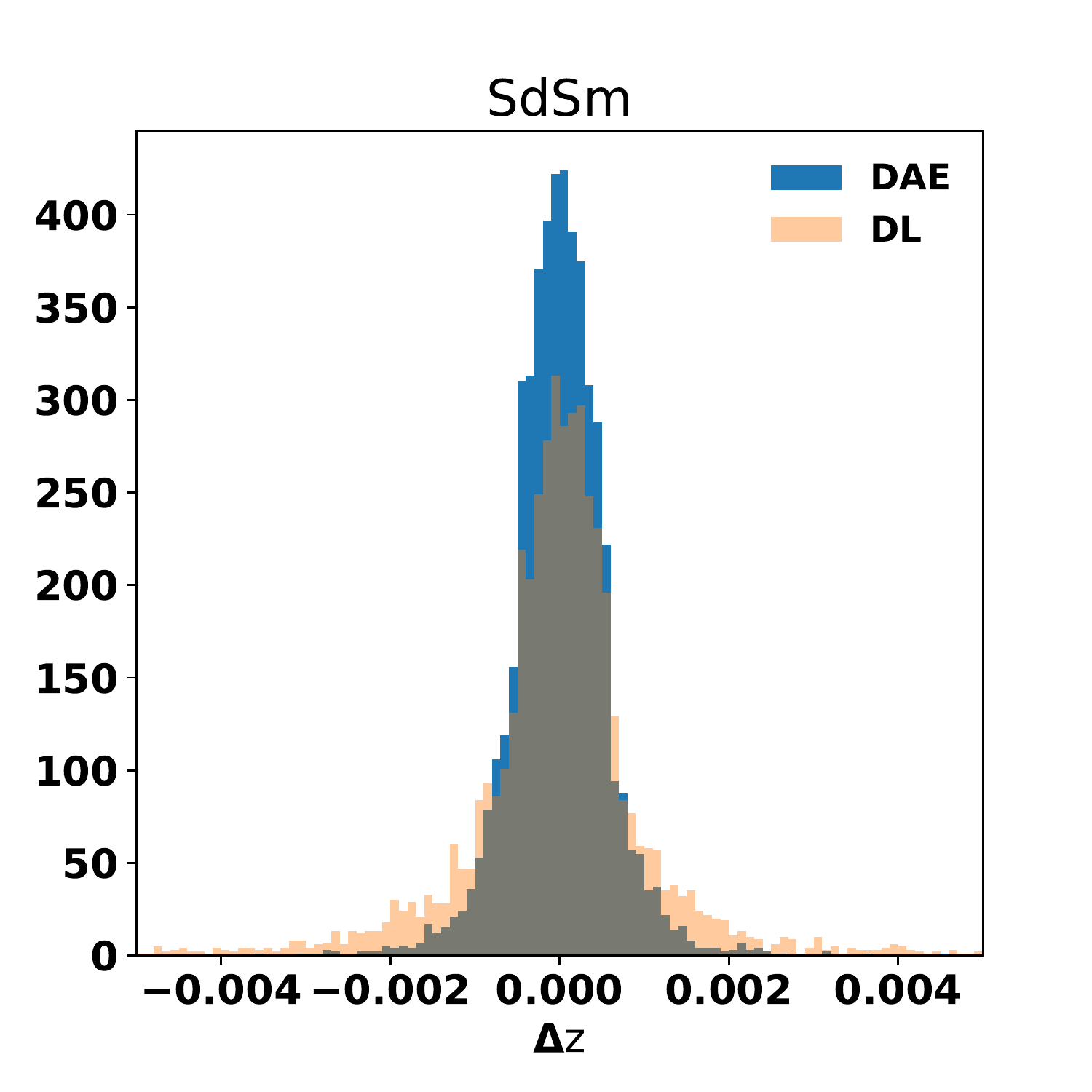}
\includegraphics[width=0.24\textwidth]{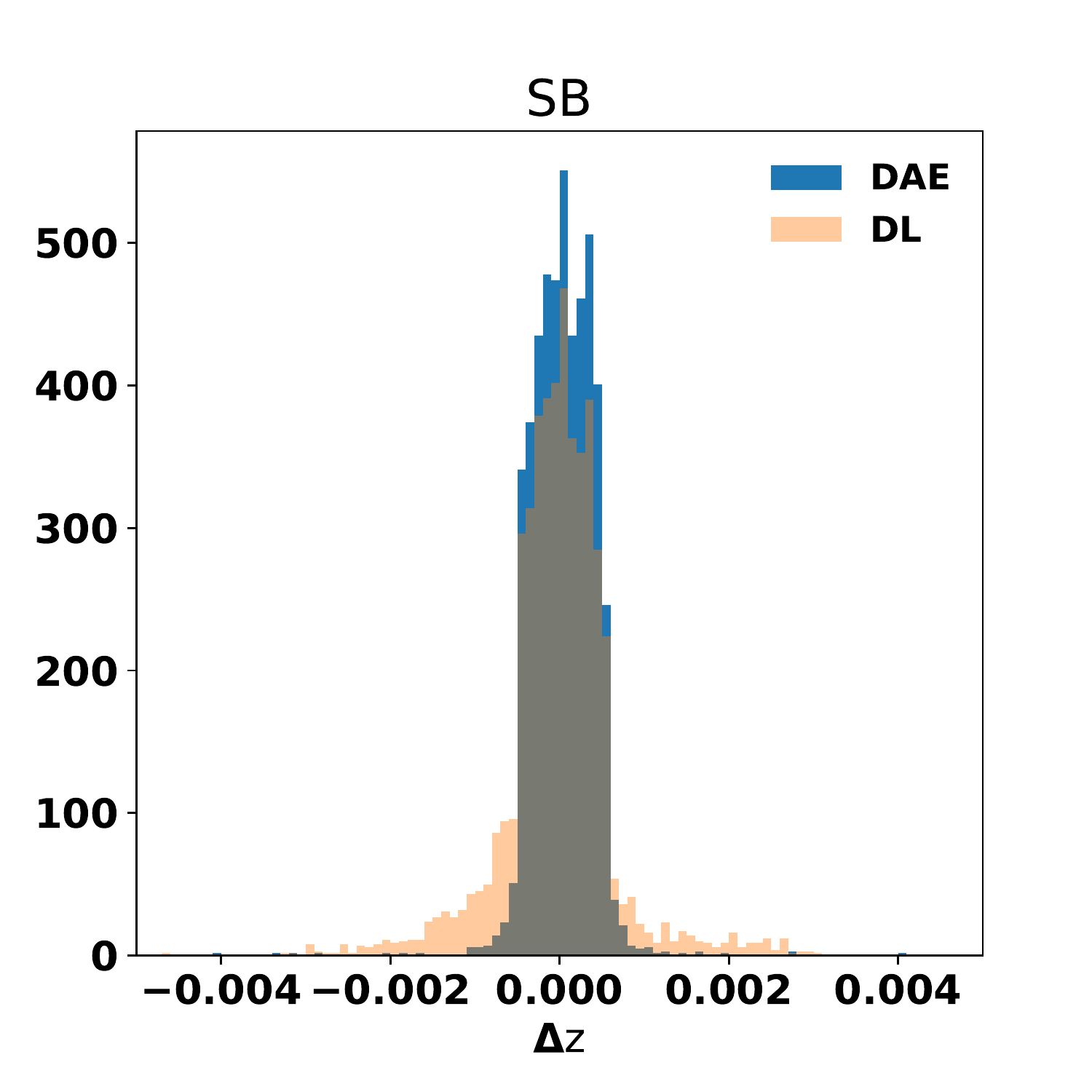}\\
\includegraphics[width=0.24\textwidth]{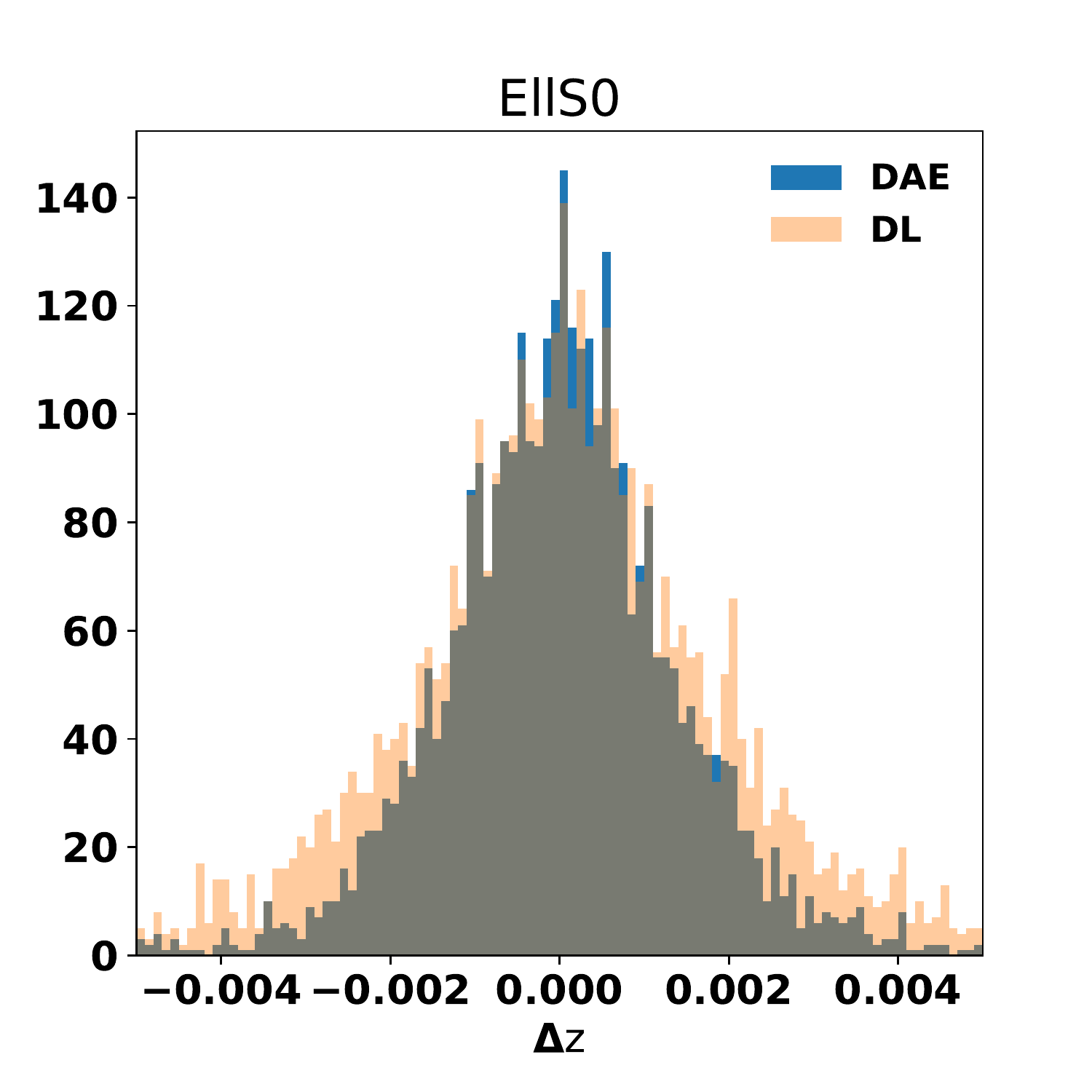}
\includegraphics[width=0.24\textwidth]{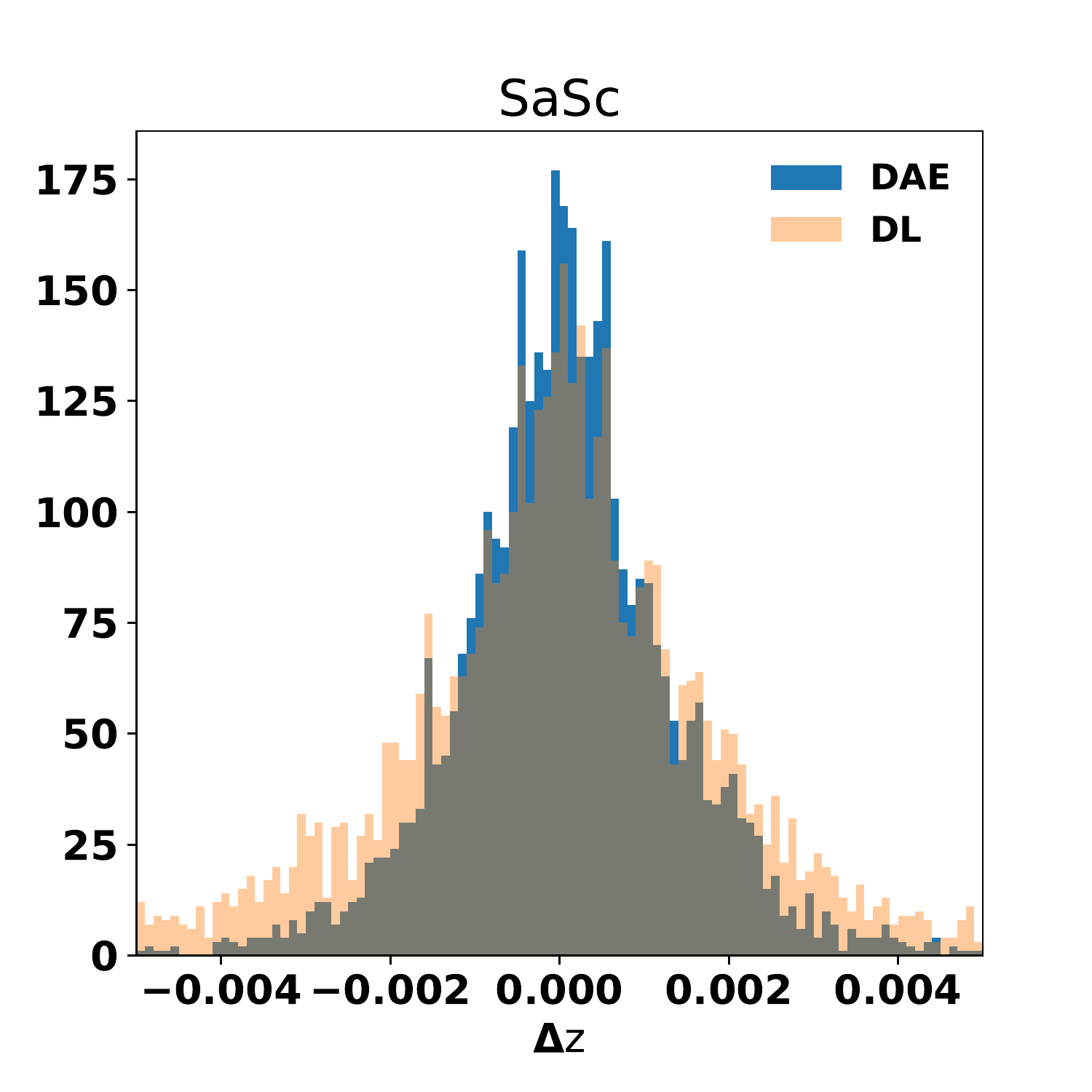}
\includegraphics[width=0.24\textwidth]{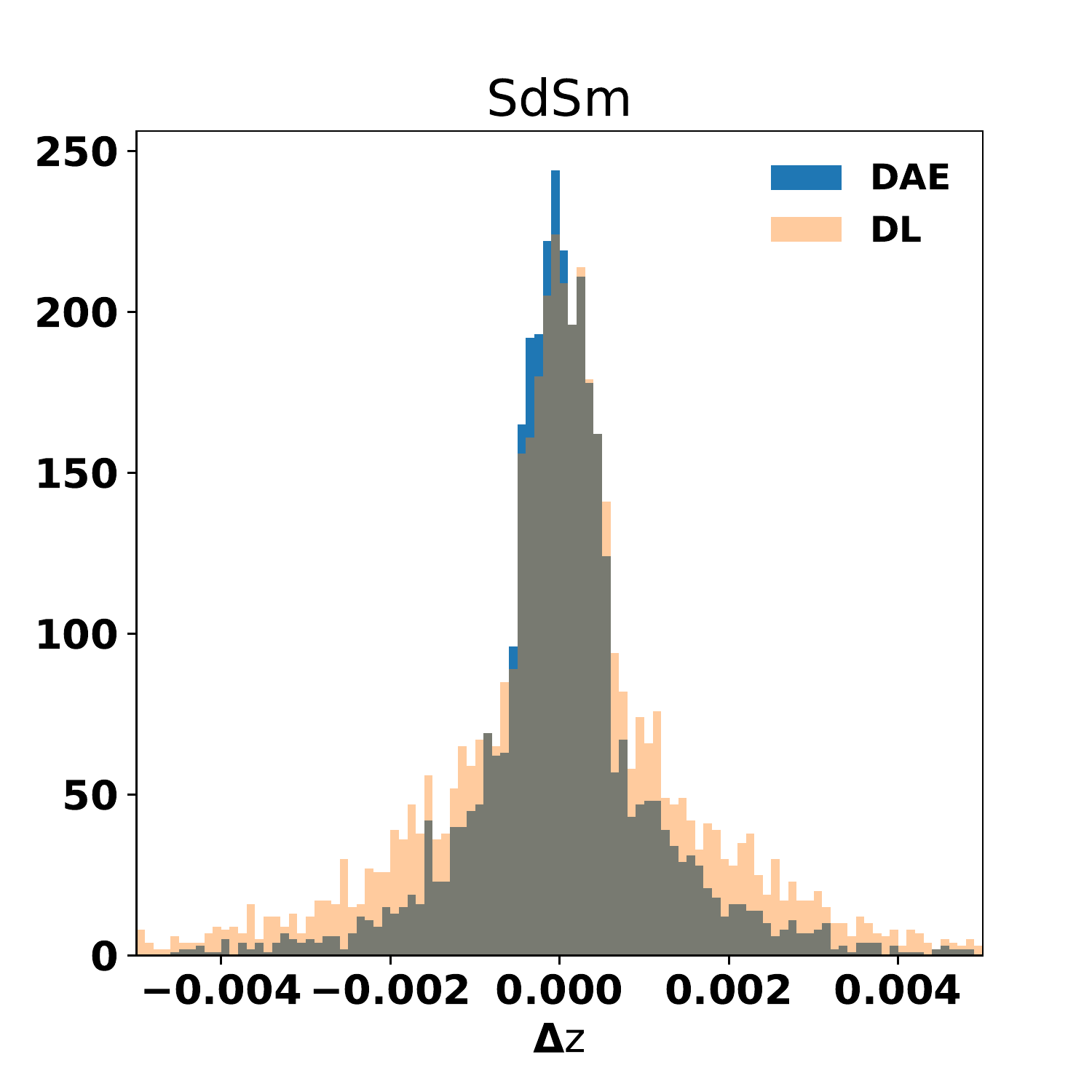}
\includegraphics[width=0.24\textwidth]{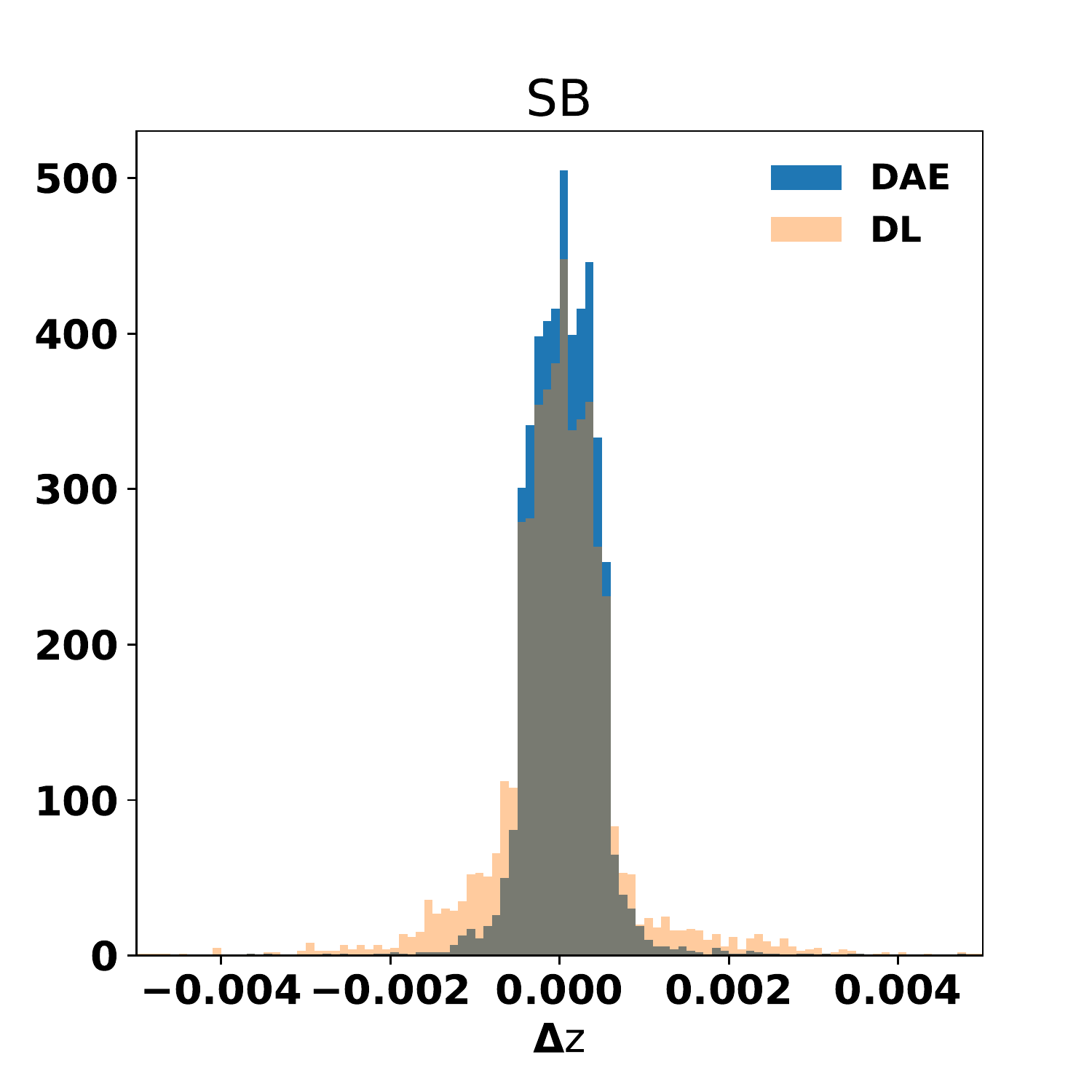}
\caption{Dispersion between true and estimated redshift values for all galaxy types in the SNR = 5 (top row) and SNR = 2 (bottom row) cases. Denoising autoencoders (blue histograms) have generally smaller dispersion than dictionary learning (orange histograms).}
\label{fig:dispersion_snr5_snr2}
\end{figure*}

We now turn to investigating the dispersion of redshift estimates around the true values after excluding catastrophic outliers from consideration. Figure \ref{fig:dispersion_snr5_snr2} compares the dispersion distribution of DAE and DL for all galaxy types in the SNR = 5 and SNR = 2 cases. Yet again, the presence of sharp features clearly influence the performance of both methods, with dispersion of estimated redshifts decreasing from elliptical to starburst galaxies. Additionally, DAE is more precise (i.e. smaller $\sigma$) than DL (except in the particular case of ellipticals at SNR = 5, where performance is comparable). As could be expected, overall dispersion of the estimated values is higher for SNR = 2 than SNR = 5, but this is the only quantitative difference between the two noise levels. In summary, inclusion of continuum for DL allows to obtain a more consistent redshift estimation when noise increases compared to DAE (lower confusion), but adding this essentially low frequency information also degrades the precision in estimating the redshift, since modeling errors on continuum (giving low precision in redshift estimation) may dominate over modeling errors of line features (giving high precision in redshift estimation).

The results described in this section suggest a clear strategy for leveraging the strengths of the different methods: for cases with high signal-to-noise in the continuum, DAE redshifts are more precise and contain less catastrophic outliers. With lower signal-to-noise, DL remains more robust and should be preferred. These results may depend on the noise characteristics, among other SED properties, and should be reevaluated for each separate application.

\subsection{Defining a 'best' redshift from a DAE/DL combination}
\label{sec:zbest}
As we discussed in the past subsections, the DAE and DL algorithms show complementary performance, indicating that a method for combining their results based on observational properties can increase the accuracy of redshift estimation. In this subsection, we will devise an algorithm to take advantage of their strengths, and assess the performance of the resulting redshifts. The two main observational properties that we will consider are the estimated redshifts from each method and the galaxy types. The latter are not an observational property \textit{per se}. Nevertheless, a broad division in four types, such as the one we are using, can be approximated by color cuts in broadband magnitudes from the optical galaxy targeting surveys that serve as a base for spectroscopic surveys. We will postpone a more realistic analysis of this particular aspect to future work.

\begin{algorithm}
\label{Algo:zbest}
\caption{Method to choose a best redshift estimate $z_{\,\mathrm{best}}$.}
\begin{algorithmic}
\ForAll{galaxies}
  \If{DAE/DL redshifts agree}
     \State Take DAE $z_{\,\mathrm{est}}$.
  \Else
     \For{DAE/DL methods}
        \State Associate galaxy to a type/$z_{\,\mathrm{est}}$ bin.
        \State Get catastrophic outlier rate $f_{\mathrm{c}}$ in that bin.
     \EndFor
     \State Take $z_{\,\mathrm{est}}$ of method with lower $f_{\mathrm{c}}$.
  \EndIf
\EndFor
\end{algorithmic}
\end{algorithm}

\begin{figure*}
 \includegraphics[width=0.33\textwidth]{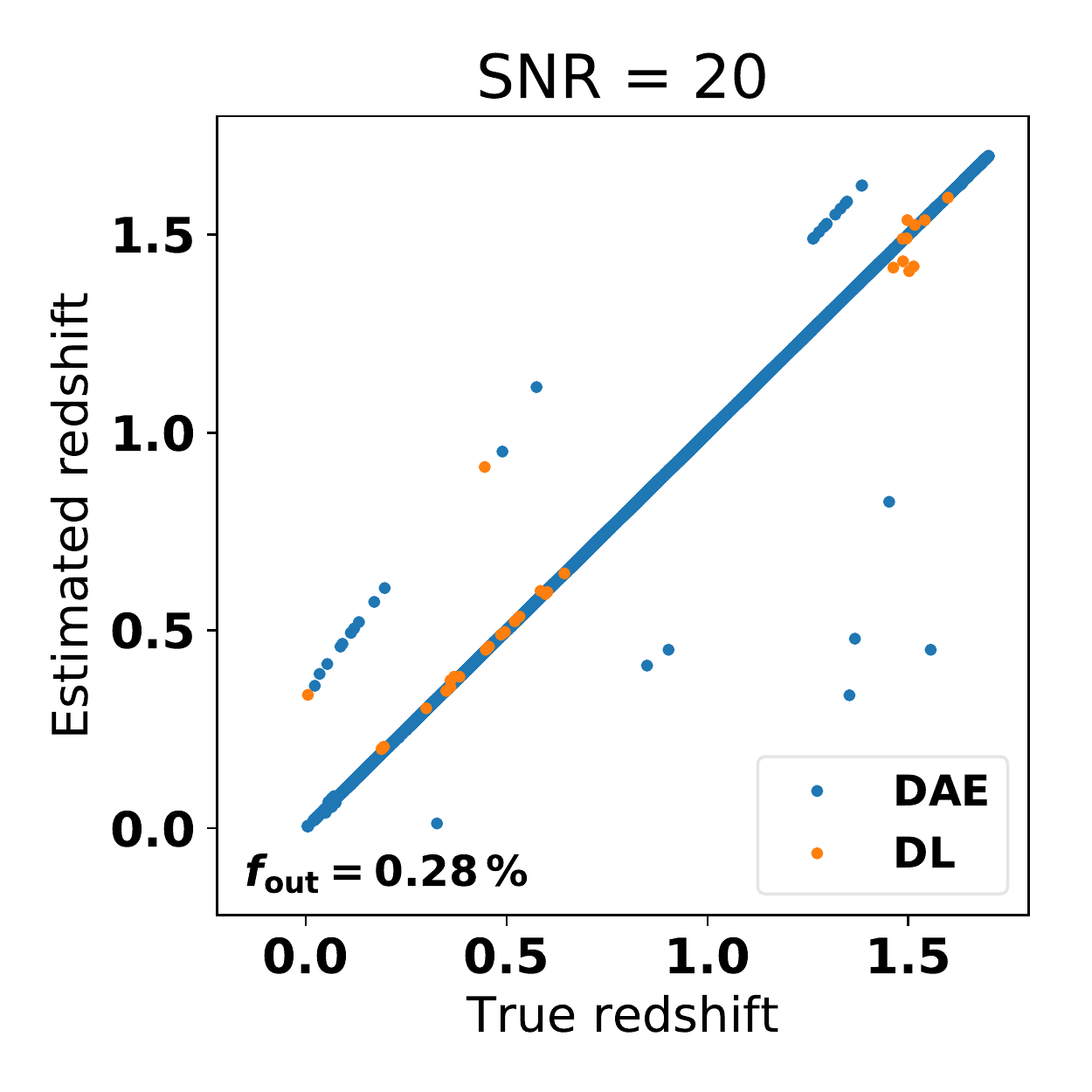}
 \includegraphics[width=0.33\textwidth]{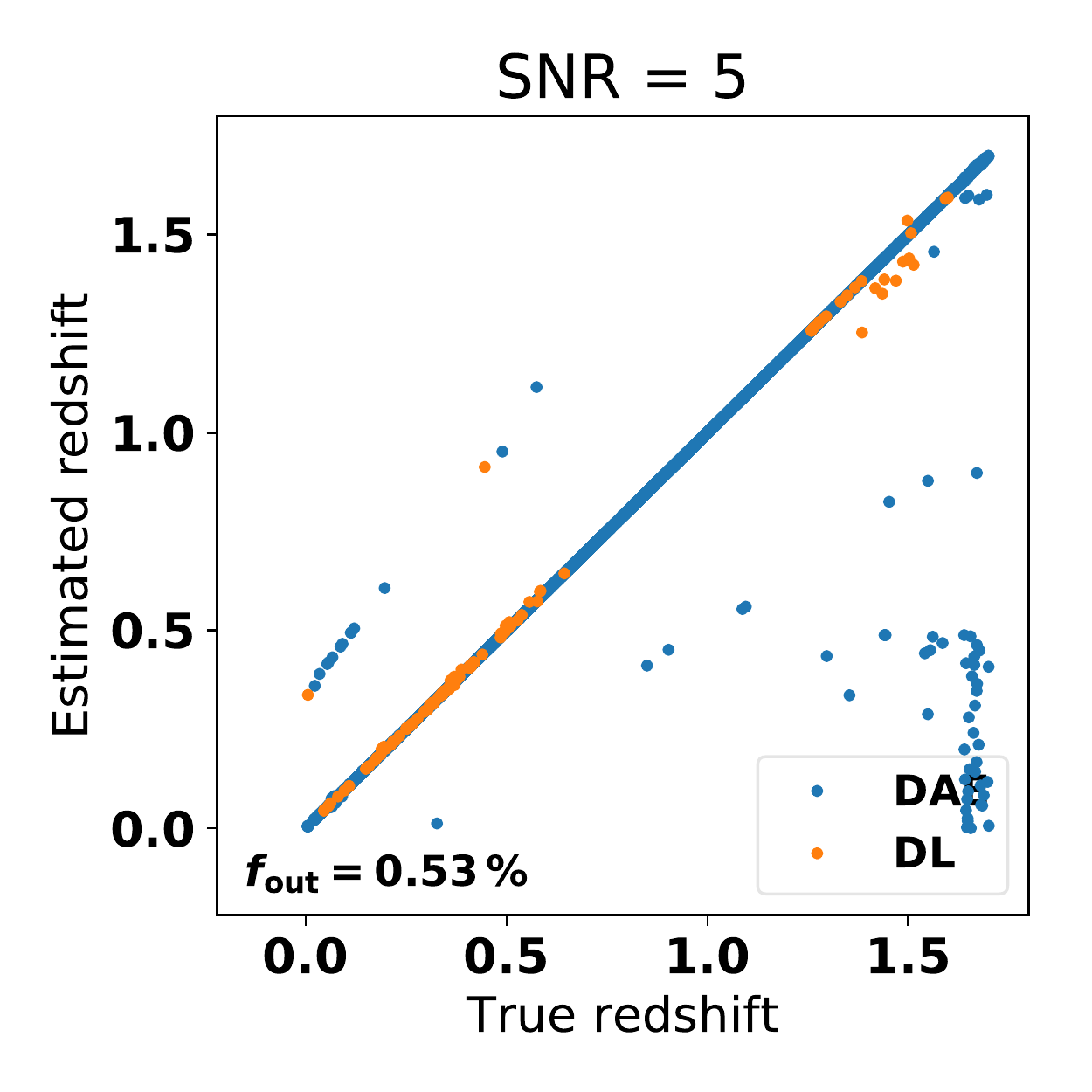}
 \includegraphics[width=0.33\textwidth]{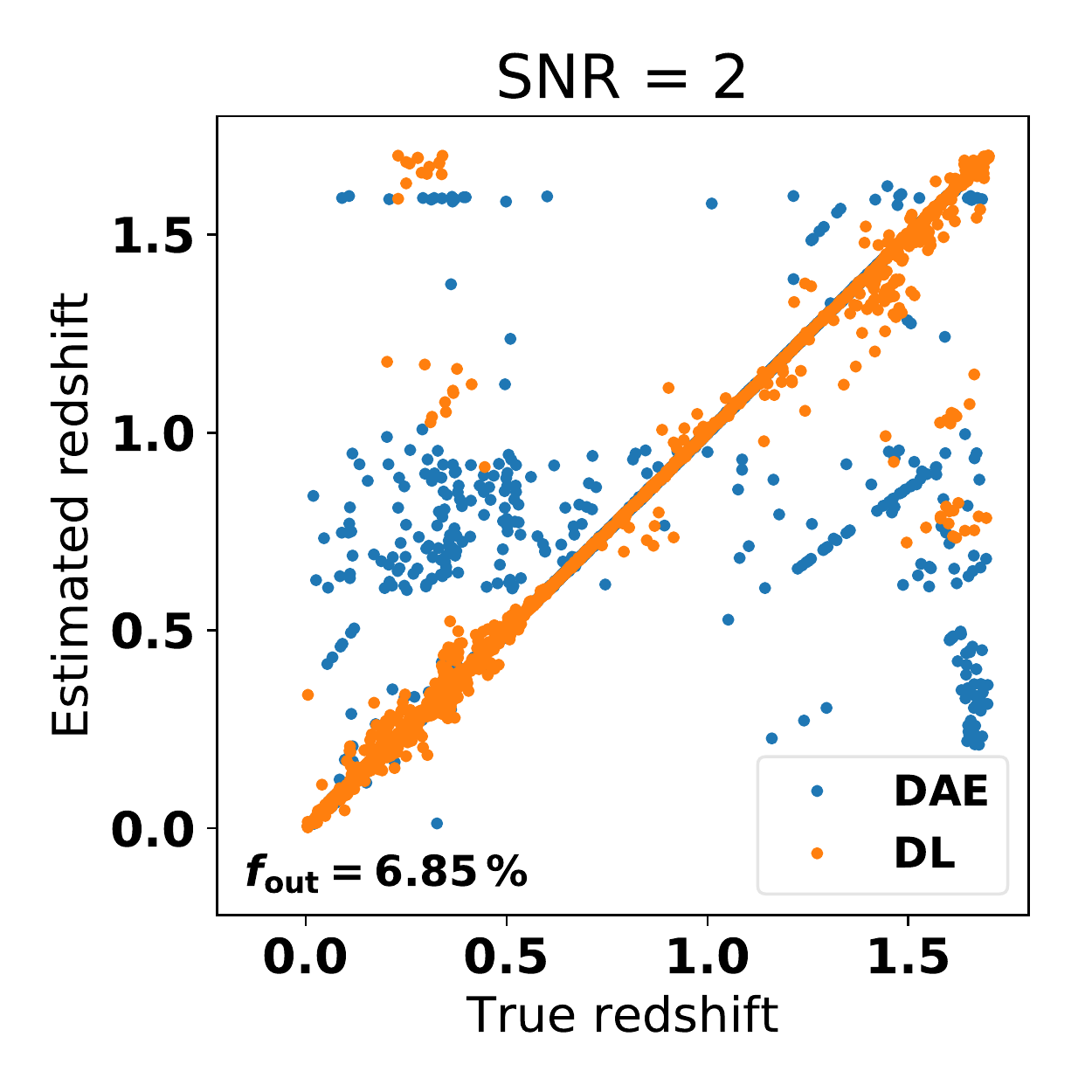}
 \caption{Best estimated redshift vs true redshift for three SNR cases. Blue/Orange dots indicate that the redshift was chosen from the DAE/DL method. In the higher SNR cases, most estimated redshifts come from the DAE method, due to its almost perfect accuracy. At lower SNR values, where the method starts failing often, our algorithm increases the proportion of DL values, which are more robust but with higher variance.}
 \label{fig:zbest_scatter_all}
\end{figure*}

The method we propose is described in Algorithm 3. For each galaxy, we assess whether DAE and DL estimated redshifts agree within a precision threshold. We define it as $\Delta z = 0.003$, which is more strict than the catastrophic outlier rate threshold due to the absence of a true-redshift dependance. Whenever those values agree, we choose the DAE redshift, which has been shown to have smaller dispersion around the true redshifts. If the redshift values don't agree, we resort to using true redshifts to define catastrophic outlier rates in \textit{estimated} redshift and type bins. For each galaxy, we locate it in a redshift-type bin, and then choose the method for which this bin has a lower catastrophic outlier rate. Note that the catastrophic outlier rates are defined with estimated, not true, redshifts in those cases, meaning that there will be two different binning schemes.

\begin{figure*}
\centering
\includegraphics[width=0.33\textwidth]{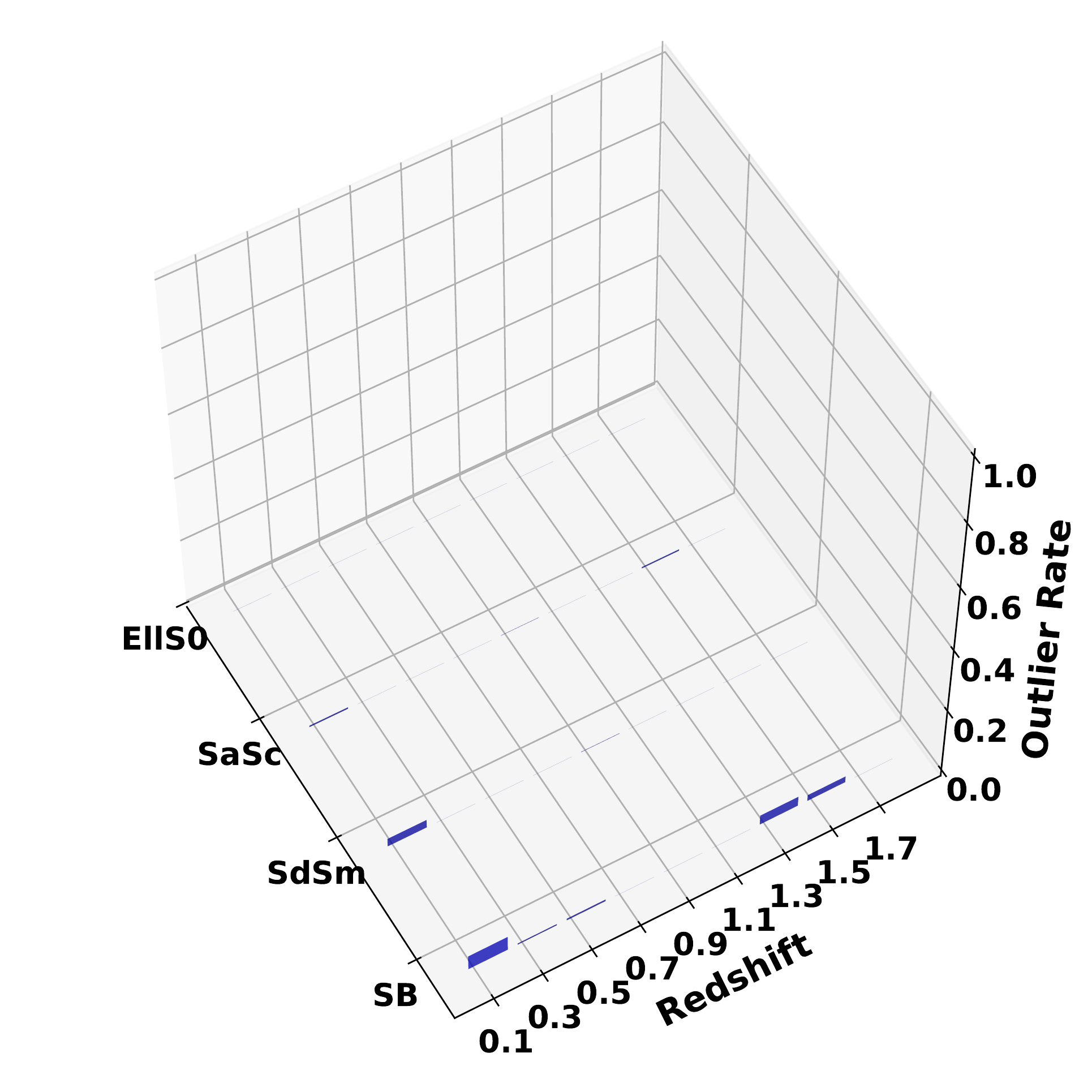}
\includegraphics[width=0.33\textwidth]{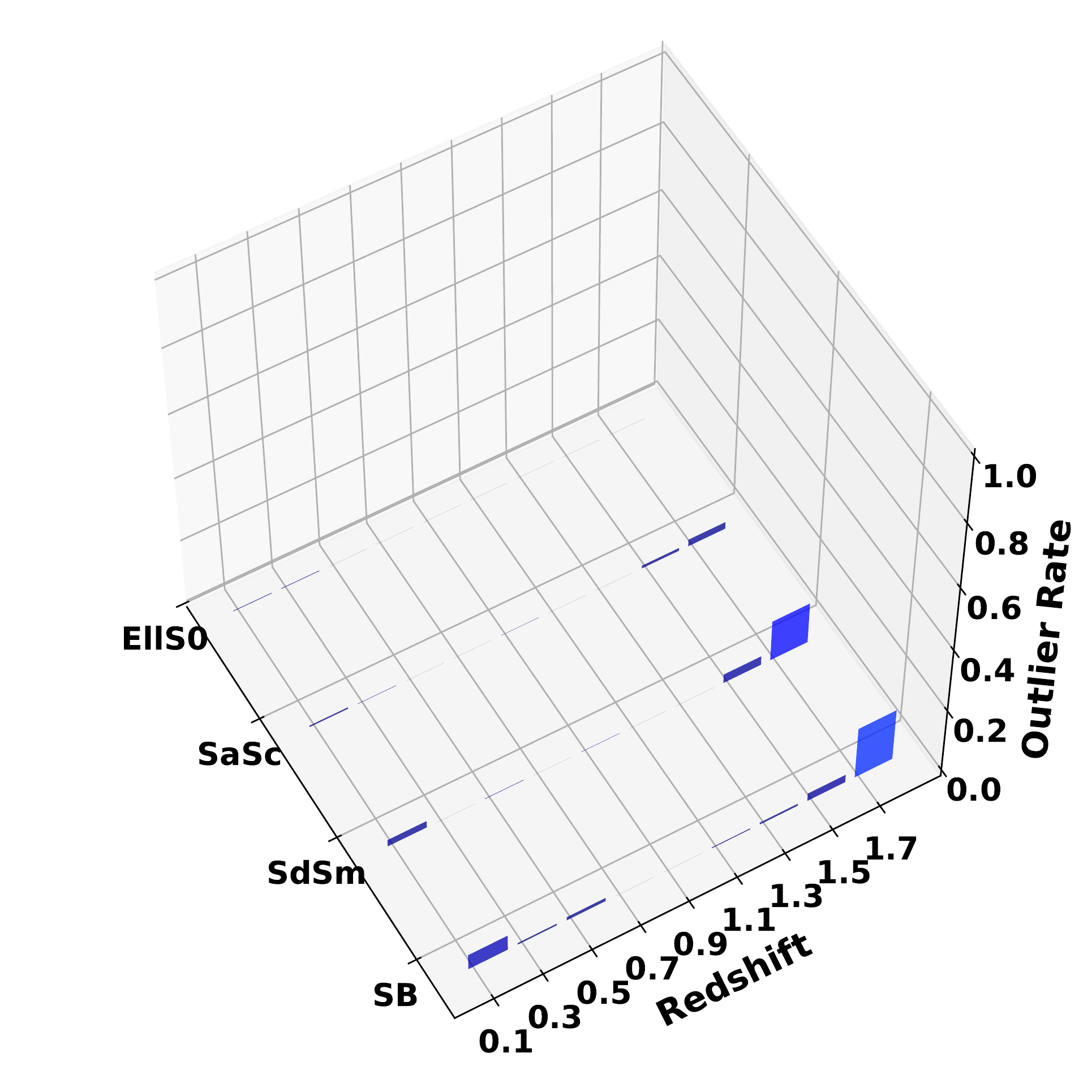}
\includegraphics[width=0.33\textwidth]{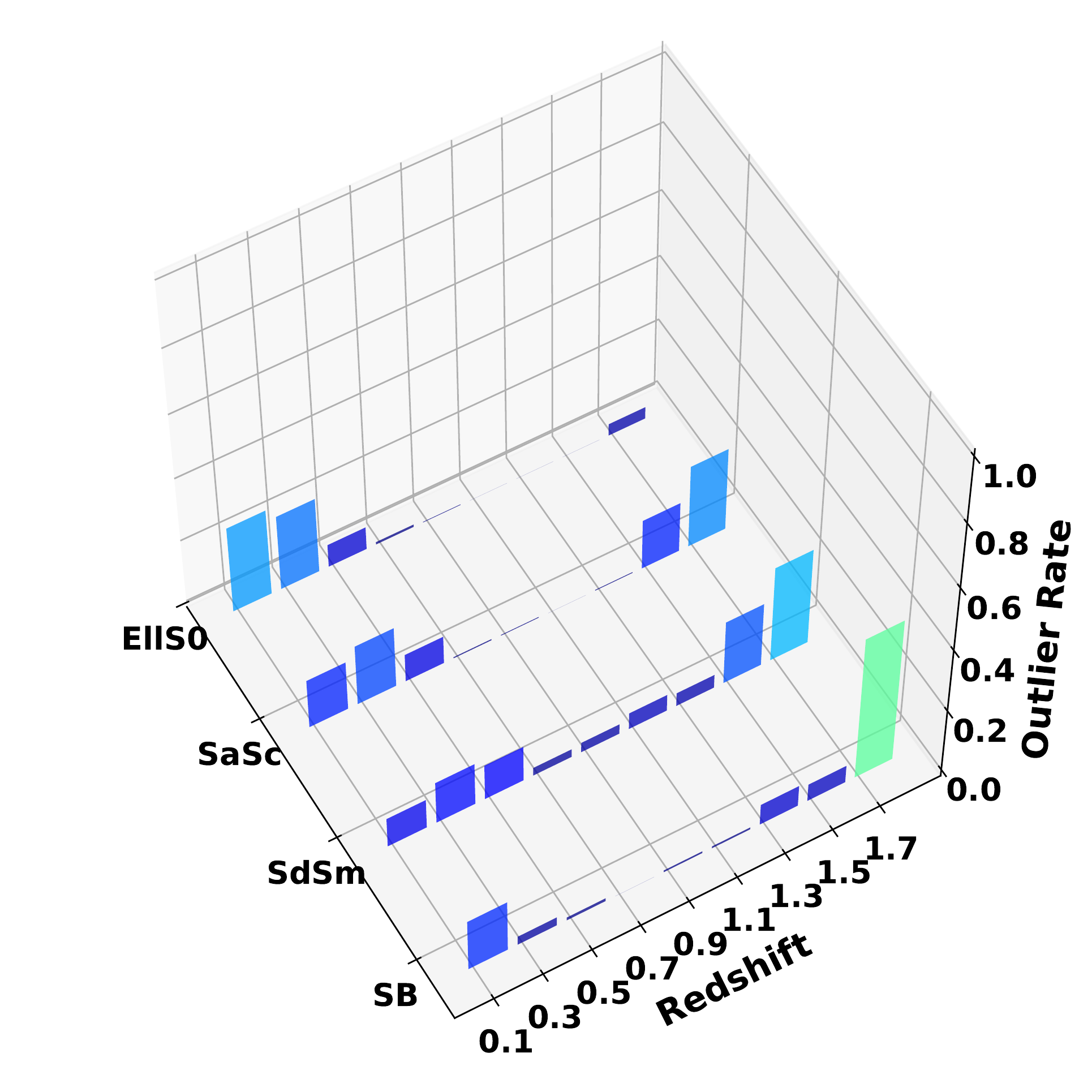}
\caption{Catastrophic failure rate (measured as $|\Delta z / (1+z)|>0.003$) by galaxy type and redshift bins for the best redshift estimation algorithm, in the SNR = 20, 5 and 2 cases (left, middle and right columns, respectively). Degradation of performance at all noise levels has mostly been remedied due to the complementarity of DL and DAE strengths.}
 \label{fig:zbest_3Dgaltypes}
\end{figure*}

Figure \ref{fig:zbest_scatter_all} shows the results of applying this algorithm to the 3 different galaxy samples. Compared to the results of the individual methods shown in Figure \ref{fig:df_scatter_all}, the improvement is manifest. In the SNR = 20 and SNR = 5 cases, where DAE redshifts are very accurate, only a handful of galaxies is selected with DL, which brings marginal improvements to the global catastrophic outlier rates. In the SNR = 2 case, however, the improvement is non-negligible . $71.4\,\%$ of the galaxies have redshifts in agreement with each other. After combination, we obtain a global success rate of $93.15\,\%$. If we had only adopted DL redshifts, the global success rate would have been $89.1\,\%$ - a few percentage points lower - with the larger DL scatter. If instead we wanted to use DAE redshifts to retain a lower scatter, we would be restricted to only $77.7\,\%$ of the sample. Figure \ref{fig:zbest_3Dgaltypes} shows the catastrophic outlier rates for each galaxy type and redshift bin, for each SNR. While SNR = 20 and SNR = 5 are mostly equivalent to DAE results, the SNR = 2 figure shows more specifically how the combination of results helps to reduce the number of catastrophic outliers. In particular, the intermediate redshift elliptical galaxy failures by DL are replaced by DAE better redshifts; conversely, in the lower and higher redshift regimes for the other galaxy types, where DAE fails much more frequently, DL redshifts are retained, bringing catastrophic outlier rates significantly down in those bins when compared to DAE redshift performance. These results clearly demonstrate the advantage of combining the two methods. The exact values of the improvement will depend on the specifics of each simulation or data, but the complementarity is related to the different algorithms and should be qualitatively similar in other data settings.

\section{Conclusion}
\label{sec:conclusion}
In this paper, we introduced two new methods of spectroscopic redshift estimation, and benchmarked them on simulated data against a reference method based on line feature estimation and cross-correlation with eigentemplates. Both new methods rely on deriving an efficient representation for the data then used for redshift estimation. The first one uses the MOD dictionary learning technique to obtain a sparse representation for the full galaxy spectra (continuum and lines), which is then used to estimated redshifts from noisy spectra by searching for the lowest sparse approximation error among all tested redshift values. The second method applies denoising autoencoders for non-linear unsupervised feature extraction, learning the features from rest-frame spectra, and deriving the best-fit redshift value by passing the test spectra through the autoencoder and minimizing the reconstruction error on the input signal.

Both methods show significant improvement over the original Darth Fader pipeline, being able to recover redshift values with high accuracy and precision at high-SNR regimes, with markedly less line confusion. Moreover, the more pronounced the line features on SEDs - as characterized by galaxy type -, the more precise results are. As SNR is reduced, measurement dispersion and catastrophic outlier rates increase as expected. In all sub-cases investigated, denoising autoencoders achieve smaller dispersion around the true redshift value. However, the catastrophic outlier rate increases rapidly as SNR is lowered. On the other hand, the catastrophic outlier rate from sparse dictionary learning is more resilient to the effects of noise, outperforming denoising autoencoders for SNR = 2 and below.

Given those complementary strengths, we design an algorithm to combine DAE and DL results in an optimal way. If they both measure the same correct redshift value, we favor DAE values due to their smaller intrinsic scatter. If they don't agree, we use catastrophic outlier rates  - calibrated with true values - to decide which value to pick. This strategy yields much-improved results: in the SNR = 2 regime, we ensure that $71.4\,\%$ of the galaxies can be identified whose redshifts agree, and this robust sample has a $0.5\,\%$ catastrophic outlier rate. If completeness is a priority, we obtain a global galaxy sample with $6.85\,\%$ catastrophic outlier rate, which is an improvement of $\sim\!\!5\,\%$ over the DL method alone.

These results are encouraging. Fully-automated spectroscopic redshift estimation methods that perform in a robust manner would be of great benefit to upcoming large-scale spectroscopic galaxy surveys such as DESI and Euclid, especially if they work in low SNR regimes. A next step would be to investigate the performance of the algorithms in simulations that fully reproduce the expected data quality from those surveys. Additionally, a combination of methods for different regimes, reliant solely on observed properties, can potentially produce extremely clean and robust redshift catalogs, although this is dependent on the specific properties of the data and noise, and will need to be investigated further.

\begin{acknowledgements}
This work is funded by the DEDALE project (contract no. 665044) and LENA (ERC StG no. 678282) within the H2020 Framework Program of the European Commission. BM thanks Julien Zoubian, St\'{e}phanie Jouvel, Adrienne Leonard and Paul Nathan for useful discussions.
\end{acknowledgements}

{\small 
\bibliographystyle{aa}
\bibliography{refs}}

\begin{thebibliography}{48}
\expandafter\ifx\csname natexlab\endcsname\relax\def\natexlab#1{#1}\fi

\bibitem[{Abadi {et~al.}(2016)Abadi, Barham, Chen, Chen, Davis, Dean, Devin,
  Ghemawat, Irving, Isard, {et~al.}}]{abadi2016tensorflow}
Abadi, M., Barham, P., Chen, J., {et~al.} 2016, in OSDI, Vol.~16, 265--283

\bibitem[{Abell {et~al.}(2009)Abell, Allison, Anderson, Andrew, Angel, Armus,
  Arnett, Asztalos, Axelrod, Bailey, {et~al.}}]{abell2009lsst}
Abell, P.~A., Allison, J., Anderson, S.~F., {et~al.} 2009
  [\eprint[arXiv]{0912.0201}]

\bibitem[{Aharon {et~al.}(2006)Aharon, Elad, \& Bruckstein}]{Aharon06}
Aharon, M., Elad, M., \& Bruckstein, A. 2006, 54, 4311--4322

\bibitem[{Alain \& Bengio(2014)}]{alain2014regularized}
Alain, G. \& Bengio, Y. 2014, The Journal of Machine Learning Research, 15,
  3563

\bibitem[{{Alam} {et~al.}(2017){Alam}, {Ata}, {Bailey}, {Beutler}, {Bizyaev},
  {Blazek}, {Bolton}, {Brownstein}, {Burden}, {Chuang}, {Comparat}, {Cuesta},
  {Dawson}, {Eisenstein}, {Escoffier}, {Gil-Mar{\'{\i}}n}, {Grieb}, {Hand},
  {Ho}, {Kinemuchi}, {Kirkby}, {Kitaura}, {Malanushenko}, {Malanushenko},
  {Maraston}, {McBride}, {Nichol}, {Olmstead}, {Oravetz}, {Padmanabhan},
  {Palanque-Delabrouille}, {Pan}, {Pellejero-Ibanez}, {Percival}, {Petitjean},
  {Prada}, {Price-Whelan}, {Reid}, {Rodr{\'{\i}}guez-Torres}, {Roe}, {Ross},
  {Ross}, {Rossi}, {Rubi{\~n}o-Mart{\'{\i}}n}, {Saito}, {Salazar-Albornoz},
  {Samushia}, {S{\'a}nchez}, {Satpathy}, {Schlegel}, {Schneider},
  {Sc{\'o}ccola}, {Seo}, {Sheldon}, {Simmons}, {Slosar}, {Strauss}, {Swanson},
  {Thomas}, {Tinker}, {Tojeiro}, {Maga{\~n}a}, {Vazquez}, {Verde}, {Wake},
  {Wang}, {Weinberg}, {White}, {Wood-Vasey}, {Y{\`e}che}, {Zehavi}, {Zhai}, \&
  {Zhao}}]{Alam2017}
{Alam}, S., {Ata}, M., {Bailey}, S., {et~al.} 2017, \mnras, 470, 2617

\bibitem[{{Baldry} {et~al.}(2004){Baldry}, {Glazebrook}, {Brinkmann},
  {Ivezi{\'c}}, {Lupton}, {Nichol}, \& {Szalay}}]{Baldry04}
{Baldry}, I.~K., {Glazebrook}, K., {Brinkmann}, J., {et~al.} 2004, \apj, 600,
  681

\bibitem[{{Bautista} {et~al.}(2018){Bautista}, {Vargas-Maga{\~n}a}, {Dawson},
  {Percival}, {Brinkmann}, {Brownstein}, {Camacho}, {Comparat},
  {Gil-Mar{\'{\i}}n}, {Mueller}, {Newman}, {Prakash}, {Ross}, {Schneider},
  {Seo}, {Tinker}, {Tojeiro}, {Zhai}, \& {Zhao}}]{Bautista2018}
{Bautista}, J.~E., {Vargas-Maga{\~n}a}, M., {Dawson}, K.~S., {et~al.} 2018,
  \apj, 863, 110

\bibitem[{Bengio {et~al.}(2013)Bengio, Courville, \&
  Vincent}]{bengio2013representation}
Bengio, Y., Courville, A., \& Vincent, P. 2013, Pattern Analysis and Machine
  Intelligence, IEEE Transactions on, 35, 1798

\bibitem[{{Blake} \& {Glazebrook}(2003)}]{BlakeGlazebrook2003}
{Blake}, C. \& {Glazebrook}, K. 2003, \apj, 594, 665

\bibitem[{{Bolton} {et~al.}(2012){Bolton}, {Schlegel}, {Aubourg}, {Bailey},
  {Bhardwaj}, {Brownstein}, {Burles}, {Chen}, {Dawson}, {Eisenstein}, {Gunn},
  {Knapp}, {Loomis}, {Lupton}, {Maraston}, {Muna}, {Myers}, {Olmstead},
  {Padmanabhan}, {P{\^a}ris}, {Percival}, {Petitjean}, {Rockosi}, {Ross},
  {Schneider}, {Shu}, {Strauss}, {Thomas}, {Tremonti}, {Wake}, {Weaver}, \&
  {Wood-Vasey}}]{Bolton2012}
{Bolton}, A.~S., {Schlegel}, D.~J., {Aubourg}, {\'E}., {et~al.} 2012, \aj, 144,
  144

\bibitem[{Bourlard \& Kamp(1988)}]{bourlardkamp88}
Bourlard, H. \& Kamp, Y. 1988, Biological Cybernetics, 59, 291

\bibitem[{{Bruzual A.} \& {Charlot}(1993)}]{Bruzual1993}
{Bruzual A.}, G. \& {Charlot}, S. 1993, \apj, 405, 538

\bibitem[{Coates {et~al.}(2011)Coates, Ng, \& Lee}]{coates2011analysis}
Coates, A., Ng, A.~Y., \& Lee, H. 2011, in International conference on
  artificial intelligence and statistics, 215--223

\bibitem[{{Coleman} {et~al.}(1980){Coleman}, {Wu}, \& {Weedman}}]{Coleman1980}
{Coleman}, G.~D., {Wu}, C.-C., \& {Weedman}, D.~W. 1980, \apjs, 43, 393

\bibitem[{Collobert {et~al.}(2011)Collobert, Weston, Bottou, Karlen,
  Kavukcuoglu, \& Kuksa}]{collobert2011natural}
Collobert, R., Weston, J., Bottou, L., {et~al.} 2011, The Journal of Machine
  Learning Research, 12, 2493

\bibitem[{Dahl {et~al.}(2012)Dahl, Yu, Deng, \& Acero}]{dahl2012context}
Dahl, G.~E., Yu, D., Deng, L., \& Acero, A. 2012, Audio, Speech, and Language
  Processing, IEEE Transactions on, 20, 30

\bibitem[{{Dawson} {et~al.}(2016){Dawson}, {Kneib}, {Percival}, {Alam},
  {Albareti}, {Anderson}, {Armengaud}, {Aubourg}, {Bailey}, {Bautista},
  {Berlind}, {Bershady}, {Beutler}, {Bizyaev}, {Blanton}, {Blomqvist},
  {Bolton}, {Bovy}, {Brandt}, {Brinkmann}, {Brownstein}, {Burtin}, {Busca},
  {Cai}, {Chuang}, {Clerc}, {Comparat}, {Cope}, {Croft}, {Cruz-Gonzalez}, {da
  Costa}, {Cousinou}, {Darling}, {de la Macorra}, {de la Torre}, {Delubac}, {du
  Mas des Bourboux}, {Dwelly}, {Ealet}, {Eisenstein}, {Eracleous}, {Escoffier},
  {Fan}, {Finoguenov}, {Font-Ribera}, {Frinchaboy}, {Gaulme}, {Georgakakis},
  {Green}, {Guo}, {Guy}, {Ho}, {Holder}, {Huehnerhoff}, {Hutchinson}, {Jing},
  {Jullo}, {Kamble}, {Kinemuchi}, {Kirkby}, {Kitaura}, {Klaene}, {Laher},
  {Lang}, {Laurent}, {Le Goff}, {Li}, {Liang}, {Lima}, {Lin}, {Lin}, {Lin},
  {Long}, {Lundgren}, {MacDonald}, {Geimba Maia}, {Malanushenko},
  {Malanushenko}, {Mariappan}, {McBride}, {McGreer}, {M{\'e}nard}, {Merloni},
  {Meza}, {Montero-Dorta}, {Muna}, {Myers}, {Nandra}, {Naugle}, {Newman},
  {Noterdaeme}, {Nugent}, {Ogando}, {Olmstead}, {Oravetz}, {Oravetz},
  {Padmanabhan}, {Palanque- Delabrouille}, {Pan}, {Parejko}, {P{\^a}ris},
  {Peacock}, {Petitjean}, {Pieri}, {Pisani}, {Prada}, {Prakash}, {Raichoor},
  {Reid}, {Rich}, {Ridl}, {Rodriguez-Torres}, {Carnero Rosell}, {Ross},
  {Rossi}, {Ruan}, {Salvato}, {Sayres}, {Schneider}, {Schlegel}, {Seljak},
  {Seo}, {Sesar}, {Shandera}, {Shu}, {Slosar}, {Sobreira}, {Streblyanska},
  {Suzuki}, {Taylor}, {Tao}, {Tinker}, {Tojeiro}, {Vargas-Maga{\~n}a}, {Wang},
  {Weaver}, {Weinberg}, {White}, {Wood-Vasey}, {Yeche}, {Zhai}, {Zhao}, {Zhao},
  {Zheng}, {Ben Zhu}, \& {Zou}}]{Dawson2016}
{Dawson}, K.~S., {Kneib}, J.-P., {Percival}, W.~J., {et~al.} 2016, \aj, 151, 44

\bibitem[{{DESI Collaboration} {et~al.}(2016){DESI Collaboration}, {Aghamousa},
  {Aguilar}, {Ahlen}, {Alam}, {Allen}, {Allende Prieto}, {Annis}, {Bailey},
  {Balland}, \& et~al.}]{Aghamousa2016}
{DESI Collaboration}, {Aghamousa}, A., {Aguilar}, J., {et~al.} 2016, ArXiv
  e-prints [\eprint[arXiv]{1611.00037}]

\bibitem[{Elad \& Aharon(2006)}]{Elad06}
Elad, M. \& Aharon, M. 2006, IEEE Transactions on Image processing, 15, 3736

\bibitem[{Engan {et~al.}(1999)Engan, Aase, \& Husoy}]{Engan99}
Engan, K., Aase, S.~O., \& Husoy, J.~H. 1999, in Acoustics, Speech, and Signal
  Processing, 1999. Proceedings., 1999 IEEE International Conference on,
  Vol.~5, IEEE, 2443--2446

\bibitem[{Frontera-Pons {et~al.}(2017)Frontera-Pons, Sureau, Bobin, \&
  Le~Floc’h}]{frontera2017unsupervised}
Frontera-Pons, J., Sureau, F., Bobin, J., \& Le~Floc’h, E. 2017, Astronomy \&
  Astrophysics, 603, A60

\bibitem[{Glazebrook {et~al.}(1998)Glazebrook, Offer, \& Deeley}]{Glazebrook98}
Glazebrook, K., Offer, A.~R., \& Deeley, K. 1998, Astrophys. J., 492, 98

\bibitem[{Hinton {et~al.}(2012)Hinton, Deng, Yu, Dahl, Mohamed, Jaitly, Senior,
  Vanhoucke, Nguyen, Sainath, {et~al.}}]{hinton2012deep}
Hinton, G., Deng, L., Yu, D., {et~al.} 2012, Signal Processing Magazine, IEEE,
  29, 82

\bibitem[{{Hinton} {et~al.}(2016){Hinton}, {Davis}, {Lidman}, {Glazebrook}, \&
  {Lewis}}]{Hinton2016}
{Hinton}, S.~R., {Davis}, T.~M., {Lidman}, C., {Glazebrook}, K., \& {Lewis},
  G.~F. 2016, Astronomy and Computing, 15, 61

\bibitem[{{Hutchinson} {et~al.}(2016){Hutchinson}, {Bolton}, {Dawson}, {Allende
  Prieto}, {Bailey}, {Bautista}, {Brownstein}, {Conroy}, {Guy}, {Myers},
  {Newman}, {Prakash}, {Carnero-Rosell}, {Seo}, {Tojeiro}, {Vivek}, \& {Ben
  Zhu}}]{Hutchinson2016}
{Hutchinson}, T.~A., {Bolton}, A.~S., {Dawson}, K.~S., {et~al.} 2016, \aj, 152,
  205

\bibitem[{Ilbert {et~al.}(2008)Ilbert, Capak, Salvato, Aussel, McCracken,
  Sanders, Scoville, Kartaltepe, Arnouts, Le~Floc'h,
  {et~al.}}]{ilbert2008cosmos}
Ilbert, O., Capak, P., Salvato, M., {et~al.} 2008, The Astrophysical Journal,
  690, 1236

\bibitem[{Jouvel {et~al.}(2009)Jouvel, Kneib, Ilbert, Bernstein, Arnouts,
  Dahlen, Ealet, Milliard, Aussel, Capak, {et~al.}}]{jouvel2009designing}
Jouvel, S., Kneib, J.-P., Ilbert, O., {et~al.} 2009, Astronomy \& Astrophysics,
  504, 359

\bibitem[{{Kazin} {et~al.}(2014){Kazin}, {Koda}, {Blake}, {Padmanabhan},
  {Brough}, {Colless}, {Contreras}, {Couch}, {Croom}, {Croton}, {Davis},
  {Drinkwater}, {Forster}, {Gilbank}, {Gladders}, {Glazebrook}, {Jelliffe},
  {Jurek}, {Li}, {Madore}, {Martin}, {Pimbblet}, {Poole}, {Pracy}, {Sharp},
  {Wisnioski}, {Woods}, {Wyder}, \& {Yee}}]{Kazin2014}
{Kazin}, E.~A., {Koda}, J., {Blake}, C., {et~al.} 2014, \mnras, 441, 3524

\bibitem[{{Kennicutt}(1998)}]{Kennicutt1998}
{Kennicutt}, Jr., R.~C. 1998, \araa, 36, 189

\bibitem[{Krizhevsky {et~al.}(2012)Krizhevsky, Sutskever, \&
  Hinton}]{krizhevsky2012imagenet}
Krizhevsky, A., Sutskever, I., \& Hinton, G.~E. 2012, in Advances in neural
  information processing systems, 1097--1105

\bibitem[{{Leauthaud} {et~al.}(2007){Leauthaud}, {Massey}, {Kneib}, {Rhodes},
  {Johnston}, {Capak}, {Heymans}, {Ellis}, {Koekemoer}, {Le F{\`e}vre},
  {Mellier}, {R{\'e}fr{\'e}gier}, {Robin}, {Scoville}, {Tasca}, {Taylor}, \&
  {Van Waerbeke}}]{leauthaud2007}
{Leauthaud}, A., {Massey}, R., {Kneib}, J.-P., {et~al.} 2007, \apjs, 172, 219

\bibitem[{Machado {et~al.}(2013)Machado, Leonard, Starck, Abdalla, \&
  Jouvel}]{Machado13}
Machado, D.~P., Leonard, A., Starck, J.~L., Abdalla, F.~B., \& Jouvel, S. 2013,
  Astron. Astrophys., 560, A83

\bibitem[{Mairal {et~al.}(2008)Mairal, Elad, \& Sapiro}]{Mairal08a}
Mairal, J., Elad, M., \& Sapiro, G. 2008, {IEEE} Transactions on Image
  Processing, 17, 53--69

\bibitem[{Mairal {et~al.}(2009)Mairal, Ponce, Sapiro, Zisserman, \&
  Bach}]{Mairal09}
Mairal, J., Ponce, J., Sapiro, G., Zisserman, A., \& Bach, F.~R. 2009, in
  Advances in neural information processing systems, 1033--1040

\bibitem[{Mallat \& Zhang(1993)}]{Mallat93}
Mallat, S. \& Zhang, Z. 1993, 41, 3397--3415

\bibitem[{{Moustakas} {et~al.}(2006){Moustakas}, {Kennicutt}, \&
  {Tremonti}}]{Moustakas2006}
{Moustakas}, J., {Kennicutt}, Jr., R.~C., \& {Tremonti}, C.~A. 2006, \apj, 642,
  775

\bibitem[{Olshausen \& Field(1996)}]{Olshausen97}
Olshausen, B. \& Field, D. 1996, Vision Research., 37, 3311--3325

\bibitem[{Pati \& Krishnaprasad(1993)}]{Pati93}
Pati, Y.~C. \& Krishnaprasad, P.~S. 1993, IEEE Transactions on Neural Networks,
  4, 73--85

\bibitem[{Rifai {et~al.}(2011)Rifai, Vincent, Muller, Glorot, \&
  Bengio}]{rifai2011contractive}
Rifai, S., Vincent, P., Muller, X., Glorot, X., \& Bengio, Y. 2011, in
  Proceedings of the 28th international conference on machine learning
  (ICML-11), 833--840

\bibitem[{{Seo} \& {Eisenstein}(2003)}]{SeoEisenstein2003}
{Seo}, H.-J. \& {Eisenstein}, D.~J. 2003, \apj, 598, 720

\bibitem[{Smee {et~al.}(2013)Smee, Gunn, Uomoto, Roe, Schlegel, Rockosi, Carr,
  Leger, Dawson, Olmstead, Brinkmann, Owen, Barkhouser, Honscheid, Harding,
  Long, Lupton, Loomis, Anderson, Annis, Bernardi, Bhardwaj, Bizyaev, Bolton,
  Brewington, Briggs, Burles, Burns, Castander, Connolly, Davenport, Ebelke,
  Epps, Feldman, Friedman, Frieman, Heckman, Hull, Knapp, Lawrence, Loveday,
  Mannery, Malanushenko, Malanushenko, Merrelli, Muna, Newman, Nichol, Oravetz,
  Pan, Pope, Ricketts, Shelden, Sandford, Siegmund, Simmons, Smith, Snedden,
  Schneider, SubbaRao, Tremonti, Waddell, \& York}]{Smee2013}
Smee, S.~A., Gunn, J.~E., Uomoto, A., {et~al.} 2013, The Astronomical Journal,
  146, 32

\bibitem[{{Starck} \& {Murtagh}(2006)}]{Starck2006}
{Starck}, J.-L. \& {Murtagh}, F. 2006, {Astronomical Image and Data Analysis}

\bibitem[{Starck {et~al.}(2015)Starck, Murtagh, \& Fadili}]{Starck2015}
Starck, J.-L., Murtagh, F., \& Fadili, J. 2015, Sparse Image and Signal
  Processing: Wavelets and Related Geometric Multiscale Analysis, 2nd edn.
  (Cambridge University Press)

\bibitem[{{Starck} {et~al.}(1996){Starck}, {Murtagh}, {Pirenne}, \&
  {Albrecht}}]{Starck1996}
{Starck}, J.-L., {Murtagh}, F., {Pirenne}, B., \& {Albrecht}, M. 1996, \pasp,
  108, 446

\bibitem[{Tuccillo {et~al.}(2017)Tuccillo, Huertas-Company, Decenci{\`e}re,
  Velasco-Forero, Dom{\'\i}nguez~S{\'a}nchez, \& Dimauro}]{tuccillo2017deep}
Tuccillo, D., Huertas-Company, M., Decenci{\`e}re, E., {et~al.} 2017, Monthly
  Notices of the Royal Astronomical Society, 475, 894

\bibitem[{Vincent {et~al.}(2008)Vincent, Larochelle, Bengio, \&
  Manzagol}]{vincent2008extracting}
Vincent, P., Larochelle, H., Bengio, Y., \& Manzagol, P.-A. 2008, in
  Proceedings of the 25th international conference on Machine learning, ACM,
  1096--1103

\bibitem[{Vincent {et~al.}(2010)Vincent, Larochelle, Lajoie, Bengio, \&
  Manzagol}]{vincent2010stacked}
Vincent, P., Larochelle, H., Lajoie, I., Bengio, Y., \& Manzagol, P.-A. 2010,
  The Journal of Machine Learning Research, 11, 3371

\bibitem[{Zhang \& Li(2010)}]{Zhang10}
Zhang, Q. \& Li, B. 2010, in Computer Vision and Pattern Recognition (CVPR),
  2010 IEEE Conference on, IEEE, 2691--2698

\end{thebibliography}

\end{document}